\documentclass{aa}
\usepackage{verbatim,float}
\usepackage{times}
\usepackage{graphics}
\usepackage{epsfig}

\begin{document}


\thesaurus{05    
              (08.01.1;  
               11.01.1;  
               11.05.2;  
               11.13.1;  
               11.19.4;  
               11.19.5)} 

\title{Age and metallicity for six LMC clusters and their surrounding field population}

\author{B. Dirsch \inst{1}, T. Richtler \inst{2},
        W.P. Gieren \inst{2}, M. Hilker \inst{3}} 
\offprints{bdirsch@astro.uni-bonn.de}
\institute{Sternwarte der Universit\"at Bonn, 
           Auf dem H\"ugel 71, 53121 Bonn, Germany
        \and
          Universidad de Concepci\'on, Departamento de F\'{\i}sica, Casilla 160-C, Concepci\'on, Chile
        \and
          P. Universidad Cat\'olica, Departamento de Astronom\'{\i}a y Astrof\'{\i}sica, Casilla 104, Santiago 22, Chile}
\titlerunning{Ages \& Metallicities of LMC stars}
\maketitle
%
%
\begin{abstract}
  We investigate, on the basis of CCD Str\"omgren photometry, the ages
  and metallicities of six LMC clusters together with their
  surrounding field population. The clusters and metallicities are:
  NGC~1651 (in the range $[Fe/H]=-0.65$ dex to $-0.41$ dex), NGC~1711
  ($-0.57\pm0.17$ dex), NGC~1806 ($-0.71\pm0.23$ dex), NGC~2031
  ($-0.52\pm0.21$ dex) and NGC~2136/37 ($-0.55\pm0.23$ dex) and
  NGC~2257 ($-1.63\pm0.21$ dex).  The metallicities for NGC~1651,
  NGC~1711, NGC~1806 and NGC~2031 have been determined for the first
  time (NGC~2031 and NGC~2136/37 are interesting for the Cepheid
  distance scale).
  
  In the cluster surroundings, we found about 650 field stars that
  were suitable to be used for a determination of an age-metallicity
  relation (AMR).  Our method is to estimate ages for individual
  stars on the basis of Str\"omgren isochrones with individually
  measured metallicities.  With this method we are able to sample the
  AMR of the field population up to 8 Gyr.
  
  Our metallicity data are incompatible with models predicting many
  metal-poor stars (G-dwarf problem). The metallicity of the field
  population increased by a factor of six, starting around 2 Gyr ago.
  The proposed AMR is consistent with the AMR of the LMC cluster
  system (including ESO 121 SC03 and three clusters with an age of 4
  Gyr). 

  The proposed AMR is incompatible with the
  recently proposed AMR by Pagel \& Tautvai\u{s}vien\.e
  (\cite{pagel98}).


\end{abstract}
%
%

\section{Introduction}
\label{chap:intro}

In spite of its enormous importance for understanding galaxy evolution
in adequate detail, the chemical enrichment process in galaxies is
still poorly known, which is especially true for the field star
component.  The Large Magellanic Cloud (LMC) is a natural target to
study the chemical evolution because of its proximity. Also, its
structure seems to be less complex than that of the Milky Way which
might imply that the chemical enrichment history can be described by a
simple global age-metallicity relationship (AMR).

First efforts to determine the AMR of LMC clusters have been made with
integrated broad band photometry of clusters (Westerlund
\cite{westerlund97}).  Recent work continuing these studies is, for
example, Bica et al. (\cite{bica98}) and Girardi et al.
(\cite{girardi95}). Another major step towards an understanding of the
LMC cluster AMR has been undertaken by Olszewski et al.
(\cite{olszewski91}), who used medium resolution spectroscopy of
individual giants to measure the metallicity for around 70 clusters,
with a quoted uncertainty of $\pm 0.2$ dex.  In addition many
photometric studies of stars in LMC clusters (e.g. with the Washington
system by Bica et al. \cite{bica98}) contributed to the unveiling of
the cluster AMR.

The current wisdom on the cluster AMR that has been established by
these studies is, that the mean metallicity of younger clusters is
distinctly higher than that of old clusters by more than 1.2~dex.
However, it is difficult to trace the AMR over the entire LMC history
with this cluster sample, since for a long time, only one cluster
(ESO121-SC03) with an age between 3~Gyr and 11~Gyr had been found
(Mateo et al. \cite{mateo86}, Bica et al. \cite{bica98}).  Recently,
Sarajedini (\cite{sarajedini98}) found three more clusters with an age
of about 4~Gyr (NGC~2121, NGC~2155 and SL~663).

The AMR as derived from LMC clusters shows a very large scatter
(Olszewski et al.  ~\cite{olszewski91}), which, if intrinsic and not
due to measurement uncertainties, would argue for a more complex
chemical enrichment history.  In addition there are hints that at
least some clusters have smaller mean metallicities than the
surrounding field population (e.g. Bica et al.~\cite{bica98}, Richtler
et al. \cite{richtler89}). Thus possibly the chemical evolution of the
cluster and field stars is to some degree decoupled. However, this is
not without contradiction (e.g. Korn et al. \cite{korn00}).  Santos
Jr. et al. (\cite{santos99}) claimed that the metallicity dispersion
of the field seems to be smaller than that of the cluster system of
similar age.

For the field population the metallicity distribution is known
primarily for the young stars since mainly F \& G supergiants have
been spectroscopically investigated (e.g. Hill et al. \cite{hillV95},
Luck \& Lambert \cite{luck92}, Russell \& Bessell \cite{russell89}). A
compilation of young LMC field stars abundances which have been
derived with high resolution spectroscopy can be found in the appendix
(Table~\ref{tab:fieldme}).  Th\'evenin \& Jasniewicz
(\cite{thevenin92}) study 9 field stars in the LMC with medium
resolution spectroscopy (5 $\AA$) and found an average abundance of
$[Fe/H]=-0.25\pm0.08$ which is higher than the mean value of field
stars that has been derived with high resolution spectroscopy
($-0.38\pm0.11$ dex). Dopita et al. (\cite{dopita97}) measured element
abundances of planetary nebulae (PNs) in the LMC and derived their age
by modelling the hot, central star. They found four PNs that are older
than $4$ Gyr. Their AMR shows only little enrichment from $15$ to $5$
Gyr ago, while the metallicity doubled in the last $2-3$ Gyr.  The
study of the older stellar field component has been limited to studies
using broad band photometry (e.g. Holtzman et al. \cite{holtzman99}
and Elson et al. \cite{elson97}).

We used a different approach and measured the metallicity of
individual stars by using the medium wide Str\"omgren filter system,
that gives a good metallicity discrimination for giants and
supergiants red-wards of $b-y = 0.4$~mag. This method has already been
used by Grebel \& Richtler (\cite{grebel92}), Hilker et al.
(\cite{hilker95b}) and Hilker et al.  (\cite{hilker95a}) to determine
age and metallicity of NGC~330, NGC~1866 and NGC~2136/37. Ardeberg et
al. (\cite{ardeberg97}) used HST observations transformed into the
Str\"omgren system to derive the SFH and the metallicity of LMC bar
stars. Their investigation differs from our approach by 
the calibration they employed which is based on bluer stars and 
includes the gravity
dependent $c1$ Str\"omgren colour index.

In the current work we investigate mainly young LMC clusters and their
surrounding fields, namely NGC~1651, NGC~1711, NGC~1806, NGC~2031, and
NGC~2257, an old cluster.  We have also re-analysed NGC~2136/37
because of the availability of Str\"omgren isochrones and a new
calibration for photometric metallicities, which improves the
calibration for more metal poor stars. This ensures the homogeneity of
the sample and also tests if systematic shifts are present between the
older investigations and the new one.  An important aspect of this new
work is exactly this homogeneity of the metallicities allowing one to
assess the real magnitude of the intrinsic dispersion among
metallicities of clusters of similar age.

Two of the clusters (NGC~2136 and NGC~2031) are particularly
interesting because they contain Cepheid variables, whose
metallicities are important to know for distance scale problems.
NGC~1866 might serve as example.  Its metallicity has been determined
by Hilker et al. (\cite{hilker95b}) via Str\"omgren photometry which
was used for the distance determination using its Cepheid members by
Gieren et al. (\cite{gieren94}).

\section{Data \& Reduction}

The data have been obtained during two observing runs with the 1.54-m
Danish telescope at La Silla, Chile. NGC~2136 and NGC~2031 were
observed during 13.11. - 15.11.1992, and NGC~1651, NGC~1711, NGC~1806,
NGC~2257 during 4.1. - 7.1.1994. The observing log is shown in
Table~\ref{tab:log} in the Appendix.
\begin{table}[t]
\caption{Coordinates of the investigated clusters}
\label{tab:clustercoo}
\begin{tabular}{c|cccc}
\hline
Cluster & $\alpha_{2000}$ & $\delta_{2000}$ & $l$ & $b$  \\\hline
NGC~1651 & $4^h 37^m 12^s$ & $-70^0 33\arcsec$ & $282.75^0$ & $-36.42^0$ \\
NGC~1711 & $4^h 50^m 36^s$ & $-69^0 59\arcsec$ & $281.61^0$ & $-35.57^0$ \\
NGC~1806 & $5^h 2^m 11^s$ & $-67^0 03\arcsec$ & $278.89^0$ & $-35.16^0$ \\
NGC~2031 & $5^h 33^m 36^s$ & $-70^0 59\arcsec$ & $281.73$ & $-31.85^0$ \\
NGC~2136 & $5^h 53^m 17^s$ & $-69^0 32\arcsec$ & $279.83$ & $-30.35^0$ \\
NGC~2257 & $6^h 30^m 24^s$ & $-64^0 17\arcsec$ & $274.10$ & $-26.51^0$ \\\hline
\end{tabular}
\end{table}
We used the UV coated Thomson THX 31560 chip, that has a field of view
of $6.5' \times 6.5'$ and a scale of 0.377\arcsec /pixel. The Danish
imaging Str\"omgren filters $v$, $b$ and $y$ were used.  The
measurements in these filters have been transformed into a Johnson $V$
magnitude, the colour $b-y$ and the colour index $m1=(v-b)-(b-y)$.

Except for NGC~1711 each cluster has been observed on at least two
different nights to have photometrically independent measurements. The
reduction included bias subtraction, flat-field correction and the
elimination of CCD defects. We used DaoPhot II in the MIDAS and IRAF
environments for the photometry.

For the calibration we employed six different E-region standard stars
from the list of J{\o}nch-S{\o}rensen (\cite{jonchsorensen93}) and
four fields with secondary Str\"omgren standards measured by Richtler
(\cite{richtler90}), namely M~67, SK~-66~80, NGC~2257 and NGC~330.
The heavily crowded field of NGC~330 with the secondary standards
measured with photoelectric photometry is problematic since slight
differences in centring the aperture on a standard star lead to
deviations in the obtained magnitude on the order of $0.02$~mag. Thus
several stars in this field have been removed from the calibration.
The photometric error of a single measurement is given by the standard
deviation of the standard stars: $\sigma_y = 0.031$, $\sigma_{b-y} =
0.028$ and $\sigma_{m1} = 0.036$. We obtained the errors by averaging
the residuals of the calibration stars in all nights.  The calibration
errors for the 1992 run are smaller: $\sigma_y = 0.016$, $\sigma_{b-y}
= 0.023$ $\sigma_{m1} = 0.025$. However, since standard stars are
always measured in the central region of the CCD chip, this error does
not include flat-field errors which are much harder to quantify
(especially also due to the problematic field concentration in
telescopes with focal reducers (Andersen et al. \cite{andersen95}).
From inspection of the sky background the accuracy of the flat-field
is $\simeq 1-2\%$ and we therefore assigned an additional error of
$0.015$ to each magnitude.  The photometric standards were measured
with apertures and thus it was necessary to determine the aperture -
PSF shift carefully. The remaining uncertainty is of the order of
$0.03$~mag.  Even if several nights have been averaged this
calibration error has been kept, since the calibrations in each night
are not truly independent: the colour terms have been determined using
the standard star observations from all nights together. Thus we
overestimated the calibration error for the clusters by a factor
$<\sqrt{nights}$ if observations from several nights have been
averaged.

\begin{figure}
\resizebox{\hsize}{!}{\includegraphics{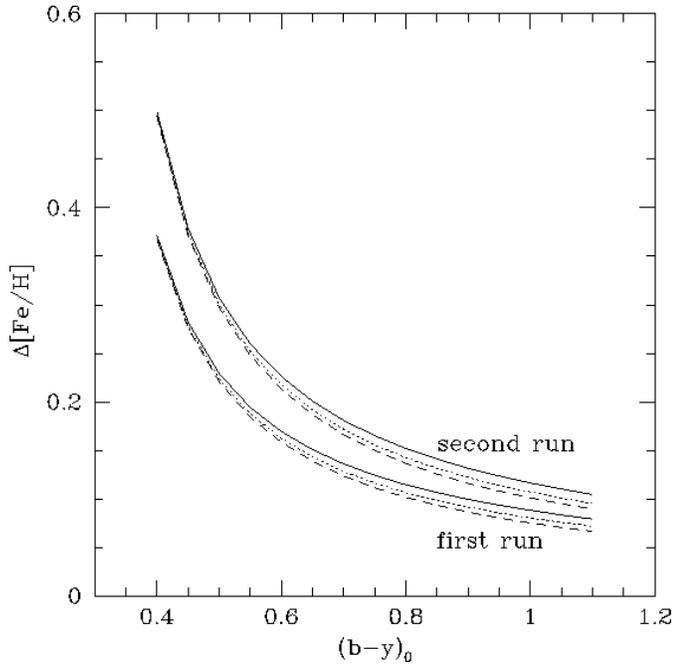}}
\caption{
  The calibration error results in an error of the measured
  metallicity that depends on the colour of the star. In this graph we
  plotted the metallicity error due to this calibration error versus
  the star's colour, each curve sample corresponds to one run (1994 \&
  1992).  Lines are drawn for 0, $-1$ and $-2$ dex stars (solid lines,
  short dashed lines, long dashed lines)}
\label{fig:error}
\end{figure}

The calibration error causes the deviation of the measured metallicity
from the ``true'' metallicity of a star to be a function of its
colour (shown in Fig.~\ref{fig:error}). The corresponding metallicity
error is larger for blue stars since lines of constant metallicity
approach each other on the blue side (see Fig.~\ref{fig:1711cme} for
illustration). In the following stars bluer than $b-y = 0.6$ are
excluded from the metallicity determination because of this strong
rise of the metallicity uncertainty.

\section{Metallicity determination via Str\"omgren colours}

The major advantage of the Str\"omgren system compared to broad band
photometric systems is the ability to get the metallicity of a star
nearly independent of its age The reason for this independence is
  the minor luminosity effect in the metallicity determination, which
  amounts to less than $\pm 0.1$~dex over a luminosity interval of $-4
  < M_V < 3$.

We have used a new metallicity calibration of the Str\"omgren
$m1-(b-y)$ two-colour relation by Hilker (\cite{hilker99}) which is
valid in the colour range $0.5 < b-y < 1.1 $.  For redder stars the
calibration breaks down due to the onset of absorption by TiO and MgH
molecules in the $y$ band. The used calibration equation is
\begin{center}
\[\left[\frac{Fe}{H}\right] = \frac{m1_0+a1\!\cdot\!(b-y)_0+a2}{a3\!\cdot\!(b-y)_0+a4}\]
\end{center}
with 
\begin{eqnarray*}
a1\!=\!-1.277\pm0.050,\; a2\!=\!0.331\pm0.035\;& &\\ 
a3\!=\!0.324\pm0.035,\; a4\!=\!-0.032\pm0.025 & &
\end{eqnarray*}

This calibration has been derived using primarily giant stars, however
as investigated by Grebel~\&~Richtler (\cite{grebel92}) it should also
apply to supergiants.  For stars bluer $b-y=0.7$ this has been
predicted by Gustafsson~\&~Bell (\cite{gustafsson79}).  We will
discuss this question in greater detail in Sect.~3.4.

The reason for the metal sensitivity is the line blocking in the $v$
filter, which is best measurable for G and K stars. The measured flux
depends largely on the strength of the Fe I lines, but also CN and CH
bands contribute.  Systematic deviations of less than $0.1$ dex are
expected from theoretical isochrones due to a small luminosity
dependence.

\subsection{The CN anomaly}

A severe problem in the interpretation of Str\"omgren colours is the
contribution of the CN molecule absorption (band head at $421.5$ nm)
in the $v$-filter ($410$ nm, width $20$ nm) to the line blocking. CN
variations have been observed in several galactic globular clusters,
however, the exact mechanism is not yet fully understood.  An
increased CN abundance leads to an increased photometric metallicity
and thus to a decreased age if it is derived via isochrones. As a rule
of thumb we estimated with the aid of Geneva isochrones that an
increased metallicity of $0.2$~dex will decrease the age by $\approx
20\%$.  A recent investigation using Str\"omgren photometry of two
globular clusters, of which one has CN anomalous stars, the other not,
illustrates the effect of CN anomaly on the Str\"omgren metallicity
(Richter et al. \cite{richter99}).

We cannot account for this CN anomaly. In this study we have to live
with this uncertainty, but there is evidence that this effect is only
modest: in the nearby giant sample (see next section) three stars were
assigned to be CN enriched, but they do not deviate within the
standard deviation from the CN-normal stars, however most probably due
to the uncertain reddening correction.  Pilachowski et al.
(\cite{pilachowski96}) found that for population II halo stars the CN
anomaly does not play an important role, in contrast to M 13 for
example. This cannot be explained by simple selection effects.
Mc~Gregor~\&~Hyland (\cite{mcgregor84}) found a general CN deficiency
by weaker CO bands in the LMC than in galactic supergiants of the same
temperature.  Concerning our LMC field stars we found a good agreement
for the young stars with spectroscopic analyses as well as for the
older field population with observed clusters (see below).  Therefore
anomalous CN abundances should not play a devastating role.
Ultimately this can only be checked with a spectroscopic investigation
of the CN behaviour of cool LMC giants and supergiants.

\subsection{The influence of reddening uncertainties}

\begin{figure}[t]
\resizebox{\hsize}{!}{\includegraphics{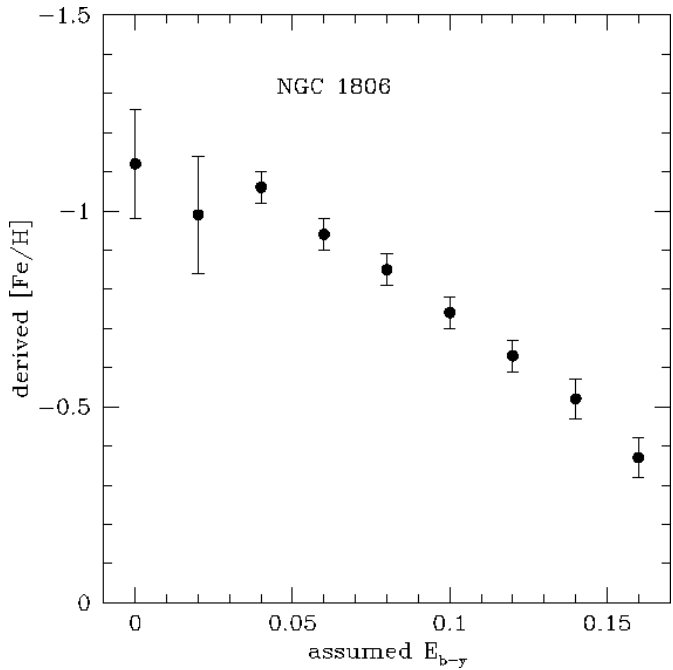}}
\caption{
  The metallicity of NGC~1806 derived from the $m1$,$(b-y)$ diagram is
  plotted as a function of the applied reddening. The errors have been
  derived by dividing the standard deviation of the mean metallicity
  with the square-root of the number of stars used for the metallicity
  determination.}
\label{fig:1806redme}
\end{figure}

Photometrically measured metallicities are very sensitive to reddening
errors, which is a major error source of the determined metallicities.
For example an underestimation of the reddening of 0.02 mag in
$E_{b-y}$ leads to an average underestimation of the metallicity by
$\Delta [Fe/H] = 0.1$ dex for a fully populated RGB with an age around
$10^{9.0}$ yr. This problem is symptomatic for photometric
investigations. For example Bica et al. (\cite{bica98}) who used
Washington photometry to derive ages and metallicities of old LMC
clusters and the field, stated that {\it ``an increase of the assumed
  reddening by E(B-V)=0.03 decreases the derived metallicity by 0.12
  dex''}. The degeneracy between reddening and abundance becomes a
severe problem for old clusters and field stars, while for young
clusters it is possible to determine the reddening quite accurately
because the colour of the hot, bright main sequence stars is nearly
independent of temperature and thus of metallicity. The dependence of
the derived metallicity on the assumed reddening is illustrated in
Fig.~\ref{fig:1806redme}, with NGC~1806 as an example (we note that
this figure greatly exaggerates the realistic uncertainty of the
reddening for this particular cluster and just serves to illustrate
the trend).

Differential reddening is another aspect of this problem. We cannot
exclude it, however there are also no hints in favour of strong
differential reddening. Olsen (\cite{olsen99}) investigated four
fields in the LMC (three in the bar, one in the inner disk), where the
reddening is expected to be larger than in our further outside lying
fields. However, they detected strong differential reddening only
around NGC~1916. For the other fields it is not significant.

We estimated with the aid of Monte Carlo simulations, that as long as
the differential reddening is less than $E_{B-V}=0.03$ (peak to peak),
the uncertainty is small compared to the photometric uncertainty. In
any case differential reddening results in a broadening of the
metallicity and hence the age distribution.

\subsection{Unresolved binaries and blending}

The observation of unresolved binaries, which is likely to be the case
for a considerable fraction of stars, leads to a change in the
photometric metallicity.  Fortunately this effect plays a negligible
role for stars on the RGB where the mass-luminosity relation is steep
and even small differences in the initial mass result in large
differences in luminosity. This has been checked with the aid of
synthetic CMDs and two-colour diagrams, which is described in Sect.~12
below.

In crowded fields blending of stars is a related problem which is
nicely illustrated in Fig.~1 in the work of Ardeberg et al.
(\cite{ardeberg97}). The first correction is to exclude the most
crowded inner part of the clusters thus we excluded stars within
$\approx19\arcsec$ from the cluster center (details in the cluster
sections).  The most probable, but unimportant, case is blending with
a faint, red main sequence star: they are not luminous enough to
change the photometric metallicity of a RGB star.  To estimate the
probability of blending with other stars we used the following
approach: we assumed that blending takes place if the luminosity
centres of gravity of two stars are nearer than the PSF radius divided
by $\sqrt{2}$ (sampling theorem). Since the PSF radius was always
around $3$ pixels the area in which only one unblended star can be is
$(2\cdot3/\sqrt{2})^2\cdot\pi$. Next we counted the stars on the blue
and red side of the observed CMD (bluer and redder than $(b-y)=0.4$)
in luminosity intervals fainter than the main star and calculated the
probability that one of these fainter stars lie within the area of the
main component. We label the blending with a star ``strong blending'',
if the luminosity difference of this star and the main component is
less than $2$ mag. Blending with a star that is between $2$ and $4$
magnitudes fainter is called "weak blending" and corresponds to a
luminosity ratio of at least $6$ that results in shifts of $<0.2$ dex.
The probability of weak blending is underestimated since
incompleteness has not been considered.  However, weak blending
primarily results in a broadening of the metallicity distribution of
$<0.15$ dex. Also the strong blending is slightly underestimated since
a truly blended star is counted in the observed CMD only once, but
since the probability for strong blending is well below $10\%$ (see
below) we regard this approximation to be justified for our fields.
Blending with a main sequence star results in a shift of the combined
pair towards bluer colour and and smaller m1. The shift in $(b-y)$
dominates and thus the resulting metallicity is in general larger for
this pair than for the individual RGB star. Very strongly blended
stars even leave the selected colour range and thus our selection also
ensures the exclusion of heavily blended stars.

The fields of NGC~1711 and NGC~1806 shall serve as examples for the
expected blending probability.  These fields are rather crowded
compared to the fields around NGC~1651, NGC~2257 and NGC~2136, but
comparable with the crowding around NGC~2031. The strong blending
probability with a blue main sequence star decreases from $8\%$ for a
$18.5$ mag RGB star to less than $1\%$ for a $15$ mag star in the
field of NGC~1711. The probability for strong blending with a red star
decreases from $3\%$ to $<1\%$.  Around NGC~1806 the strong blending
with red stars is more probable: it decreases from $10\%$ to $<1\%$
for a luminosity of the main component between $18.5$ mag and $15$
mag. The probability of weak blending is for a $17$ mag RGB stars
$\approx 10\%$ around NGC~1806 and $\approx 5\%$ around NGC~1711. We
conclude that $\approx 5\%$ of our selected stars have a Str\"omgren
metallicity that deviates by more than $0.2$ dex from its ``true''
value due to blending. Around $10\%$ of the stars are blended with a
resulting shift of $<0.2$ dex.

The deviations in age due to blending are more difficult to determine:
on the one hand the increasing luminosity would result in an
underestimation of the age, on the other hand for strong blending with
a blue star, the metallicity would be underestimated, which generally
leads to an overestimation of the age. We conclude that also for the
age, blending results primarily in a broadening of the age
distribution, however, with a distribution that is more extended
towards younger ages, i.e. it is more probable to underestimate than
to overestimate the age (this has been found with the aid of the
mentioned simulation).  In the discussion in Sect.~14 we will give an
additional argument that the blending in clusters result in shifts of
less than $0.1$ dex compared to the field, assumed that cluster and
field population of the same age have the same metallicity based on
the observations.

\subsection{AGB versus RGB stars}

AGB stars are in the age range of $10^{8.4}$ yr to $10^{9.0}$ yr the
dominating giant stars. These stars are potentially problematic since
their surface abundances might have changed considerably compared to
the initial composition. However, in M~13 Pilachowski et al.
(\cite{pilachowski96}) found that the AGB stars are less CN enriched
than the RGB stars.

Frogel \& Blanco (\cite{frogel90}) identified several AGB stars around
some of our clusters, for which Olszewski et al. (\cite{olszewski91})
obtained the metallicity. However, these stars are always too red to
allow a photometric metallicity determination.

\subsection{Are the Str\"omgren metallicities independent of the luminosity class?}

The re-calibration of the Str\"omgren metallicity of red stars by
Hilker (\cite{hilker99}) is based on giant stars for the lower
metallicity range and approach the calibration of Grebel \& Richtler
(\cite{grebel92}) for higher metallicities (however also below 0 dex).
In the earlier work by Grebel \& Richtler (\cite{grebel92}) no
difference between giant and supergiant stars has been found.  >From
the theoretical point of view, only a very small luminosity effect is
expected (Bell \& Gustafsson \cite{bell78} and the used isochrones by
Grebel \& Roberts \cite{grebel95a}).

\begin{figure}[t]
\resizebox{\hsize}{!}{\includegraphics{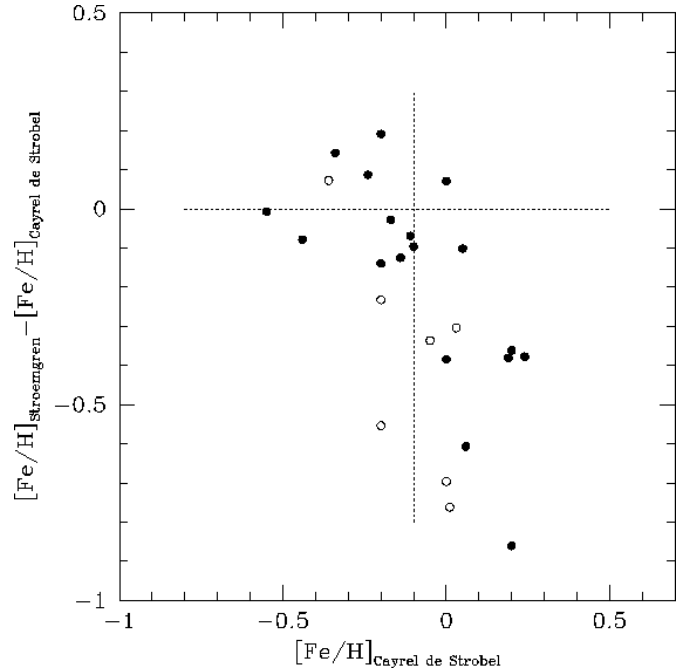}}
\caption{
  Difference between the Str\"omgren metallicity of galactic
  supergiants (luminosity class I \& II) and the metallicities given
  by Cayrel de Strobel (\cite{strobel97}). Solid dots denote "normal"
  stars while open dots are used for variable supergiants.  The
  reddening correction employed is $E_(B-V) = 0.03$. With such a small
  average reddening the metallicity difference of the normal stars
  which are metal poorer than $0$ dex is $\Delta [Fe/H] =
  -0.01\pm0.12$ dex. }
\label{fig:SGmetal}
\end{figure}

To reinvestigate observationally the dependence of the Str\"omgren
metallicity on the luminosity class, we selected supergiants
(luminosity class I \& II) with metallicity measurements and
Str\"omgren photometry from the compilation of Cayrel de Strobel et
al. (\cite{strobel97}) and SIMBAD, respectively. The major problem of
this approach is the largely unknown reddening towards these galactic
field supergiants. To exclude stars that are most probable highly
reddened we took only supergiants with a galactic latitude $|l|>20^0$
and brighter than $6$ mag in $V$ into account. The remaining
supergiants are shown in Fig.~\ref{fig:SGmetal}, where the difference
of the Str\"omgren metallicity to the measured metallicity (taken from
the list of Cayrel de Strobel et al. \cite{strobel97}) versus the
measured metallicity is displayed.  Fig.~\ref{fig:SGmetal} open
circles are used for variable stars, and open star symbols for carbon
stars and stars with a CN anomaly.  We did not attempt to correct for
the individual reddening, which explains partially the considerable
scatter. The vertical lines shows the location of the literature
metallicity of $-0.1$~dex and the horizontal line indicates where
spectroscopic and photometric metallicities are equal.

It can be seen in Fig.~\ref{fig:SGmetal} that the calibration seems to
hold for supergiants with a metallicity of less than $-0.1$ dex (we
have corrected for a systematic shift of $-0.1$ dex which can easily
be explained with an average reddening of $E_{B-V}=0.02$). For larger
metallicities the Str\"omgren metallicities seem to underestimate the
``true'' metallicity. However, the calibration has been made only for
stars of subsolar metallicity, thus the deviation of more metal rich
stars is not surprising.

\section{The age determination}

To determine ages of stars and clusters we employed isochrones
provided by the Geneva (Schaerer et al. \cite{geneva93}) and Padua
(Bertelli et al. \cite{padua94}) groups transformed into the
Str\"omgren system by Grebel \& Roberts (\cite{grebel95a}); they are
called "Geneva" and "Padua" isochrones in the following. The
isochrones show a zero point difference to the empirical calibration
in the sense that an isochrone of a given metallicity is too red
and/or $m1$ is too low compared to the calibration, which is of the
order of $0.15$~dex.  To solve this discrepancy we increased the $m1$
values of the isochrones by $0.04$ mag to bring them into accordance
with the empirical calibration. We have chosen the $m1$ colour index,
since it contains the $v$ filter, which is the most critical one in
the filter-band integration of the model spectra due to it's fairly
short wavelength. For this central wavelength the applied stellar
atmospheres of red giants and supergiants are not very precise
(Bressan priv. comm.).  Unfortunately, only Geneva isochrones with
metallicities of more than $-1.4$ dex and ages of less than $10^{9.9}$
yr were available. Padua isochrones on the other hand covered only the
age range between $10^{7.0} - 10^{9.0}$ yr and $10^{10.0} -
10^{10.24}$ yr, thus most of our results are based on the Geneva
isochrones.

The very red part of the RGB is not red enough to describe the
location of the observed RGB stars correctly, the isochrones are
slightly too steep for $(B-V)>1.1$. With the observed RGBs of
NGC~1651, NGC~1806 and the field surrounding NGC~1711, we introduced
an empirical linear colour term to bring the isochrones into agreement
with these stars (for $(b-y)>0.7$ the applied shift is:
$0.6\cdot((b-y)-0.7)$, the maximum correction is $\Delta (b-y)=0.12$).
After we applied this correction the isochrones fit also to the
younger clusters, which is a hint that the colour term is valid for
all gravities. However, nothing can be said for old stars ($>10^{9.5}$
yr) since no clear RGB with such an age was available to test the
empirical colour term (NGC~2257 is too metal poor and too old for this
purpose). The $m1$ had to be changed according to the colour term in
$b-y$, which has been performed on the basis of the empirical
metallicity calibration.

The isochrones show a small age-metallicity degeneracy: a substantial
age difference between a $10^{7.5}$ yr and a $10^{9.9}$ yr isochrone
leads only to a difference in the photometric metallicity of $\Delta
[Fe/H] = 0.2$, in the sense that younger stars would appear more metal
rich.  Since we adjusted a $10^{9.0}$ yr isochrone (via the $m1$
shift) to the empirical metallicity calibration, the resulting
metallicity uncertainty is less than $0.1$~dex.  Throughout the paper
we assumed a distance modulus of 18.5 for the LMC based on surface
brightness analysis of Cepheids (Gieren et al. \cite{gieren98}),
results from SN1987A (Panagia et al. \cite{panagia91}) and on the
recent revision of the "classical" Cepheid distance calibration
(Madore \& Freedman \cite{madore98}). A distance uncertainty has a
direct effect on the age determination in the sense that a smaller
distance to the LMC would result in lower ages.

\section{Selecting cluster and field stars}

\subsection{Cluster stars}

The first step in measuring the metallicity of a cluster is to
separate its members from the surrounding field population. We
performed this mainly by selecting stars within a certain radial
distance from the cluster.  The selection radius is defined as the
radius where the cluster star density starts to be higher than $2
\sigma$ over the background star density, which has been derived with
a radial density profile of the stars in the frame. The innermost part
($<20\arcsec$) of the clusters has been excluded because it is
impossible to derive reliable photometry for stars in this crowded
region, especially due to blending.

For young clusters also the luminosity of a star is a good criterion
to separate cluster and field stars, since it is easy to distinguish
bright, young cluster stars from old field RGB stars.  Clearly this
criterion does not separate field and cluster stars of similar age.
The lower luminosity limit that was used to exclude RGB field stars in
this approach has been determined by visual inspection of the CMD. It
is straightforward for NGC~1711, NGC~2031 and NGC~2136/37, where the
cluster stars are much brighter than the field RGB, however, this
criterion could not be applied for NGC~1651, NGC~1806 and NGC~2257.

We did not perform a statistical field star subtraction for two
reasons: we wanted to have the most reliable cluster stars and
including stars from a larger radius might lead to a bias towards
field stars that have a similar age and metallicity, since the
statistical field star subtraction has to work in chunks of colour and
luminosities. If the colour and luminosity range of the bins in which
the field stars are subtracted are chosen too small than the Poisson
error is large, if they are chosen too large, then the resulting
distribution is not ``cleaner'' than the one in our approach.
Moreover, since the numbers of stars are frequently much smaller than
for the field population, the error in the number of stars would
heavily depend on the field star population, especially if the
incompleteness varies strongly with radial distance from the cluster.
This varying incompleteness would lead to a considerable uncertainty.
The disadvantage of our approach is certainly that there will always
be some field stars left.

Stars with a photometric error (DaoPhot) of more than $\Delta
(b-y)=0.1$ and $\Delta m1 =0.1$ have been discarded. This selection
ensures more reliable results.

Finally, only stars redder than $(b-y)_0=0.6$ have entered the
metallicity measurement to reduce the systematic shifts in the derived
metallicity due to a possible error in the applied reddening
correction, which is illustrated in Fig.~\ref{fig:1806redme}. Also
stars being redder than $(b-y)=1.1$ have been discarded due to
additional lines in the $y$ filter.

From the remaining sample, individual stars have been excluded if they
deviate strikingly from the mean metallicity or from the mean RGB
location of the other stars, because still a few field and foreground
stars might be present, as mentioned above.

\subsection{Field stars}

Field stars have been selected with a radial selection criterion as
well: we regard as field stars those with a radial distance of more
than $30\arcsec$ plus the radius used for selecting the corresponding
cluster The additional $30\arcsec$ have been added to ensure that
cluster stars are a minor fraction among the field stars. Only field
stars with a relative photometric error of less than $\Delta
(b-y)=0.1$ and $\Delta m1 =0.1$ have been kept. In case of the young
clusters NGC~1711, NGC~2031 and NGC~2136/37 we used all RGB stars
having a distance of more than $70\arcsec$ from the cluster center,
since they clearly do not belong to the cluster.

To minimise the influence of photometric and calibration errors on the
derived metallicity, it is necessary to introduce the colour criterion
$(b-y)_0>0.6$, as it has been done in the case of clusters.  However,
it is dangerous to limit the sample just in colour since this
introduces a large bias towards metal poor stars with larger
metallicity errors \footnote{$\Delta m1$ is the dominating error in
  the Str\"omgren two-colour diagram. Therefore, the metallicity error
  will be larger for more metal poor stars than for more metal rich
  stars of the same colour.}.  This is not a big problem for younger
clusters, where the giants extend far into the red and therefore the
metallicity measurement does not depend so severely on the blue stars.
To circumvent this problem and to have nevertheless a reasonable
homogeneous selection criterion, we included only stars that are
redder than an inclined line in the $m1-(b-y)$ diagram that is nearly
perpendicular to $[Fe/H] = -1$ dex, a metallicity which is in the
middle of the expected metallicity range in the LMC. This line is
shown for example in the two-colour plot of the field population
around NGC~1711 (Fig.~\ref{fig:1711fme}).

\subsection{Galactic foreground stars}

Foreground stars of our own Galaxy contaminate the field and cluster
sample in the observed fields. Most of the foreground stars are red
clump stars which show up as a broad vertical strip at $b-y \approx
0.4$. Ratnatunga \& Bahacall (\cite{ratnatunga85}) presented a galaxy
model and give the amount of galactic foreground stars in luminosity
and colour bins. Their results are compiled in
Table~\ref{tab:Nforeground}. With these numbers in mind it is obvious
that we expect only few galactic foreground stars in the colour and
luminosity range we used for the metallicity and age determination.

\begin{table}[t]
\caption{Expected foreground stars (Ratnatunga \& Bahcall \cite{ratnatunga85}) per $44 (\arcmin)^2$ 
(the size of our CCD field) for
the LMC.}
\label{tab:Nforeground}
\begin{tabular}{c|ccc}
\hline
$V$ &$(b-y)<0.4$ & $0.5<(b-y)<0.8$ & $(b-y)<0.8$ \\\hline
$13-15$ & $1.3$ & $0.7$ & $0.1$ \\
$15-17$ & $4$   & $3.4$ & $0.7$ \\
$17-19$ & $4$   & $7.0$ & $5.0$ \\
$19-21$ & $8.4$ & $5.7$ & $17.2$\\\hline
\end{tabular}
\begin{list}{}{}
\item[]{To transform $(B-V)$ into the Str\"omgren $(b-y)$ we
used $(B-V)=-0.055+1.707 (b-y)$, that has been derived with
the available isochrones.}
\end{list}
\end{table}

\section{The reddening towards individual clusters and the surrounding field}

Because of the large influence of the reddening on the measured
metallicity, as described in Sect.~3, it is necessary to get a hand on
the reddening correction towards the observed regions. For this
purpose we used the theoretical upper main sequence ($(b-y)_0 < 0.1$
and $M_V < 0$) for the reddening determination, because of the
negligible metallicity metallicity effects. It is essential not to fit
the isochrone to the centre of the main sequence, since the isochrones
are calculated for non-rotating stars and evolutionary effects on the
upper main sequence. Rotation shifts a star redwards and thus one has
to fit the isochrone more to the blue border of the observed main
sequence. Also unresolved binaries on the main sequence are redder
than the observed isochrones. However, since photometric errors are
also present the fit of a blue envelope would be exaggerated.

For the extinction correction the relations of Crawford \& Barnes
(\cite{crawford70}) have been employed ($E_{b-y}=0.7 E_{B-V}$ and
$E_{m1} = -0.3 E_{b-y}$). Since the reddening is derived on the
assumption that the colour of the isochrone is correct for the very
bright blue main sequence we only give the possible uncertainty in the
adjustment of the isochrones to the main sequence as a reddening
error. However, our reddening is always smaller than the reddening
given by Schlegel et al. (\cite{schlegel98}), which is a hint that
there is a zero point shift in $b-y$ between either our calibration or
the isochrones on the order of $\Delta {b-y}=0.03$. It might of course
be a zero point shift in the Schlegel et al. values as well,
especially when considering that the given reddening values are
frequently larger than the one by other authors (see Sect.~6 -
Sect.~11).

\section{NGC~1711}

\subsection{The cluster}

Several attempts have been made to determine the age of NGC~1711. One
of them used also isochrone fitting (Sagar \& Richtler
\cite{sagar91}), however, with an {\it assumed} metallicity of $-0.4$
dex. The previous results on the age of NGC~1711 are compiled in
Table~\ref{tab:1711lit} in the appendix. No metallicity measurement
for this cluster has been published yet. The CMD of the entire CCD
field is shown in Fig.~\ref{fig:cmd}.

Using the upper main sequence ($V<18.5$) we have deduced a reddening
of $E_{b-y} = 0.06 \pm 0.02$, relative to the isochrone, which
corresponds to $ E_{B-V} = 0.09 \pm 0.05$ (the calibration error has
already been included).  The determined reddening agrees, within the
errors, with $E_{B-V} = 0.14$ given by Cassatella et al.
(\cite{cassatella96}) and with the reddening derived from measurements
by Schwering \& Israel (\cite{schwering91}) ($E_{B-V}=0.11$).
Burstein \& Heiles (\cite{burstein82}) gives $E_{B-V}=0.12$.

\begin{figure}[t]
\resizebox{\hsize}{!}{\includegraphics{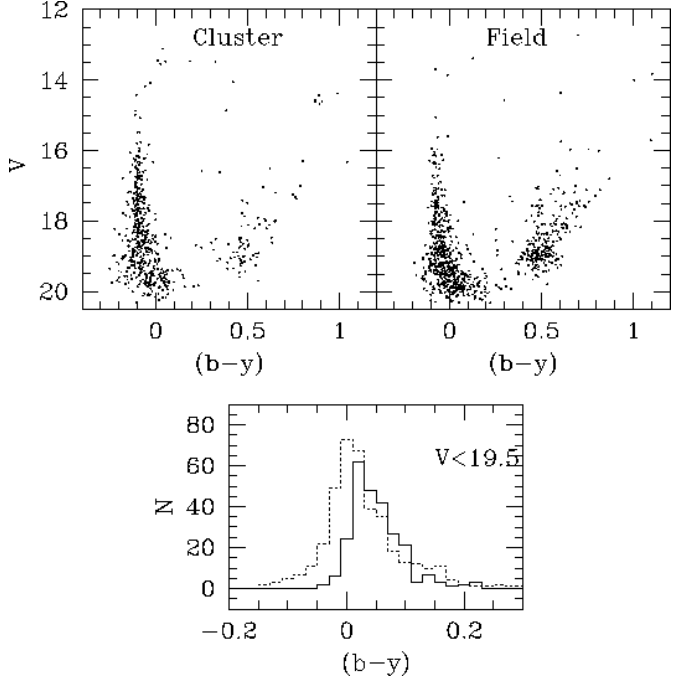}}
\caption{
  Illustration of the reddening difference of $0.02$ mag between
  NGC~1711 and its surrounding field stars.  In the upper panel the
  CMDs of cluster and field are shown. In the lower panel the colour
  histogram of main sequence stars brighter than $V = 19.5$ mag are
  plotted with a dotted line for the cluster and with a solid line for
  the field. For illustrative purposes, the radial selection for the
  field was chosen to result in approximately the same number of field
  and cluster stars. }
\label{fig:1711fld}
\end{figure}

Concerning the reddening, it is important to note that the surrounding
field, where one can observe young stars with a similar age, shows a
$0.02$ mag higher reddening, indicating that NGC~1711 is located in
front of the LMC disk.  Fig.~\ref{fig:1711fld} illustrates this
difference. This also holds for a small concentration of brighter
stars south-east of NGC~1711. Unfortunately, the number of stars in
this group is not large enough to allow a reliable age or metallicity
determination. However, because of the same colour difference between
field and cluster, it might form a binary cluster with NGC~1711, a
configuration, which seems to be common in the LMC \footnote{10\% of
  the LMC clusters are thought to be paired} (e.g. Dieball \& Grebel
\cite{dieball98} and references therein).

\begin{figure*}
\setlength{\unitlength}{1cm}
\begin{picture}(18,25)
\put(0,16.5){
\begin{minipage}[t]{8.5cm}
\resizebox{\hsize}{!}{\includegraphics{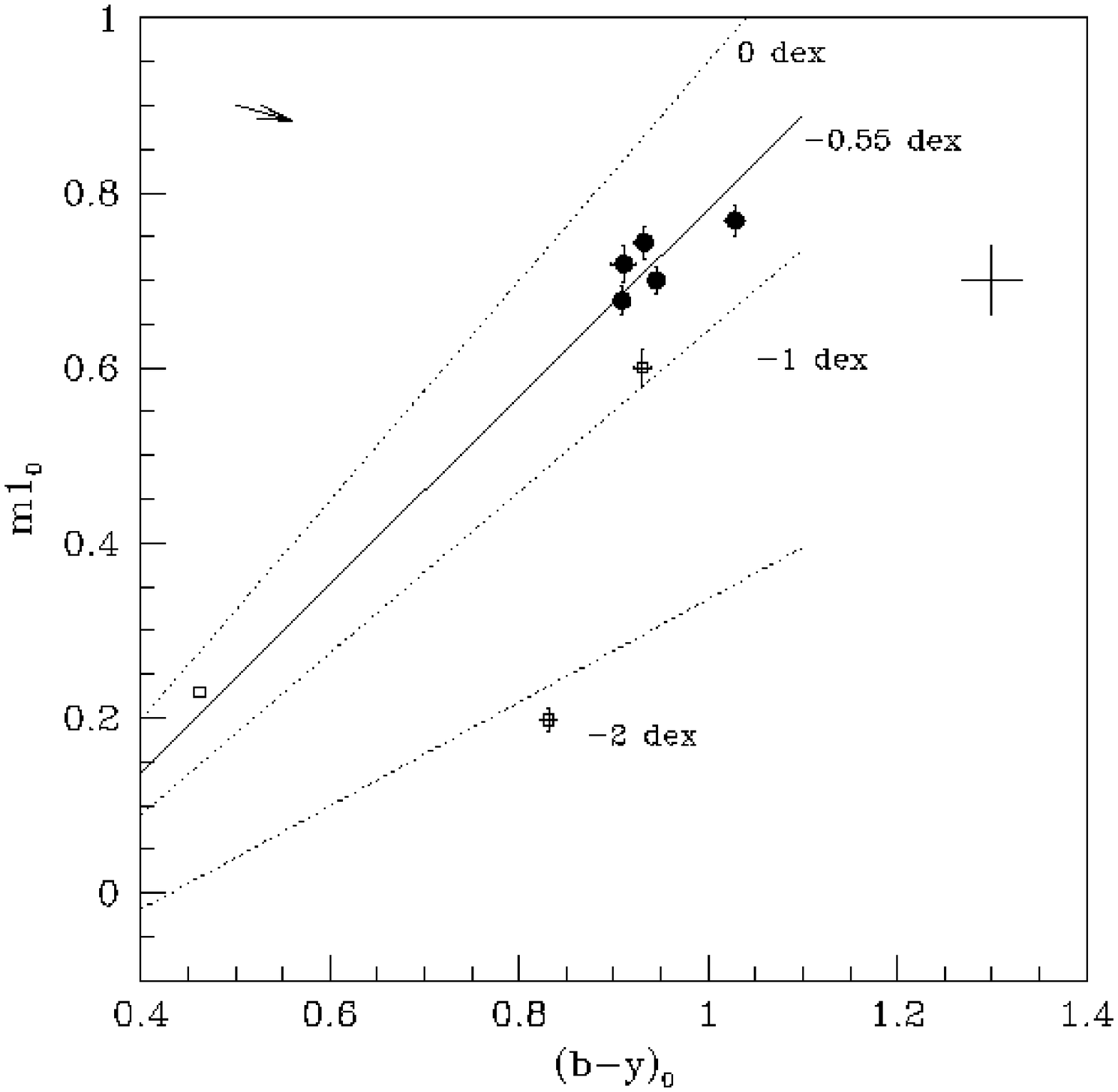}}
\caption{
  Reddening corrected two-colour diagram for the stars selected in the
  CMD of NGC~1711 with a radial ($r<90\arcsec$) and a luminosity
  criterion ($V<15$). (Fig.~\ref{fig:1711cage}). The open squares are
  used for excluded stars, while the filled circles show stars which
  entered the metallicity determination. The apparently metal poor
  star is most probable a Galactic halo foreground star. The cross at
  the right border shows the calibration error.}
\label{fig:1711cme}
\end{minipage}
}
\put(0,3){
\begin{minipage}[t]{8.5cm}
\resizebox{\hsize}{!}{\includegraphics{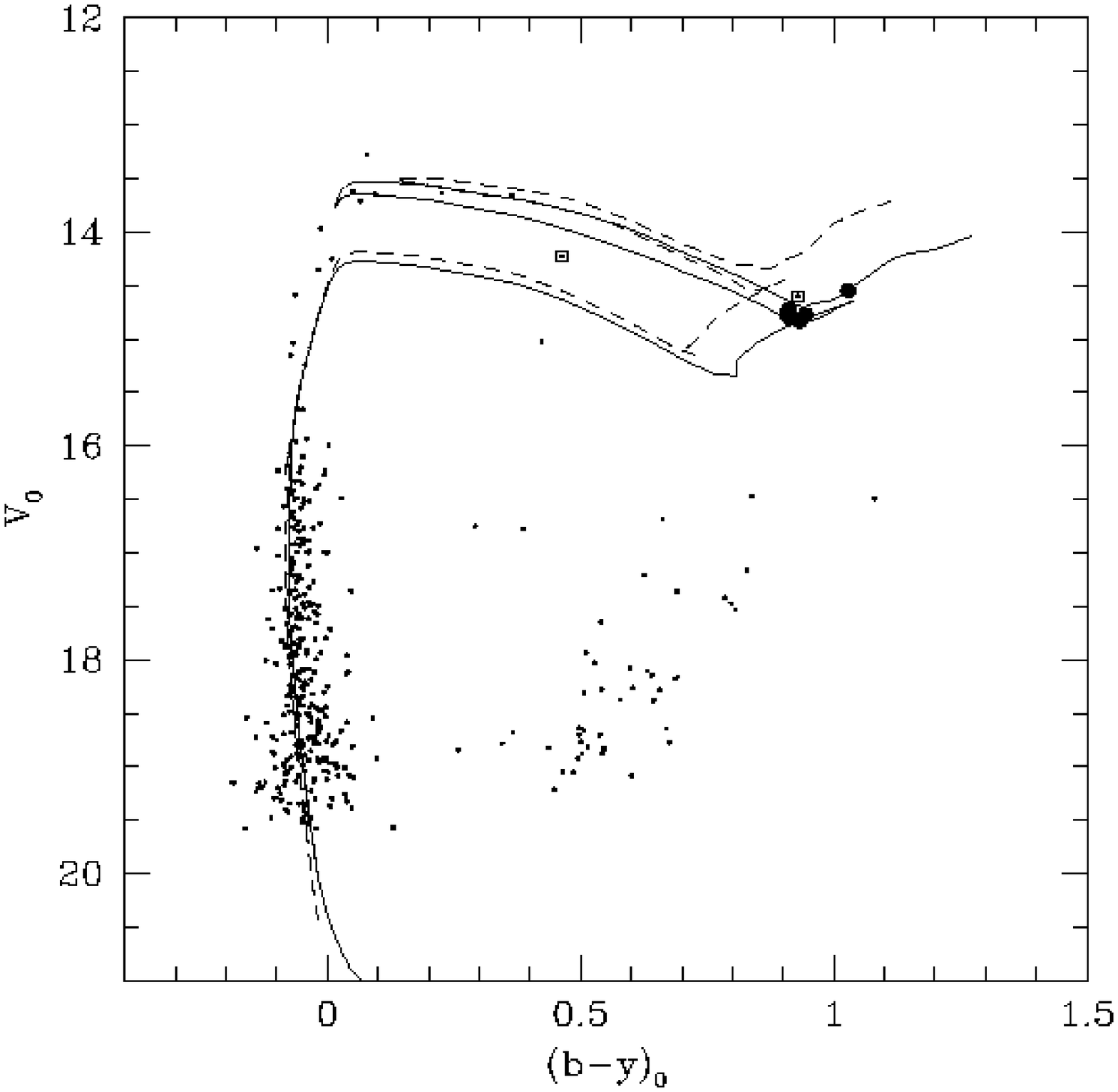}}
\caption{
  CMD for the cluster area of NGC~1711. The symbols are the same as in
  Fig.~\ref{fig:1711cme}.  Geneva isochrones are overlaid with $-0.4$
  dex (full line) and $-0.7$ dex (dashed line) with an age of
  $10^{7.7}$ yr.}
\label{fig:1711cage}
\end{minipage}
}
\put(9,16.5){
\begin{minipage}[t]{8.5cm}
\resizebox{\hsize}{!}{\includegraphics{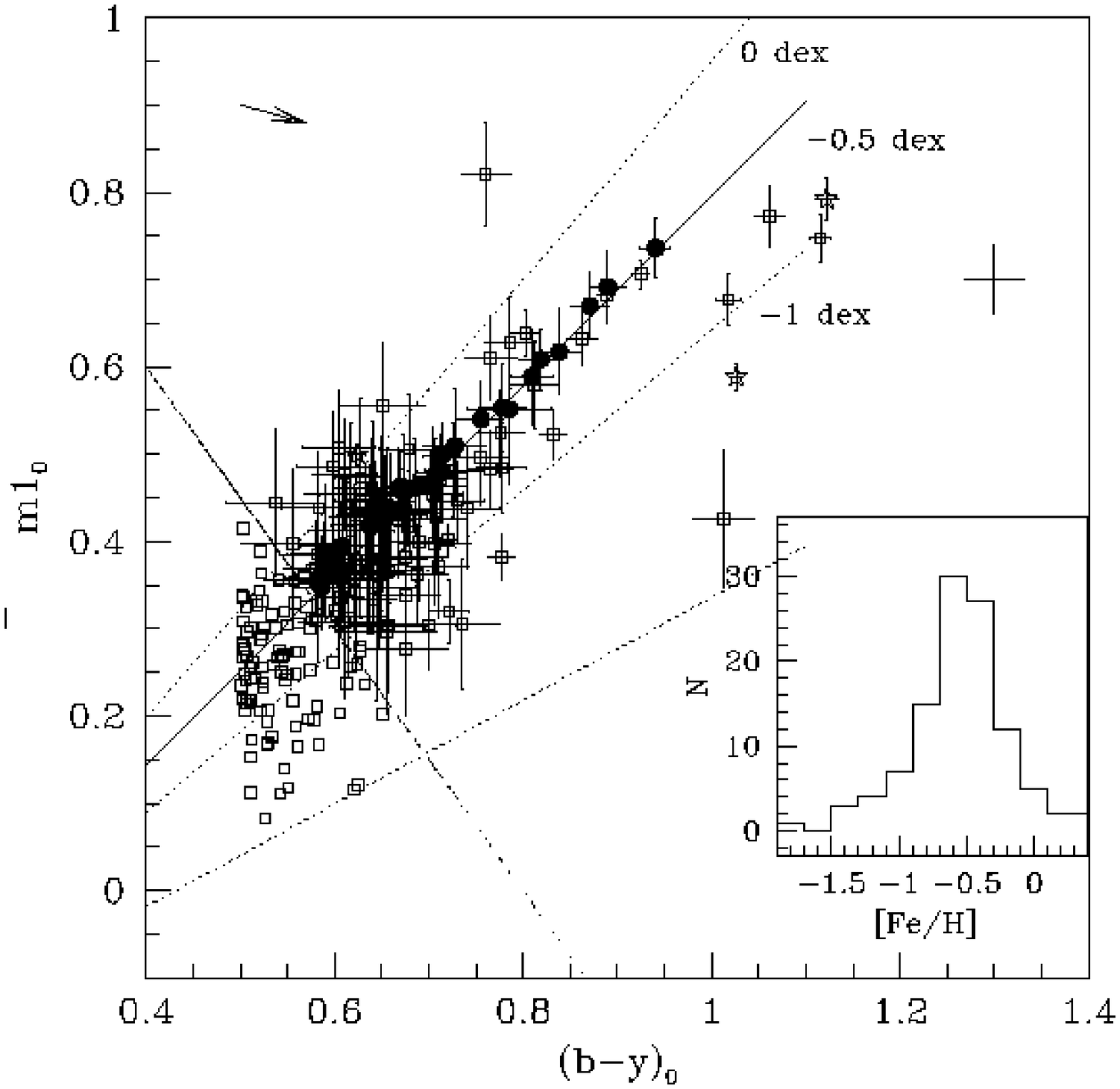}}
\caption{
  Two-colour diagram for the stars which are further than $110\arcsec$
  away from the cluster centre of NGC~1711. The inclined line from the
  upper left to the lower right marks the selection criterion applied
  to the colour: only stars redder than this line have been used to
  calculate the mean and standard deviation of the field star
  metallicity.  The open squares redder than the selection line are
  stars fainter $V=15$, the stars brighter than $V=15$ are shown as
  open stars. Filled circles represent stars being fainter $V=15$ and
  in the metallicity range $-0.75<[Fe/H]<-0.45$. The other open
  squares mark all stars for which the calibration by Hilker
  (\cite{hilker99}) is valid. The insert shows the metallicity
  distribution of all stars redder than the selection line.}
\label{fig:1711fme}
\end{minipage}
}
\put(9,3){
\begin{minipage}[t]{8.5cm}
\resizebox{\hsize}{!}{\includegraphics{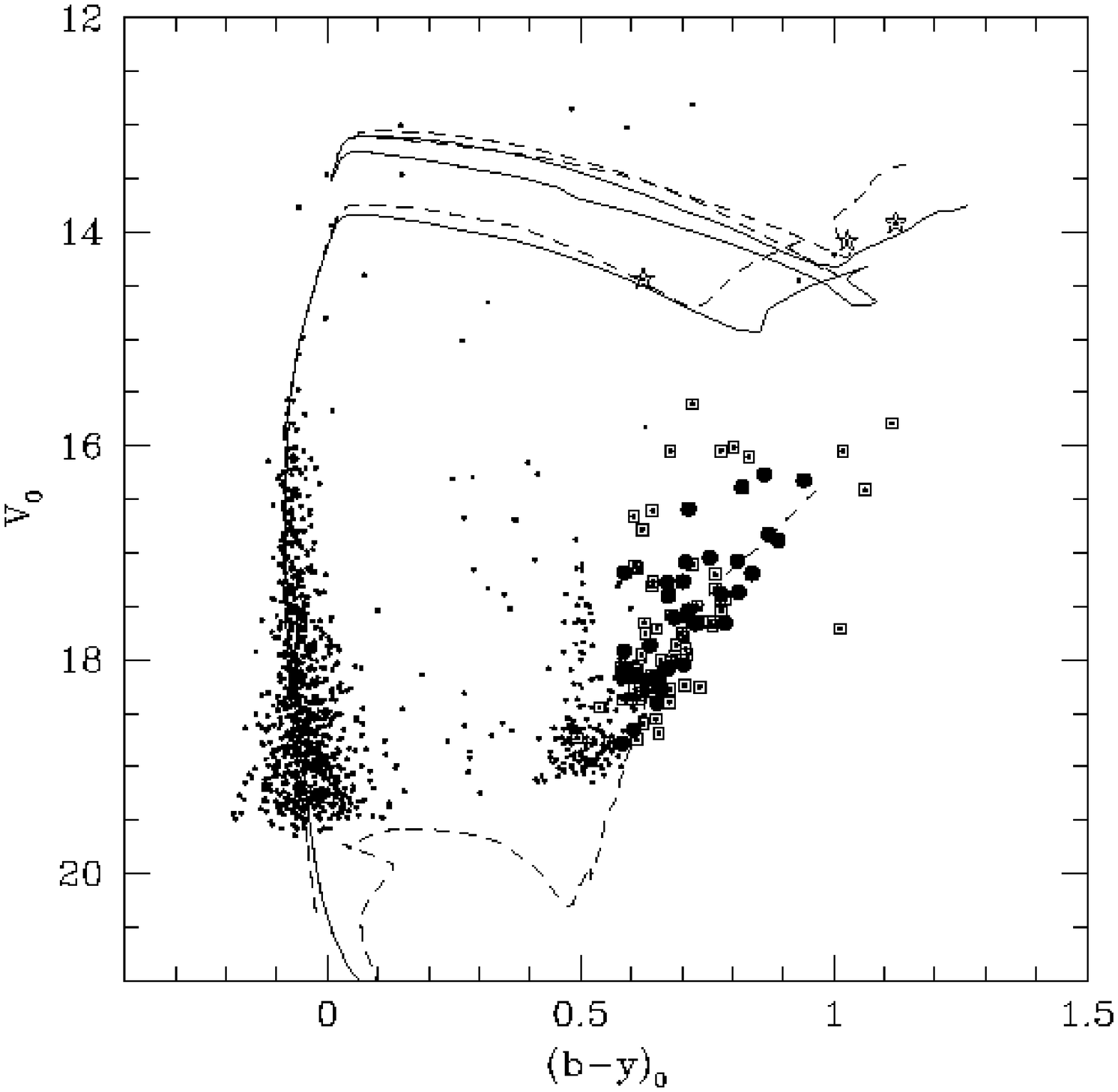}}
\caption{
  CMD of the field stars around NGC~1711. Only stars redder than the
  colour selection line have been marked with symbols. The symbols are
  the same as those used in the two-colour diagram of the field stars
  (Fig.~\ref{fig:1711fme}).  The younger isochrones are Geneva
  isochrones with a metallicity of $-0.4$ dex and $-0.7$ dex and an
  age of $10^{7.6}$ yr, the older one has a metallicity of $-0.7$ dex
  and an age of $10^{9.2}$ yr.}
\label{fig:1711fage}
\end{minipage}
}
\end{picture}
\end{figure*}

By inspecting the radial number density of stars around NGC~1711, we
have found that cluster stars begin to dominate ($2 \sigma$ level) the
field stars at a radial distance of $90\arcsec$. This radius has been
used for the radial selection. To exclude older field RGB stars, we
regarded only stars brighter than $V_0 = 15.5$ as potential cluster
members for the metallicity analysis. The CMD and two-colour diagram
of NGC~1711 is presented in Fig.~\ref{fig:1711cage} and
Fig.~\ref{fig:1711cme}, respectively. In the two-colour diagram we
plotted the calibration error separately and assigned only the
photometric error to the individual stars. The effect of these two
errors is completely different: the metallicity error due to the
photometric errors decrease with increasing size of the sample, while
in contrast the metallicity error due to the calibration can only be
decreased if observations from different nights are averaged.

We measured a metallicity of $[Fe/H] = -0.57 \pm 0.06$ dex for
NGC~1711. The error is the standard deviation of the individual stars
divided by the square root of the number of used stars ($5$).
Reddening and calibration error account for an additional error of
$0.16$ dex, thus we finally obtained $[Fe/H] = -0.57 \pm 0.17$.  To be
able to fit the red supergiants one needs an isochrone with a
metallicity of at least $-0.4$ dex, despite the measured metallicity.
We showed in Sect.~3.2, that the calibration is valid for supergiants,
as long as they have a metallicity below $0$ dex. From
Fig.~\ref{fig:SGmetal} one could estimate that a supergiant with a
metallicity around $0.2$ dex could be mistaken for a $-0.5$ dex star.
However, we prefer a different explanation: as described in Sect.~3, a
slight age dependence exist accounting for $\pm 0.1$ dex, in the sense
that younger stars appear more metal poor. With this in mind we
derived a metallicity of $[Fe/H]=-0.45\pm0.2$ (we assigned an
additional error of $0.05$ because of the uncertainty of the shift).
A third possibility is that the isochrones does not sufficiently
extend towards the red for these bright stars, despite the empirical
correction. This is mainly a problem in the treatment of overshooting
and a common problem for red supergiants (Bressan priv. comm.).  We
arrived at an age of $10^{7.70 \pm 0.05}$ yr using Geneva isochrones.
The isochrone is overlayed in Fig.~\ref{fig:1711cage}.

\subsection{The surrounding field  population}

We used only bright ($V<16$) stars that are more than $110\arcsec$
away from the cluster centre and all RGB stars with a distance of
$>50\arcsec$.

The field population consists of a main sequence slightly more
reddened and an older population clearly distinguishable by its giant
branch. A remarkable feature is the vertical extension of the red
clump (VRC), which has been observed in other areas of the LMC as
well. This has been interpreted by Zaritsky \& Lin (\cite{zaritsky97})
as a signature of an intervening population towards the LMC, but
Beaulieu \& Sacket (\cite{beaulieu98}) showed that also normal stellar
evolution could lead to such a feature, if a $10^{8.5}$ yr to
$10^{9.0}$ yr old population is present. Even the fainter extension
(or fainter second red clump), might be present (Girardi
\cite{girardi99}, Piatti et al. \cite{piatti99}), however the numbers
are definitely too small to allow an unambiguous identification.

With a reddening of $E_{B-V} = 0.11$ we can derive the metallicity of
the field stars which is illustrated in Fig.~\ref{fig:1711fme}. The
average metallicity of the field population is $[Fe/H]=-0.53$ dex and
the standard deviation $0.42$ dex.

The unambiguously young field stars, which are marked with star
symbols in Fig.~\ref{fig:1711fme} and Fig.~\ref{fig:1711fage}, have a
mean metallicity of $-0.56\pm0.27$ which is not systematically larger
than the one of the older population, even when accounting for the
slight age dependence of the metallicity.  Stars in the narrow
metallicity range $-0.75<[Fe/H]<-0.45$ (filled circles) do not exhibit
a uniform age. An upper age limit of the field stars is $10^{8.9}$ yr.
This means that between $10^{7.7}$ yr and approximately $10^{8.9}$ yr
no clear age-metallicity dependence can be seen in this field.

\section{NGC~1806}

\subsection{The cluster}

\begin{figure*}
\setlength{\unitlength}{1cm}
\begin{picture}(18,25)
\put(0,16.5){
\begin{minipage}[t]{8.5cm}
\resizebox{\hsize}{!}{\includegraphics{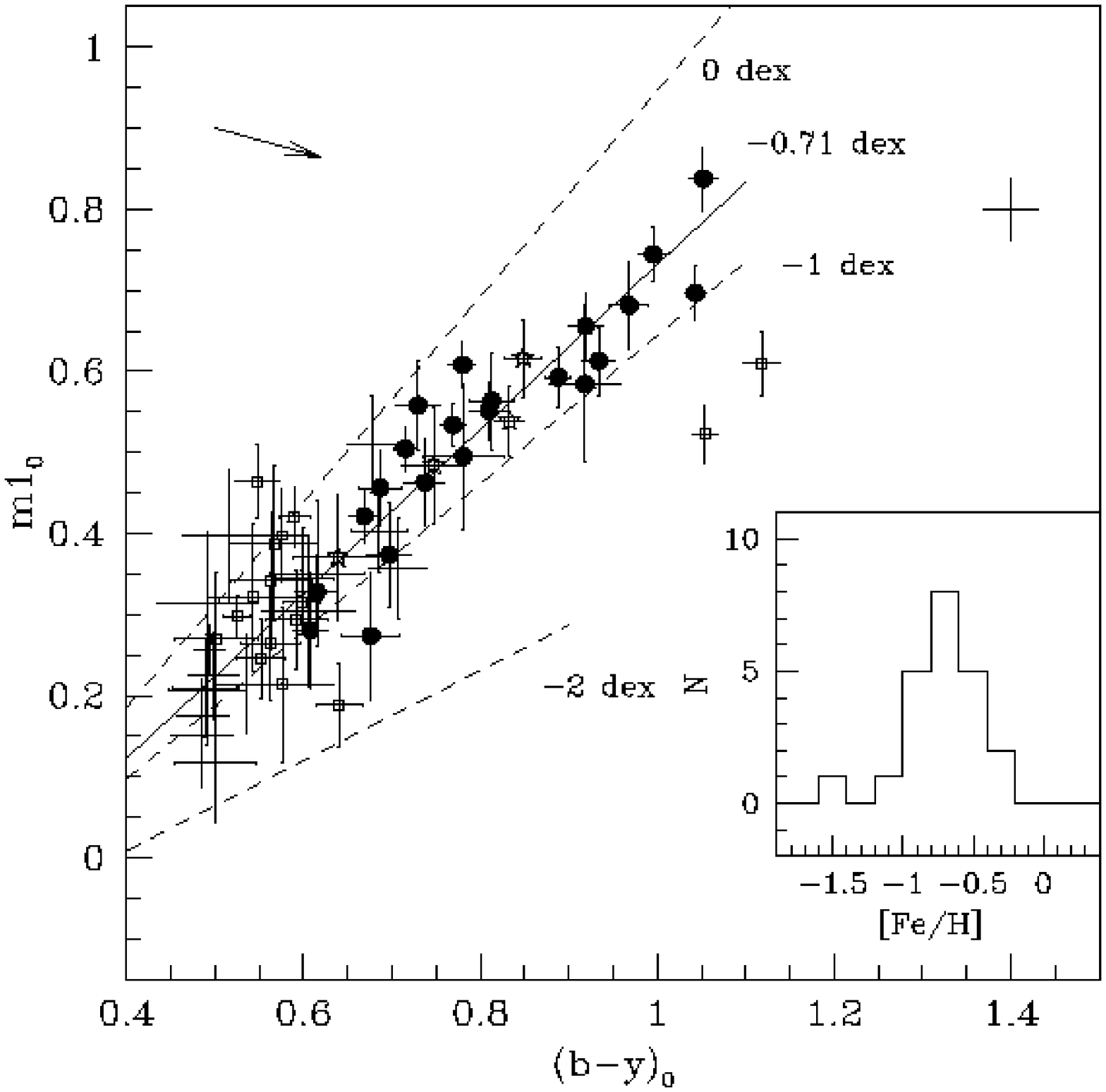}}
\caption{
  Two-colour diagram of NGC~1806. The filled circles refer to stars
  with a distance of less than $60\arcsec$ from the cluster centre.
  In addition to the usual selection criteria we have excluded stars
  that are brighter or fainter than the average RGB as indicated in
  Fig.~\ref{fig:1806cage}.  These stars are marked with open star
  symbols. The stars marked with filled circles have been used for the
  metallicity determination.  The applied reddening correction is
  shown as an arrow in the upper left corner.}
\label{fig:1806cme}
\end{minipage}
}
\put(0,4){
\begin{minipage}[t]{8.5cm}
\resizebox{\hsize}{!}{\includegraphics{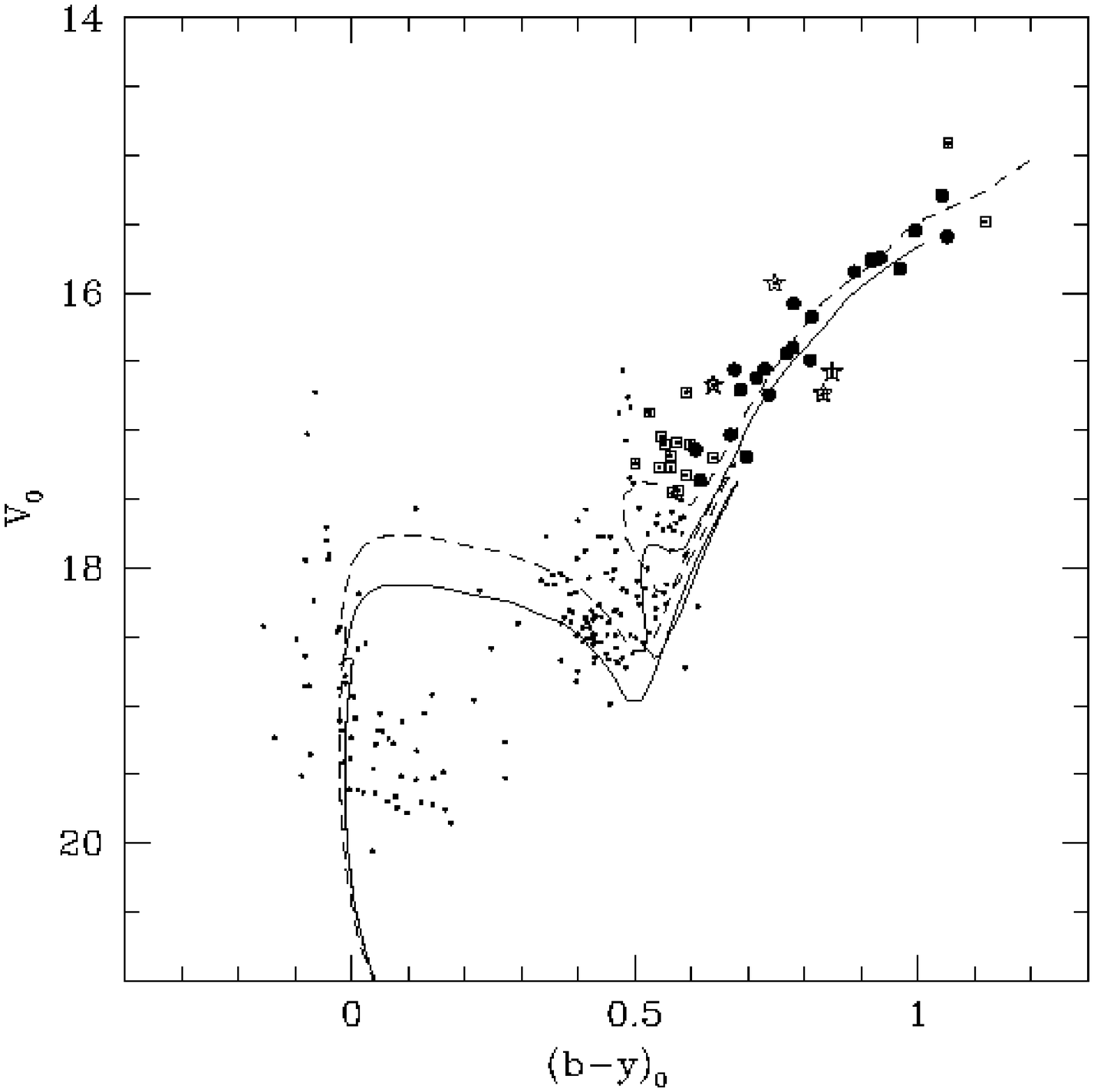}}
\caption{
  CMD of the NGC~1806 cluster stars. The overplotted Geneva isochrones
  have a metallicity of -0.7 dex and ages of $10^{8.7}$ yr (solid
  line) and $10^{8.6}$ y (dashed line). The symbols are analogous to
  the ones described in the two-colour diagram for NGC~1806
  (Fig.~\ref{fig:1806cme}).}
\label{fig:1806cage}
\end{minipage}
}
\put(9,16.5){
\begin{minipage}[t]{8.5cm}
\resizebox{\hsize}{!}{\includegraphics{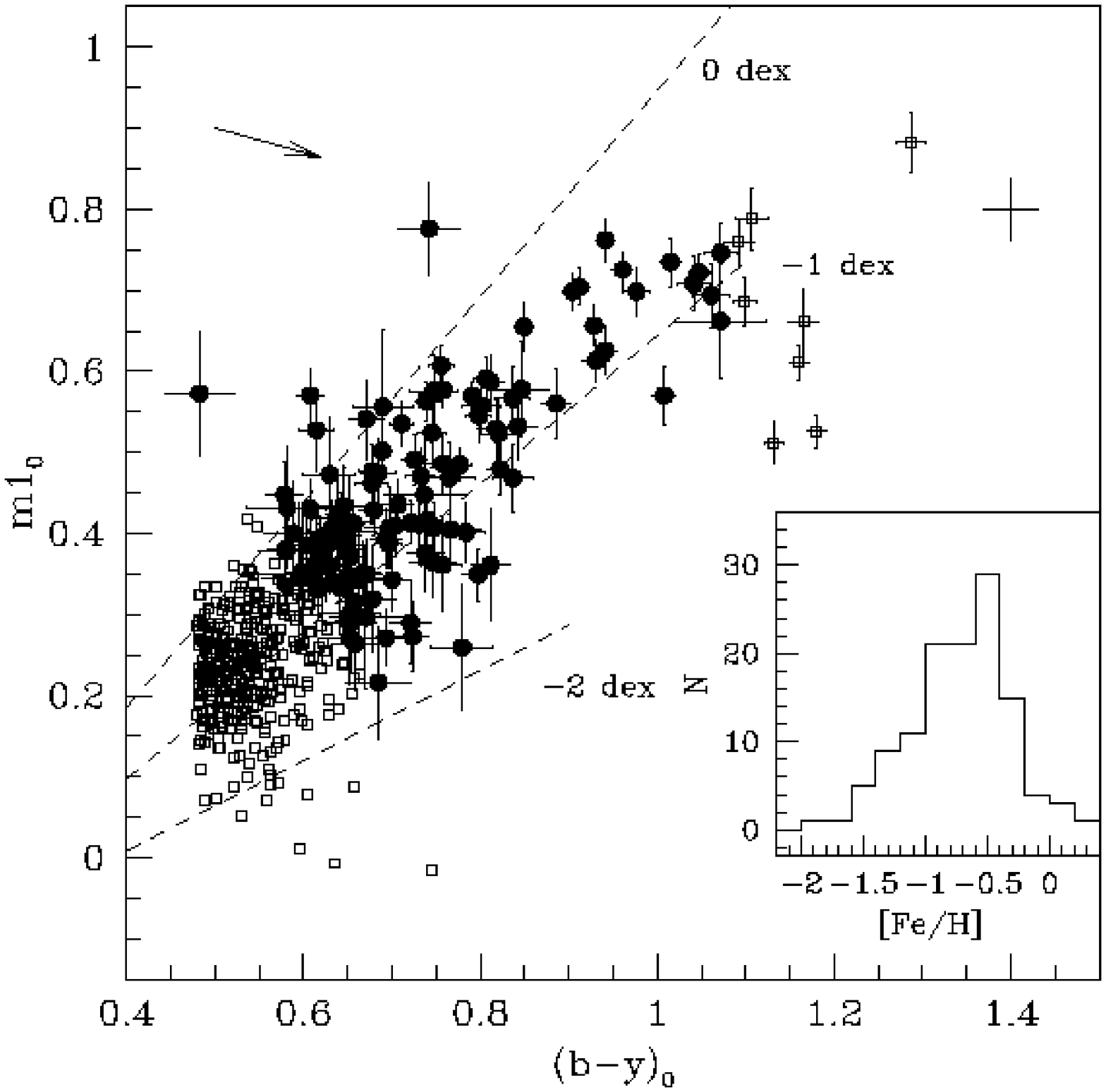}}
\caption{
  Field population around NGC~1806 (radial distance larger than
  $120\arcsec$).  The filled circles are stars that have been used to
  determine the mean metallicity of the field and are used for the
  metallicity histogram shown in the small panel.  The open squares on
  the blue side are stars for which the calibration is valid, but have
  been excluded according to our colour selection. The open squares on
  the red side of the two-colour diagram show stars beyond $(b-y)_0 =
  1.1$, for which the calibration is not valid any more.  }
\label{fig:1806fme}
\end{minipage}
}
\put(9,4.5){
\begin{minipage}[t]{8.5cm}
\resizebox{\hsize}{!}{\includegraphics{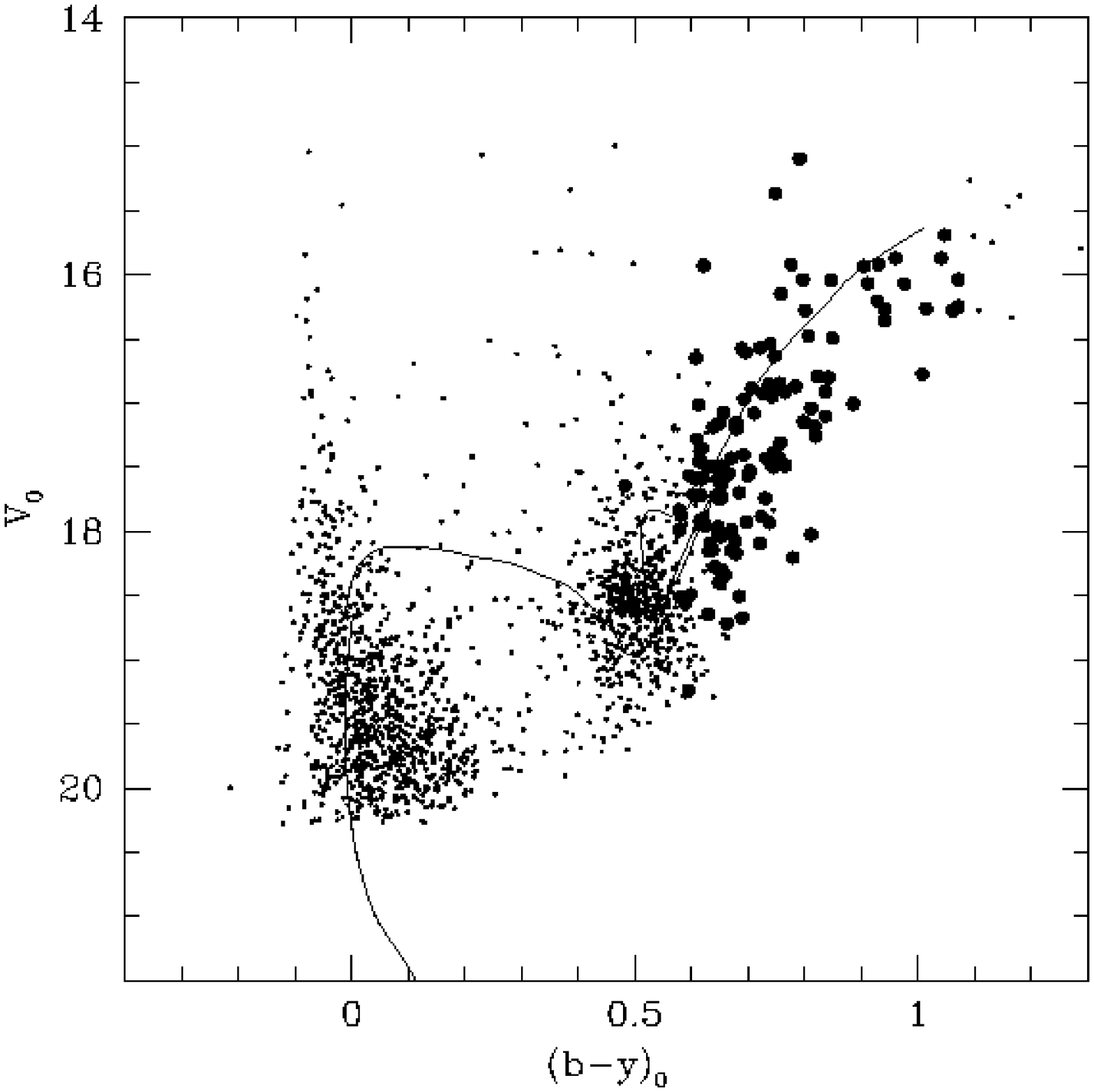}}
\caption{
  The CMD of the field stars around NGC~1806. The filled circles are
  stars which have been used for calculating the mean metallicity. The
  overplotted isochrone is the same solid one as in
  Fig.~\ref{fig:1806cage} and serves to illustrate the location of the
  cluster RGB.}
\label{fig:1806fage}
\end{minipage}
}
\end{picture}
\end{figure*}

NGC~1806 is older than NGC~1711, which can immediately be seen from
its pronounced RGB (see Fig.~\ref{fig:1806cage}). No previous CCD CMD
is available in the literature.  A faint main sequence of a younger
field population is visible in the field around NGC~1806
(Fig.~\ref{fig:cmd}).  This population has been used to determine a
reddening of $E_{b-y} = 0.12 \pm 0.02$ ($E_{B-V} = 0.17 \pm 0.03$),
which agrees, within the errors, with the values given by Cassatella
et al. (\cite{cassatella87}) ($E_{B-V} = 0.12$).  Schwering \& Israel
(\cite{schwering91}) give $E_{B-V} =0.10$ and Schlegel et al.
(\cite{schlegel98}) $E_{B-V}=0.24$.  Burstein \& Heiles
(\cite{burstein82}) derived a reddening of $E_{B-V}=0.06$ towards this
direction.  We assumed that the cluster is reddened by the same amount
as the field main sequence stars. The strong dependence of the derived
metallicity on the reddening correction has been demonstrated for this
cluster in Fig.~\ref{fig:1806redme}.

Only stars within the radial distance of $60\arcsec$ have been taken
for the metallicity determination.  In addition, some stars lying
apart from the average RGB location have been excluded. These stars
are marked with open star symbols in the cluster CMD
(Fig.~\ref{fig:1806cage}). The inclusion of these stars would not
change the derived metallicity.

We obtained a metallicity of $[Fe/H] = -0.71 \pm 0.06$ dex for
NGC~1806. The calibration and reddening uncertainty result in an
additional error of $\Delta [Fe/H] = \pm 0.23$ dex, hence $[Fe/H] =
-0.7 \pm 0.24$ dex.  The reddest cluster star (that has been excluded
because of its red colour) is an identified AGB star (Frogel \& Blanco
\cite{frogel90}, LE 6), for which Olszewski et al.
(\cite{olszewski91}) determined a metallicity of $-0.7$ dex, which is
in good agreement with our value for this cluster.

The two-colour diagram is presented in Fig.~\ref{fig:1806cme}. Some
remarkable stars are those being redder $(b-y)=1.0$ and apparently
more metal poor than the other cluster stars. If these stars were
members of a true metal poor population one would have expected to
find bluer stars of the same metallicity, which is not the case.
Because of this reason we think that these stars belong to NGC~1806,
but they possess additional absorption lines in the $y$ band compared
to the bluer RGB stars. This is theoretically expected for red stars
with solar metallicity, however the theoretical Geneva isochrones of
this metallicity do not extend far into the red regime and no Padua
isochrone with appropriate metallicity and age has been available.
>From the identified AGB star one might speculate that these deviating
stars are AGB stars in NGC~1806, which is supported by the best
fitting isochrone (see below).

The age determination is illustrated in Fig.~\ref{fig:1806cage}. The
best fitting isochrone yields an age of $10^{8.7 \pm 0.1}$ yr. This is
much younger than the age of $10^{9.6 \pm 0.1}$ yr derived by Bica et
al.  (\cite{bica96}) using the SWB classification. The red branch of
this isochrone consists mainly of AGB stars especially in the employed
colour range for the metallicity determination.

\subsection{The surrounding field population}

The metallicity of the field population has been derived with all
stars that have a distance of at least $100\arcsec$ from the cluster
centre. With the above stated reddening the mean metallicity of the
field population can be obtained as $-0.67$ dex with a (relatively
small) standard deviation of $0.23$ dex.

A feature that is visible in the two-colour diagram of NGC~1806
(Fig.~\ref{fig:1806fme}) is that stars that are redder than $b-y
\simeq 1$ seem not to follow a straight line for a given metallicity,
but rather get smaller $m1$ values with increasing $b-y$. The
  deviation is of the order $-0.5$~dex.  This behaviour is similar to
what is observed among the cluster stars of NGC~1806. Again, we argue
that these stars might deviate from the line of constant metallicity
given by the calibration.  If these stars belong to a true metal-poor
population one would expect to find more metal-poor stars with a
colour between $0.9<(b-y)_0<1.1$; this is not the case. No star is
found in the whole colour range with a metallicity of lower than -1.3
dex, while four stars are found in an even smaller colour range of
$\Delta(b-y)=0.1$. Therefore, one would expect to find at least eight
stars in the bluer colour range when assuming a homogeneously
populated RGB, which is not unreasonable since the bottom part of a
RGB around this age usually is even more populated than the upper part
of the RGB. These peculiar red stars could be foreground stars as
well, however it is intriguing that they are mixed with the other RGB
stars in the CMD. Therefore we think that they belong to the LMC. Thus
we reconfirm the statement that even low metallicity stars are only
good tracers for metallicity as long as they are bluer than $b-y =
1.1$.

\section{NGC~2136/37}

\subsection{The cluster}

NGC~2136/37 is a potential triple cluster system (Hilker et al.
\cite{hilker95a}) and thus another example of the common multiplicity
among LMC clusters. The main components have an angular separation of
$1\arcmin.34$.  We have re-investigated this cluster due to the
availability of Str\"omgren isochrones and because of the new
calibration. NGC~2136 contains at least eight Cepheids making the
knowledge of its metallicity particularly interesting for the Cepheid
distance scale and the metallicity dependence of the PLC relation.
The Cepheids have not been included in the derivation of the
metallicity.

After the inspection of the radial number distribution of stars around
the cluster centre we selected all stars with a distance of less than
$75\arcsec$. Additionally we excluded probable RGB stars of the field
population with $V>16.5$.

The reddening can be determined with the upper main sequence and we
obtained $E_{b-y} = 0.07 \pm 0.02$ ($E_{B-V}=0.10\pm0.03$). This
agrees with the reddening of $E_{B-V}=0.09$ given by Schwering \&
Israel (\cite{schwering91}). Burstein \& Heiles (\cite{burstein82})
obtained a reddening of $E_{B-V}=0.075$.

\begin{figure*}
\setlength{\unitlength}{1cm}
\begin{picture}(18,25)
\put(0,16.5){
\begin{minipage}[t]{8.5cm}
\resizebox{\hsize}{!}{\includegraphics{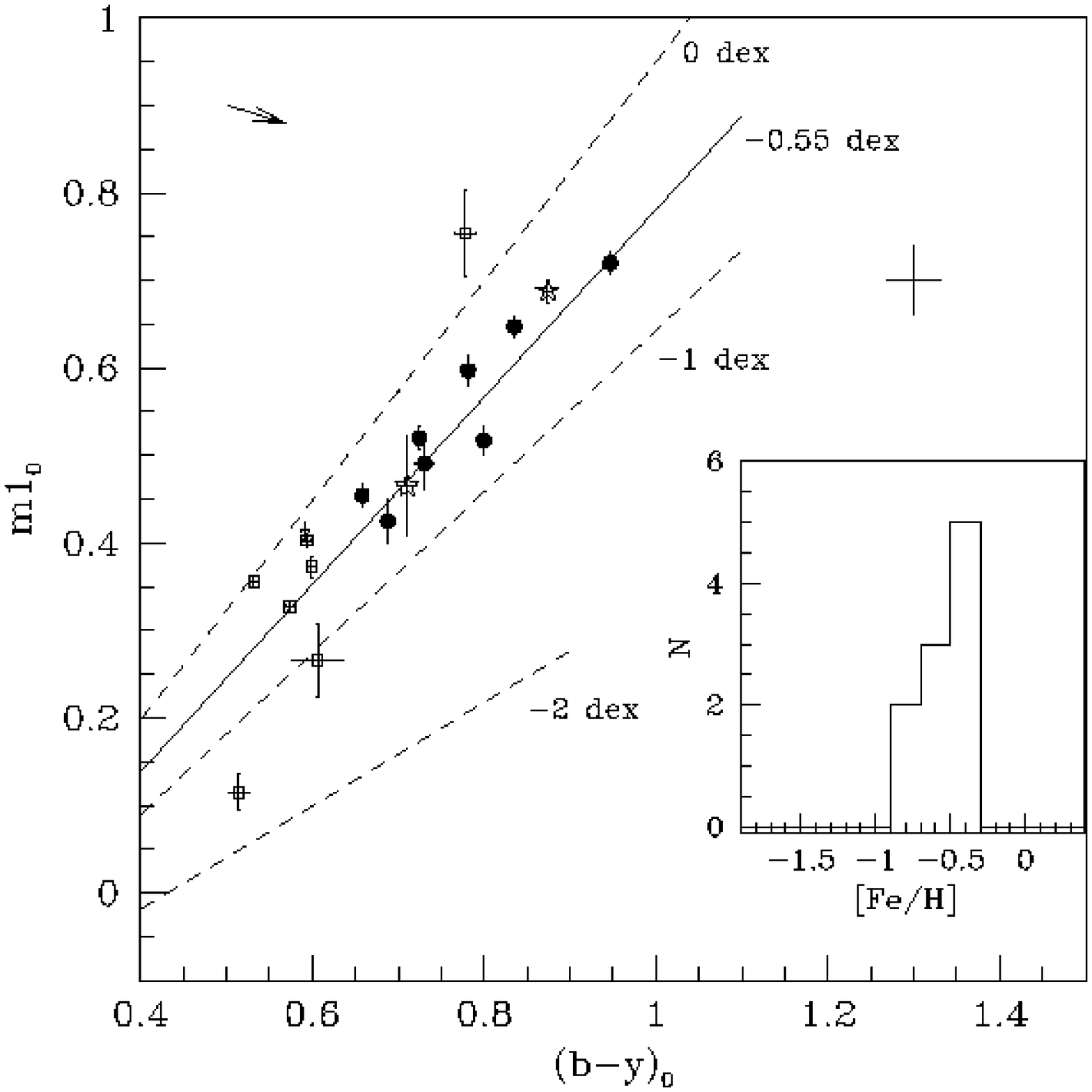}}
\caption{
  Two-colour diagram of NGC~2136/37. Solid dots are used for stars
  which have been used to determine the metallicity of NGC~2136. The
  two open star symbols show the giants that belong to NGC~2137. Open
  squares bluer $b-y=0.6$ mark stars for which the calibration holds,
  but which have been excluded. The apparently metal rich stars has
  been excluded because of its deviant location from the other giants
  in this diagram.  The histogram shows the metallicity distribution
  of the NGC~2137 and NGC~2136 giants.  }
\label{fig:2136cme}
\end{minipage}
}
\put(0,3.2){
\begin{minipage}[t]{8.5cm}
\resizebox{\hsize}{!}{\includegraphics{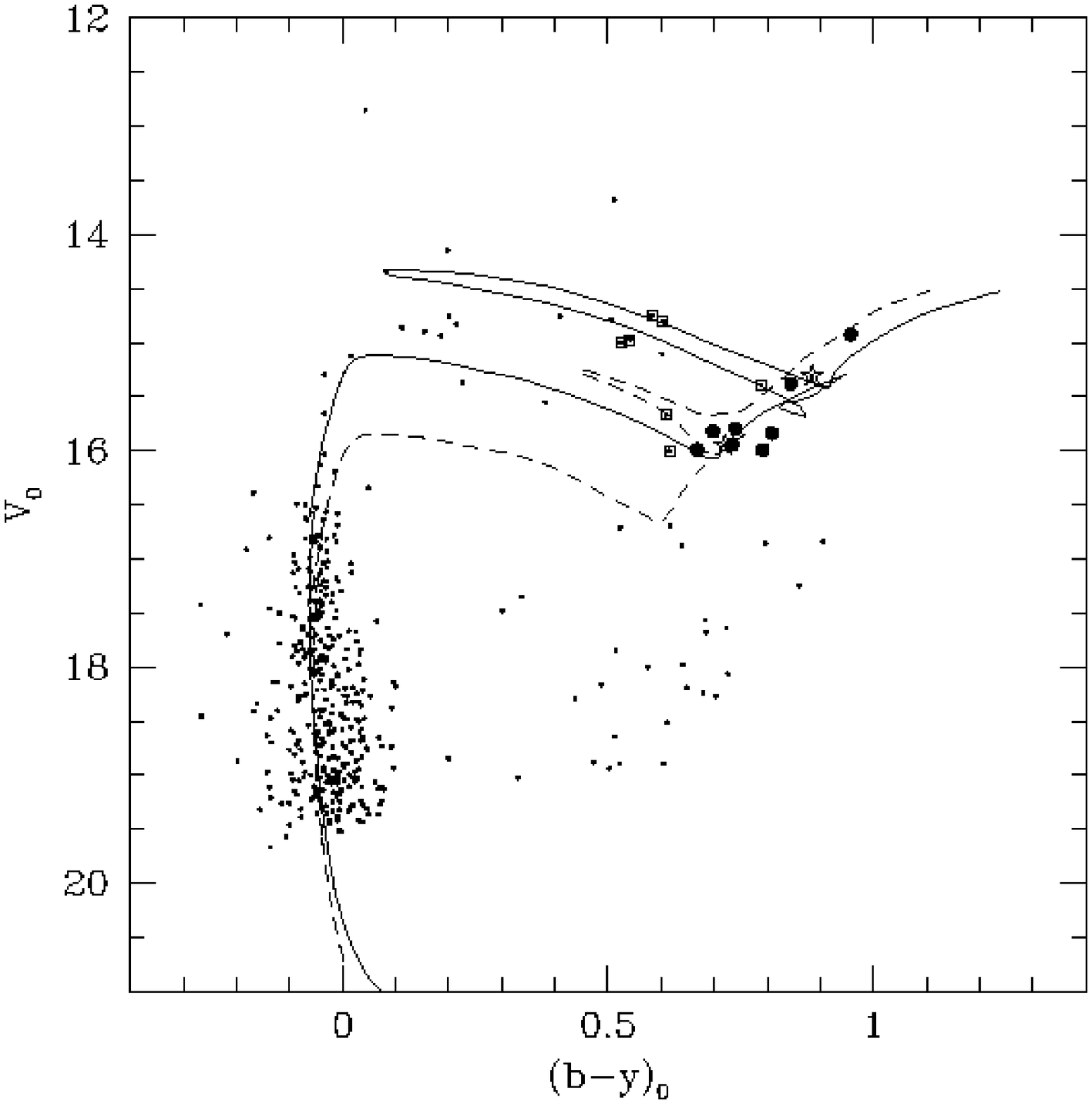}}
\caption{
  CMD of the NGC~2136/37 stars. The red giants used for the
  metallicity determination are denoted by the same symbols as in the
  two-colour diagram (Fig.~\ref{fig:2136cme}). Overlayed is a $-0.4$
  dex Geneva isochrone with an age of $10^{7.9}$ yr (solid line) and
  one with a metallicity of $-0.7$ dex and an age of $10^{8.0}$ y
  (dashed line).}
\label{fig:2136cage}
\end{minipage}
}
\put(9,16.5){
\begin{minipage}[t]{8.5cm}
\resizebox{\hsize}{!}{\includegraphics{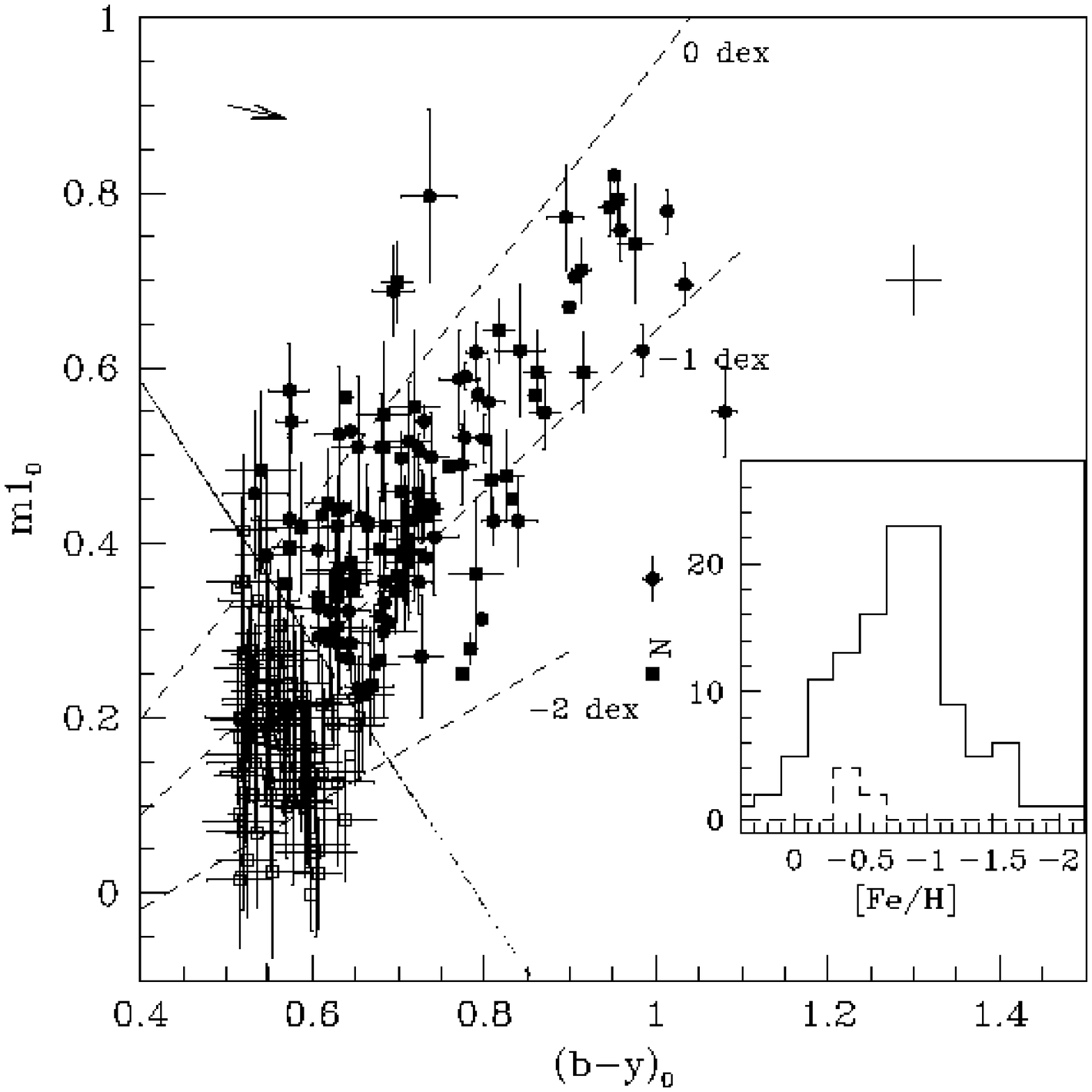}}
\caption{
  Two-colour diagram for the field stars around NGC~2136/37 (further
  than $120\arcsec$ away from the centre of NGC~2136 and further than
  $50\arcsec$ away from the centre of NGC~2137). Filled circles and
  open stars symbols mark stars that have been used to calculate the
  mean and standard deviation of the field star metallicity.  Open
  squares show stars for which the calibration is in principle valid,
  but they have been excluded due to our colour selection. In the
  insert the solid line shows the metallicity distribution of all
  stars redder than the colour selection line, the dashed line
  indicates the metallicity distribution of the younger stars which
  are indicated by the open star symbols in the CMD
  ((Fig.~\ref{fig:2136fage}).}
\label{fig:2136fme}
\end{minipage}
}
\put(9,3.2){
\begin{minipage}[t]{8.5cm}
\resizebox{\hsize}{!}{\includegraphics{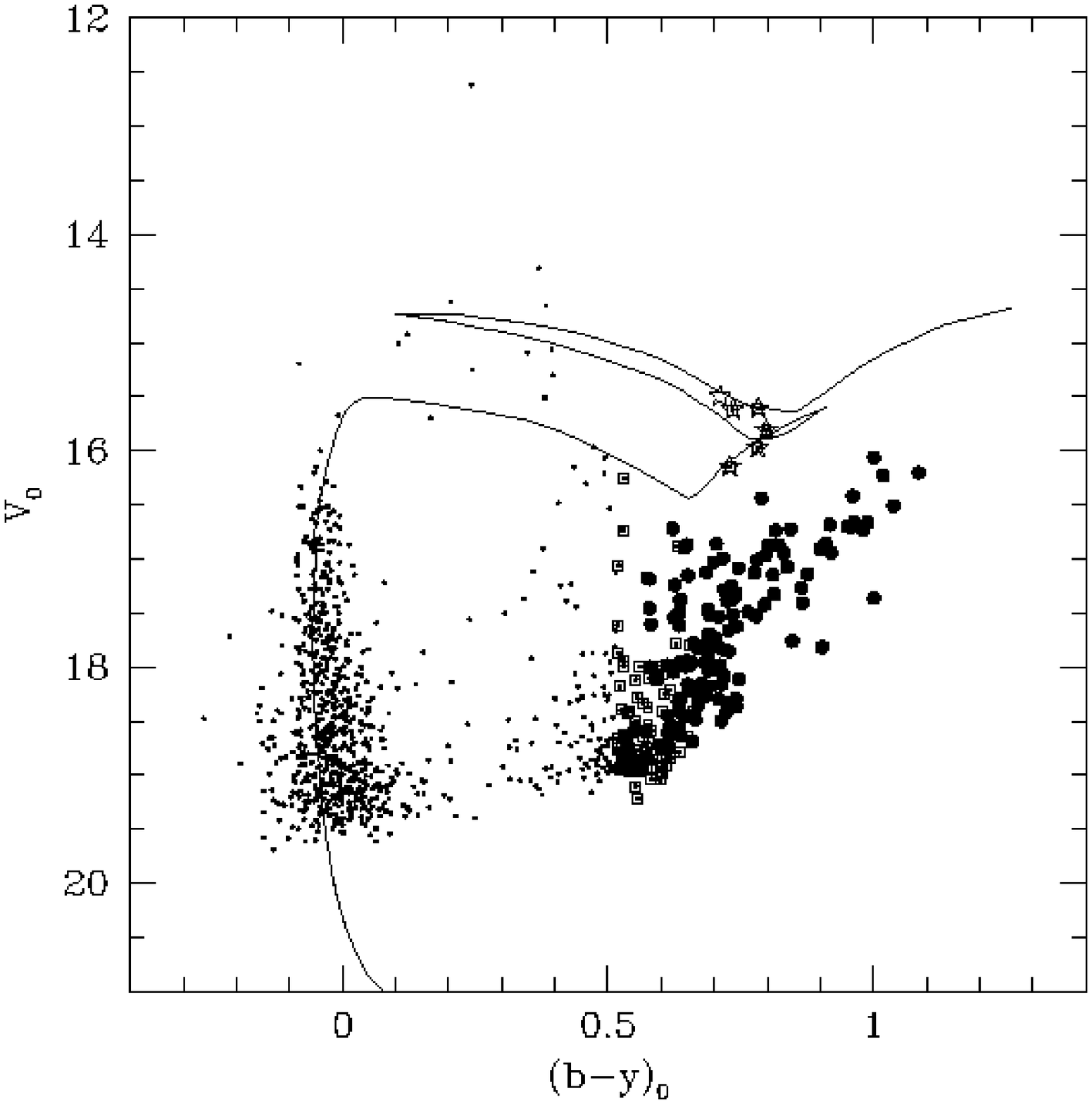}}
\caption{
  CMDs of the field population around NGC~2136/37. The symbols of the
  marked stars are the same as the two-colour diagram
  (Fig.~\ref{fig:2136fme}). The isochrone has an age of $10^{8.0}$ yr
  and a metallicity of $-0.4$ dex.  }
\label{fig:2136fage}
\end{minipage}
}
\end{picture}
\end{figure*}

The resulting metallicity of NGC~2136 is $[Fe/H]=-0.55 \pm 0.06$ dex
(see Fig.~\ref{fig:2136cme}). Two stars have been excluded, one metal
rich one with solar metallicity and a metal poor one with $\approx -1$
dex.  The more metal rich star is most probably a binary star or the
centre of a background galaxy, since its $\chi^2$ value given by the
DaoPhot PSF fitting routine is worse than for stars with comparable
luminosity. The more metal poor star might be a remaining field star.
The stars used for the metallicity determination are shown together
with the $[Fe/H]$ histogram in Fig.~\ref{fig:2136cme}. Including the
calibration and reddening error we got $-0.55 \pm 0.23 $ dex.  This is
the same value as Hilker et al. (\cite{hilker95a}) one obtained.

The age of this cluster is $10^{8.0\pm0.1}$ yr. In
Fig.~\ref{fig:2136cage} Geneva isochrones with an age / metallicity of
$-0.4$ dex / $10^{7.9}$ yr and $-0.7$ dex / $10^{8.1}$ yr are
overlayed.

For NGC~2137 we have chosen a radial selection radius of $20\arcsec$.
Within this radius two stars remain after applying the usual selection
criteria. These stars are plotted with stars symbols in the two-colour
plot and CMD of NGC~2136/37 (Fig.~\ref{fig:2136cme} and
Fig.~\ref{fig:2136cage}). The metallicities and ages of NGC~2136 and
NGC~2137 agree well. Therefore, it is plausible that these clusters
are a physical pair and not just a chance superposition, as Hilker at
al. (\cite{hilker95a}) already stated.

\subsection{The surrounding field population}

The two-colour diagram of the field star population is shown in
Fig.~\ref{fig:2136fme} and the corresponding CMD in
Fig.~\ref{fig:2136fage}.  The stars brighter than $V=16.2$ are younger
than the majority of the RGB stars and are marked with open stars in
the two-colour diagram and the CMD of the field population. Unlike the
case of the field population around NGC~1711, the younger stars have a
lower metallicity than the dominating older RGB field stars, however,
also a large fraction of RGB stars have the same metallicity. We
measured $[Fe/H] = -0.75$ dex for the mean metallicity and $0.59$ dex
for the standard deviation.  For the younger population (star symbols
in Fig.~\ref{fig:2136fage}) we derive an abundance of $[Fe/H]=-0.46$
and a standard deviation of $0.11$ dex. In Fig.~\ref{fig:2136fage} an
isochrone with a metallicity of $-0.4$ dex and an age of $10^{8.0}$ yr
is plotted that fit these stars.

\section{NGC~2031}

\subsection{The cluster}

\begin{figure*}
\setlength{\unitlength}{1cm}
\begin{picture}(18,25)
\put(0,16.5){
\begin{minipage}[t]{8.5cm}
\resizebox{\hsize}{!}{\includegraphics{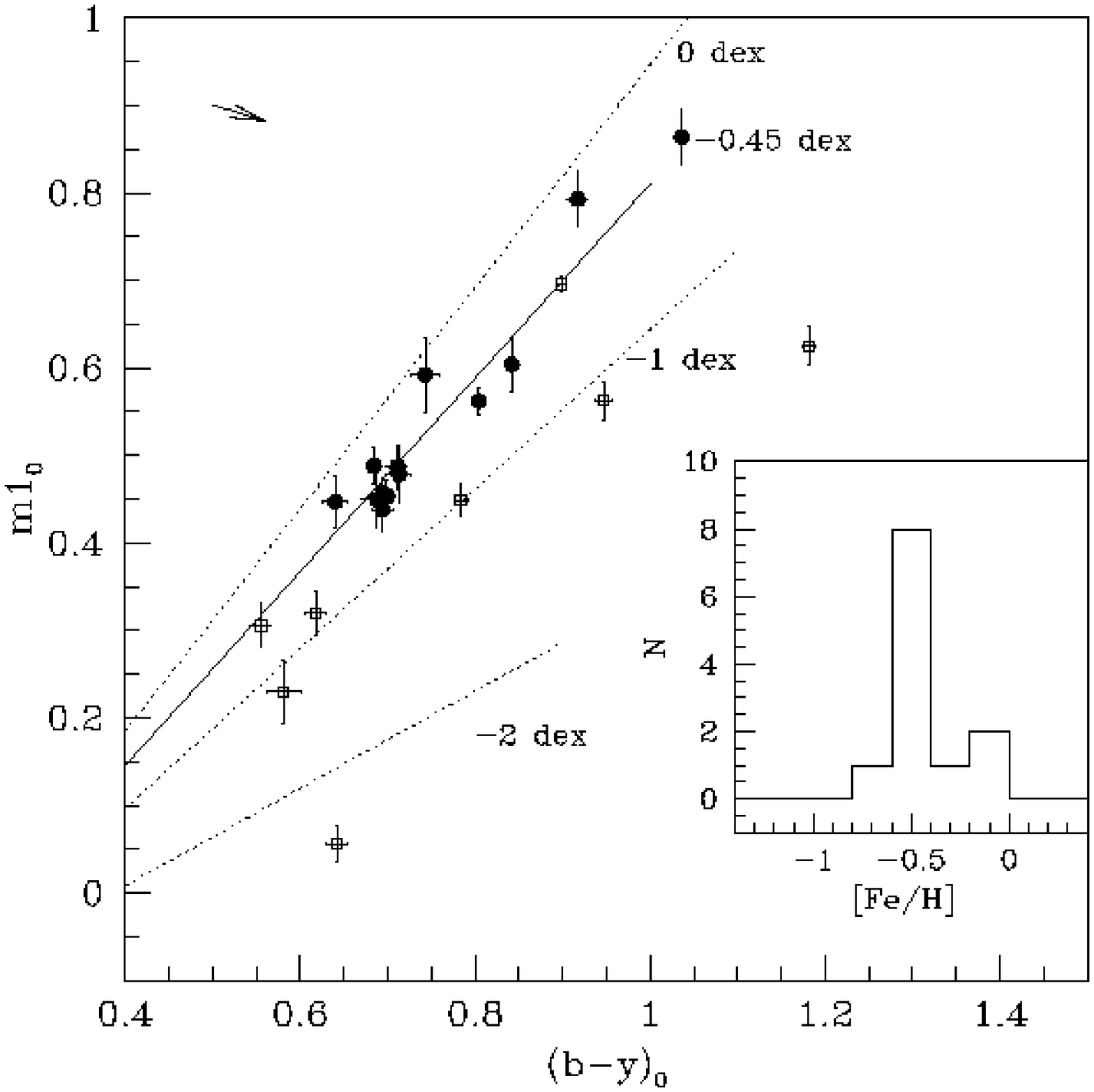}}
\caption{
  Two-colour diagram of the stars within $80\arcsec$ from the centre
  of NGC~2031.  The stars that have been used for the determination of
  the metallicity of NGC~2031 are shown with solid circles and in the
  insert with the solid line in the histogram.  Excluded stars are
  displayed using with open squares: one is redder than the used
  colour range, four are bluer and 2 stars lie well apart from the
  mean location of the other stars in the CMD shown in
  Fig.~\ref{fig:2031cage}.}
\label{fig:2031cme}
\end{minipage}
}
\put(0,4){
\begin{minipage}[t]{8.5cm}
\resizebox{\hsize}{!}{\includegraphics{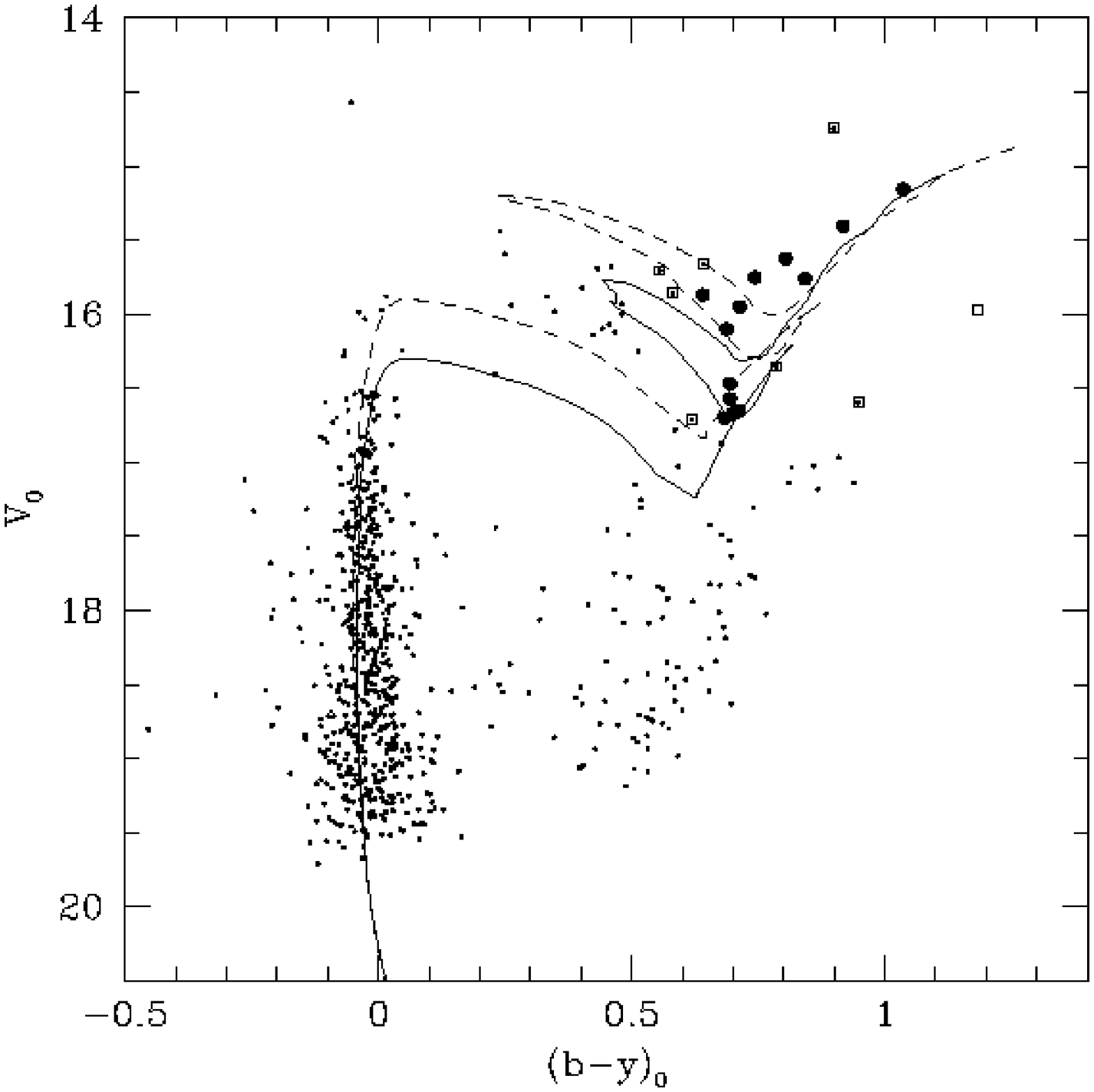}}
\caption{
  CMD of NGC~2031. The symbols are the same as those in
  Fig.~\ref{fig:2031cme}.  Overlaid are Geneva isochrones with a
  metallicity of $-0.4$ dex and an age of $10^{8.0}$ yr and $10^{8.1}$
  yr, respectively.}
\label{fig:2031cage}
\end{minipage}
}
\put(9,16.5){
\begin{minipage}[t]{8.5cm}
\resizebox{\hsize}{!}{\includegraphics{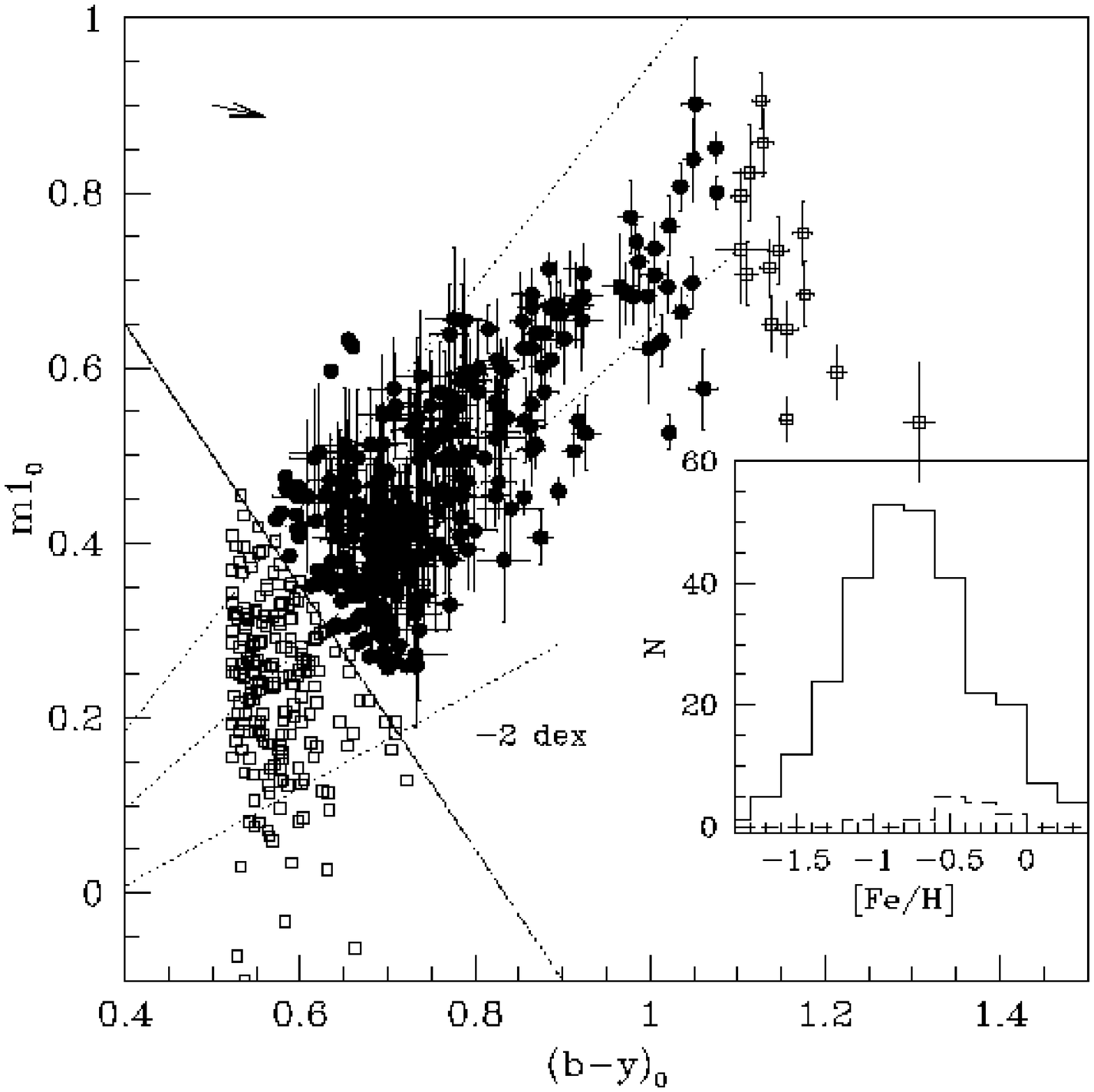}}
\caption{
  Two-colour plot for the field stars around NGC~2031 (with a radial
  distance larger than $110\arcsec$). Filled circles mark stars which
  have been used to calculate the mean and standard deviation of the
  field star metallicity. The open squares indicate excluded stars In
  the small panel the solid line shows the metallicity distribution of
  the older stars, the one of the younger stars is plotted using a
  dashed line. The younger stars are selected in the CMD
  (Fig.~\ref{fig:2031fage}) and are marked in the CMD with open star
  symbols. }
\label{fig:2031fme}
\end{minipage}
}
\put(9,4){
\begin{minipage}[t]{8.5cm}
\resizebox{\hsize}{!}{\includegraphics{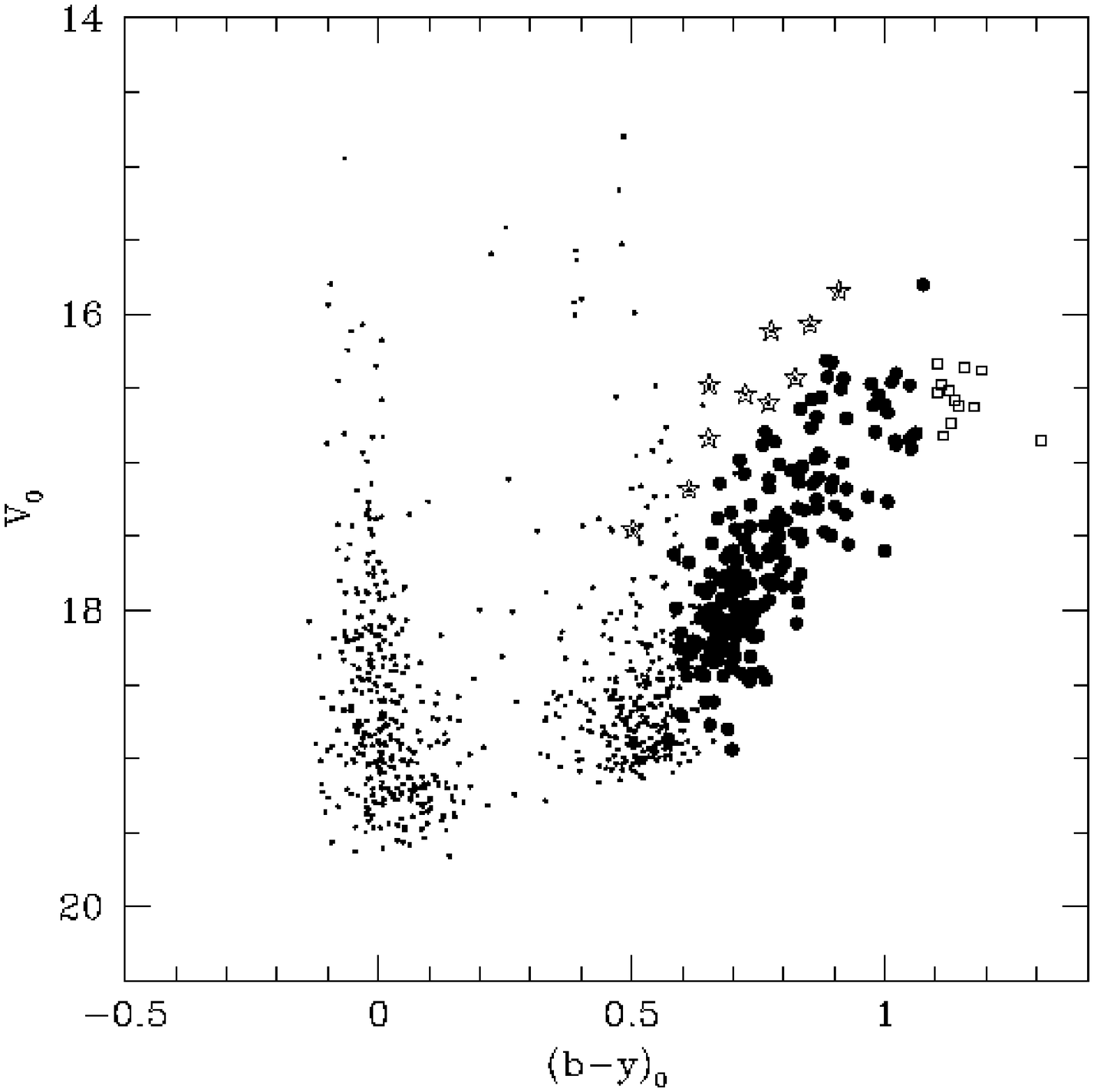}}
\caption{
  CMD of the field population surrounding NGC~2031 (they have a
  distance of more than $110\arcsec$ from the cluster centre). The
  filled circles are the older stars that have been used for the
  metallicity determination of the field.  The open star symbols have
  been selected due to their younger age, their metallicity
  distribution is shown in the insert in Fig.~\ref{fig:2031cage} with
  the dashed line.}
\label{fig:2031fage}
\end{minipage}
}
\end{picture}
\end{figure*}

To select the members of NGC~2031 we chose (from the radial density
distribution of stars) $75\arcsec$ as a good radius to separate
cluster and field stars effectively. In addition to our usual
selection criteria we have also excluded the very metal poor star with
$-2.3 \pm 0.3$ dex, which is most probably a foreground star, judging
from its very deviant metallicity. Another excluded star is slightly
above the cluster RGB. However, its metallicity fits well to the
metallicity of the cluster.  The stars used for the metallicity and
age determination are shown in Fig.~\ref{fig:2031cme} and
Fig.~\ref{fig:2031cage}.

For this cluster we have found a reddening of $E_{b-y} = 0.06 \pm
0.03$ ($E_{B-V} = 0.09 \pm 0.04$).  Mould et al. (\cite{mould93})
quote $E_{B-V} = 0.18 \pm 0.05$ based on HI measurements.  Schlegel et
al. (\cite{schlegel98}) give the even larger value of $E_{B-V}=0.3$ as
galactic foreground reddening in this direction. On the other hand
derived Schwering \& Israel (\cite{schwering91}) a reddening of
$E_{B-V}=0.1$. Burstein \& Heiles (\cite{burstein82}) derived a
reddening of $E_{B=V}=0.07$.  Using our reddening value a metallicity
of $[Fe/H] = -0.52 \pm 0.21$ can be derived (the error includes the
reddening error and the calibration error). With this metallicity we
found the best fitting isochrone to be $10^{8.1 \pm 0.1}$ yr. This
agrees well with the age determined by Mould et al. (\cite{mould93})
($10^{8.14 \pm 0.05} $ yr with $-0.4$ dex).

\subsection{The surrounding field population}

The mean metallicity of the field population for this cluster is
$[Fe/H] = -0.75$ dex and the standard deviation $0.44$ dex.
Definitely young stars are selected in the CMD
(Fig.~\ref{fig:2031fage}) and are marked with open star symbols, while
the older stars which can be used to determine a metallicity are
marked with filled circles. The metallicity distribution of the older
stars is shown in the small panel with the solid line and the
distribution of the younger stars with a dashed line. It can be seen,
that we do not find young metal poor stars, although we also find
older stars with the same metallicity as the younger ones. However,
the younger stars are in average more metal rich than the older stars.

\section{NGC~1651}

\subsection{The cluster}

The RGB of NGC~1651 merges with the RGB of the field population which
can be seen in the CMD of all stars found in the field around NGC~1651
(Fig.~\ref{fig:cmd}).  This makes the visual separation of field and
cluster stars in the CMD impossible.  It is only possible to
distinguish a few young field stars unambiguously from the mixture of
field and cluster RGB.

\begin{figure*}
\setlength{\unitlength}{1cm}
\begin{picture}(18,25)
\put(0,16.5){
\begin{minipage}[t]{8.5cm}
\resizebox{\hsize}{!}{\includegraphics{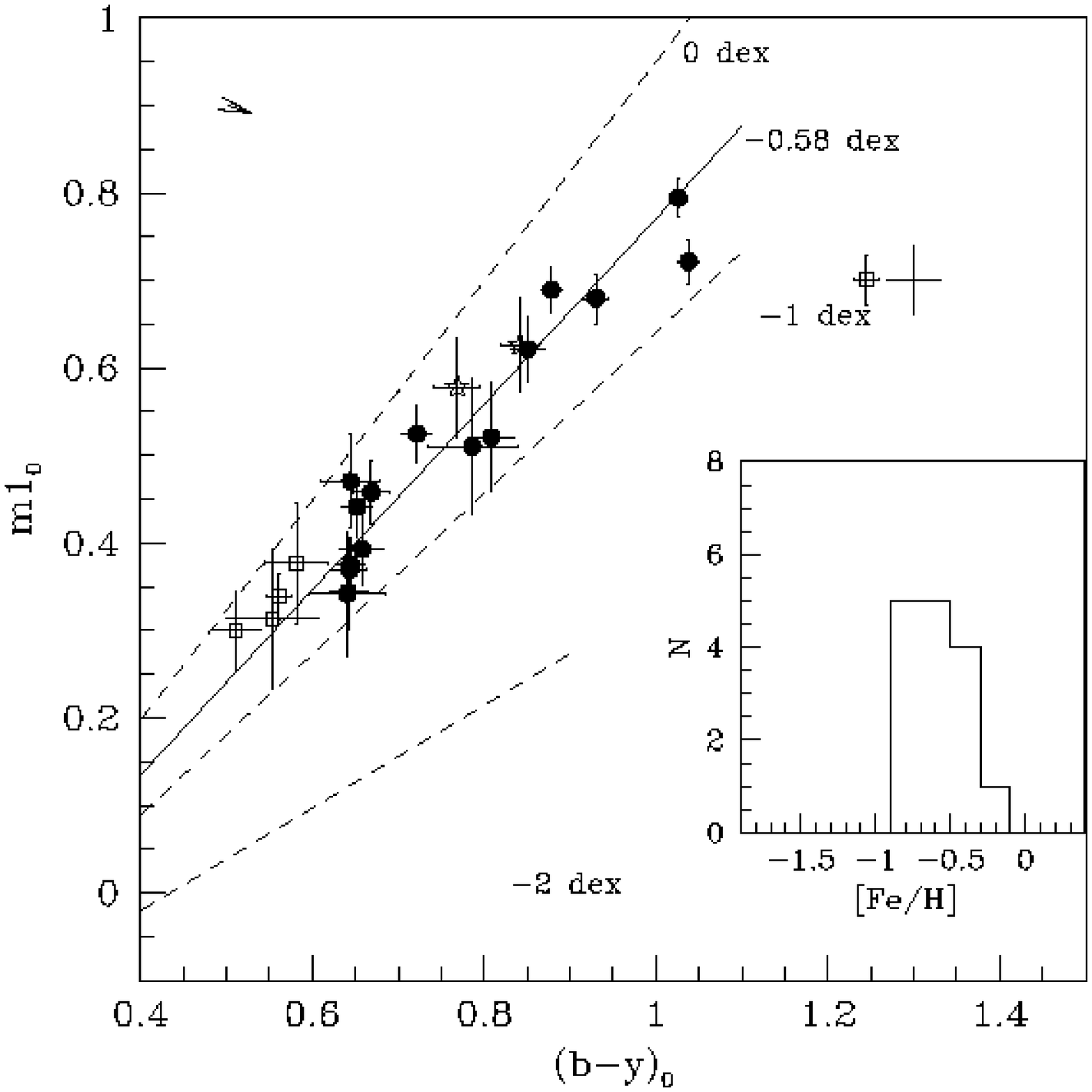}}
\caption{
  Two-colour diagram of NGC~1651. The cluster stars have a distance of
  less than $60\arcsec$ from the cluster centre. The filled circles
  denote stars which have been used for the metallicity determination.
  The stars marked with open star symbols have been excluded because
  they deviate considerably from the mean RGB location in the CMD of
  this cluster (Fig.~\ref{fig:1651cage}).}
\label{fig:1651cme}
\end{minipage}
}
\put(0,4.5){
\begin{minipage}[t]{8.5cm}
\resizebox{\hsize}{!}{\includegraphics{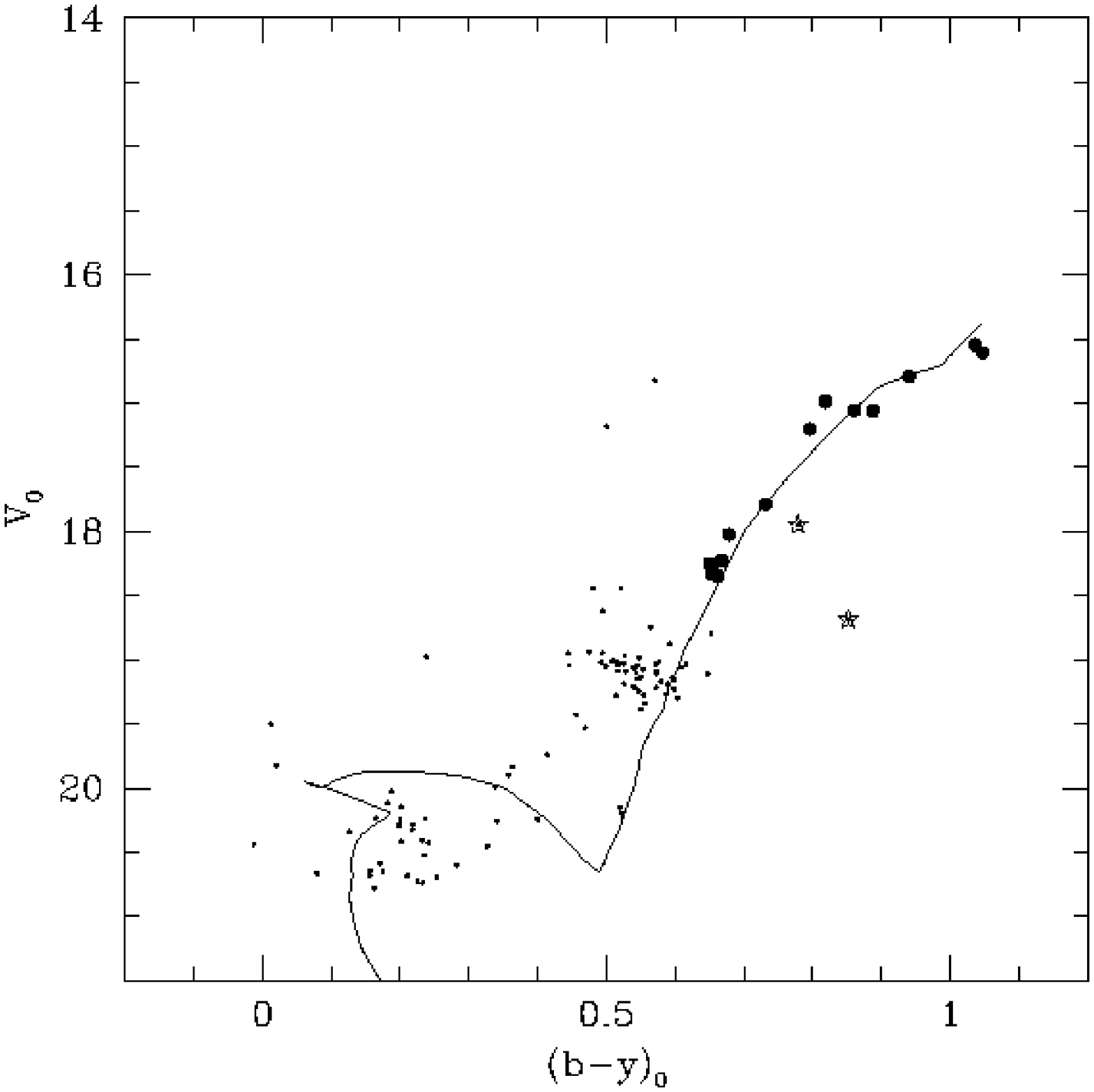}}
\caption{
  CMD of NGC~1651 with a $-0.4$ dex, $10^{9.1}$ yr (solid line)
  isochrone and a $-0.7$ dex, $10^{9.2}$ yr isochrone over-plotted.
  The filled circles are the stars that have been used for the
  metallicity measurement.}
\label{fig:1651cage}
\end{minipage}
}
\put(9,16.5){
\begin{minipage}[t]{8.5cm}
\resizebox{\hsize}{!}{\includegraphics{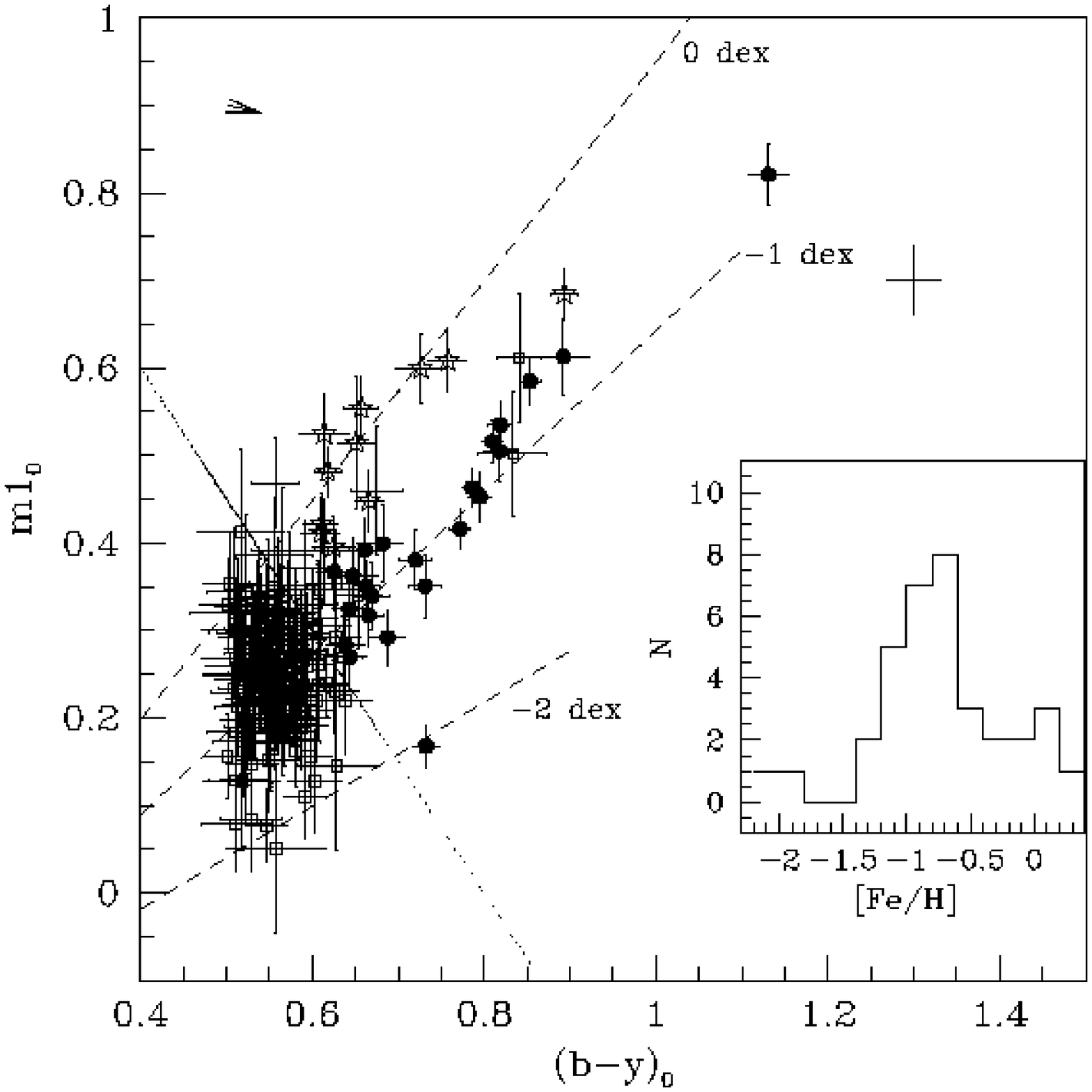}}
\caption{
  Two-colour diagram of the field population around NGC~1651. The
  stars that have been used for the metallicity determination are
  marked with filled circles and open star symbols. In the insert
  their metallicity distribution is plotted as histogram. The
  separation into stars marked by filled circles or open star symbols
  has been applied according to their location in the two-colour
  diagram (metal poor/ metal rich).}
\label{fig:1651fme}
\end{minipage}
}
\put(9,4.5){
\begin{minipage}[t]{8.5cm}
\resizebox{\hsize}{!}{\includegraphics{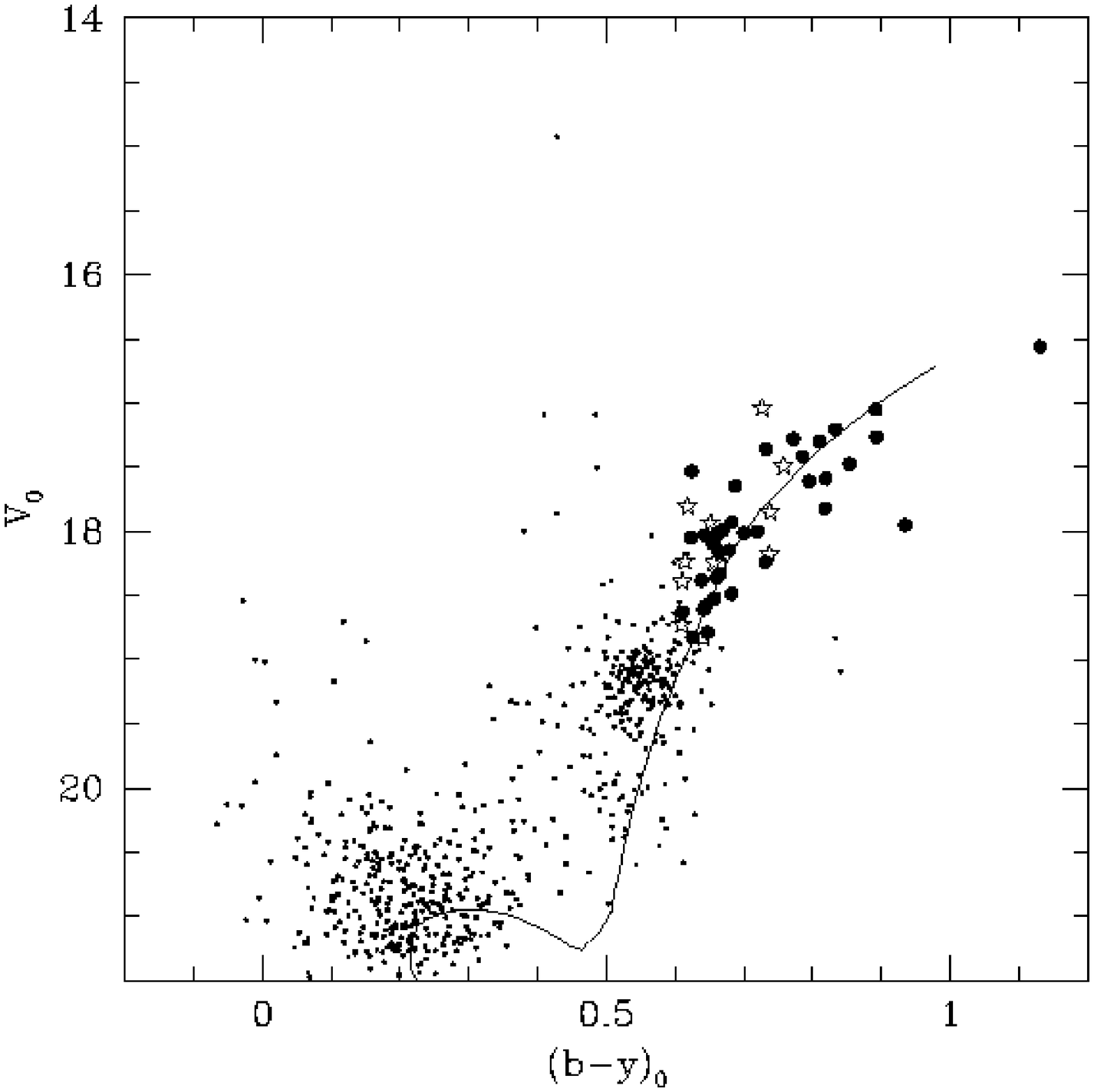}}
\caption{
  CMD of the field population around NGC~1651. The symbols correspond
  to those used in Fig.~\ref{fig:1651fme}. The isochrone has a
  metallicity of $-1$ dex and an age of $4$ Gyr.}
\label{fig:1651fage}
\end{minipage}
}
\end{picture}
\end{figure*}

We decided, after inspecting the radial number density of stars around
NGC~1651, to use a radius of $55\arcsec$ for the field and cluster
separation. In addition we excluded three stars lying well below the
cluster RGB. They are most probably foreground stars in the Galactic
halo because a LMC star with such a luminosity and metallicity would
have an unreasonable age of more than $20$ Gyr. Also two stars above
the mean RGB of the cluster stars have been excluded. All these
excluded stars have been marked in the CMD and two-colour diagram of
NGC~1651. The remaining stars used for the metallicity and age
determination are shown as filled circles in the two-colour diagram
(Fig.~\ref{fig:1651cme}) and CMD ( Fig.~\ref{fig:1651cage}).

The reddening given by Schwering \& Israel (\cite{schwering91}) is
$E_{B-V} = 0.08$, Mould et al. (\cite{mould97}) used $E_{B-V}=0.1$ for
this cluster, Schlegel et al. (\cite{schlegel98}) found $E_{B-V}=0.14$
towards this direction and Burstein \& Heiles give $E_{B-V}=0.1$.
With a reddening of $E_{B-V}=0.1$ the cluster metallicity would be
$-0.28 \pm 0.02$ dex. However, no fitting isochrone with such
reddening and metallicity can be found.  The theoretical RGB stars of
this rather large metallicity would be too red or - for younger ages -
the main sequence should be visible. The only possibility to fit an
isochrone is to use a smaller reddening and hence a lower metallicity.
Only with such a metallicity a self-consistent fit in the CMD and the
two-colour diagram can be found.

Using $E_{b-y} = 0.03$ ($E_{B-V} = 0.04$) the cluster metallicity
derived is $[Fe/H] = -0.58 \pm 0.02$ dex. The two-colour diagram for
this reddening is shown in Fig.~\ref{fig:1651cme}. The calibration
error accounts for an additional error of $0.19$ dex.  Using this
metallicity the age is $10^{9.3\pm0.1}$ yr. The isochrone is overlayed
to the cluster CMD in Fig.~\ref{fig:1651cage}.

However, this solution is not unique: also with a smaller reddening
acceptable fits are possible, resulting in larger ages and smaller
metallicities, hence we only got a range of parameters for this
cluster: $0.01 < E_{b-y} <0.05$, $-0.65<[Fe/H]<-0.45$,
$10^{9.4}>lg(Age)>10^{9.0}$.  The calibration uncertainty has to be
included into these upper and lower limits.  Three stars around this
cluster have been spectroscopically investigated by Olszewski et al.
(\cite{olszewski91}). Two have been identified, both being very red
($(b-y)>1.3$. They are also identified AGB stars (Frogel \& Blanco
\cite{frogel90}). These stars have a metallicity of $-1.33$ dex and
$-1.6$ dex, a more metal rich ($-0.37$ dex) AGB star could not be
identified.

NGC~1651 has been observed with the HST by Mould et al.
(\cite{mould97}). They derived an age of $10^{9.2 \pm 0.1}$ yr (using
$[Fe/H] = -0.4$ dex). The age is in good agreement despite the
metallicity discrepancy.

The elongated and tilted red clump of the cluster (see
Fig.~\ref{fig:1651cage}) is a feature that remains worth mentioning.
The elongation is approximately along a reddening vector, however
strong differential reddening should not cause this shape since the
RGB does not show a similar large colour spread. We consider it more
probable that this it is an intrinsic feature of an HB of a certain
age and metallicity. The red clump of the field population does not
show this elongated shape, it is rather a slightly fainter clump. Such
an elongated red clump has also been observed by Piatti et al.
(\cite{piatti99}) with in the Washington system in three of their $21$
investigated fields. In case of NGC~2209 they discuss the possibility
that an increased helium content or differential reddening could cause
such a red clump morphology.

\subsection{The surrounding field population }

The dominating field population around NGC~1651 $[Fe/H] = -0.75$ using
the low reddening of $E_{b-y}=0.04$. The two-colour diagram for the
field population is shown in Fig.~\ref{fig:1651fme} and the
corresponding CMD in Fig.~\ref{fig:1651fage}. The red clump of the
field population is not elongated and slightly fainter.

In this field, two groups of stars show up with distinct
metallicities: one group with an approximately solar abundance
($[Fe/H] = 0.03 \pm 0.03$ dex) and one group around $-1$ dex. The
metal poor stars can be fitted with a Geneva isochrone of $-1$ dex and
an age on the order of $10^{9.6}$ yr. However, it is impossible to
find a fitting isochrone for the apparent more metal rich stars
(around $0$~dex), since these stars would have an age around
$10^{8.5}$~yr and thus many more main sequence stars should be
present.  Only with the assumption of no reddening being present, one
would have derived a metallicity of $[Fe/H]=-0.21 \pm 0.19$ dex for
these stars. With such a metallicity they could have been fitted with
a $10^{8.8 \pm 0.1}$ yr Geneva isochrone.

\section{NGC~2257}

\subsection{The cluster}

The oldest cluster in our sample is NGC~2257, which can be seen from
the cluster CMD (Fig.~\ref{fig:2257cage}) that is very similar to the
CMDs of Galactic globular clusters. Especially the pronounced blue
horizontal branch (HB) is a sign for an old, metal poor population.
NGC~2257 lies $\simeq 9^0$ away from the centre of the LMC to the
north east. Because of this large distance the field is very sparsely
populated, thus we renounced a radial selection, because no radius can
be found at which the field stars dominate.

No reddening determination via a main sequence is possible. However,
only a reddening of less than $E_{b-y}=0.06$ results in consistency of
a reasonable age with a reasonable metallicity. We adopt here the
reddening used by Testa et al. (\cite{testa95}), $E_{B-V}=0.04$ and an
error of $\Delta E_{B-V}=0.04$. Schwering \& Israel's map
(\cite{schwering91}) shows a reddening of $E_{B-V}=0.03$ at position
of NGC~2257, Schlegel et al.  {\cite{schlegel98}) give $E_{B-V}=0.06$
  and Burstein \& Heiles (\cite{burstein82}) $E_{B-V}=0.04$.  With a
  reddening of $E_{B-V}=0.04$ the metallicity measurement results in
  $[Fe/H]=-1.63\pm0.21$ dex (including the calibration errors). In the
  two-colour diagram the metallicity determination is illustrated in
  Fig.~\ref{fig:2257cme}.

\begin{figure*}[t]
\setlength{\unitlength}{1cm}
\begin{picture}(18,11.5)
\put(0,3){
\begin{minipage}[t]{8.5cm}
\resizebox{\hsize}{!}{\includegraphics{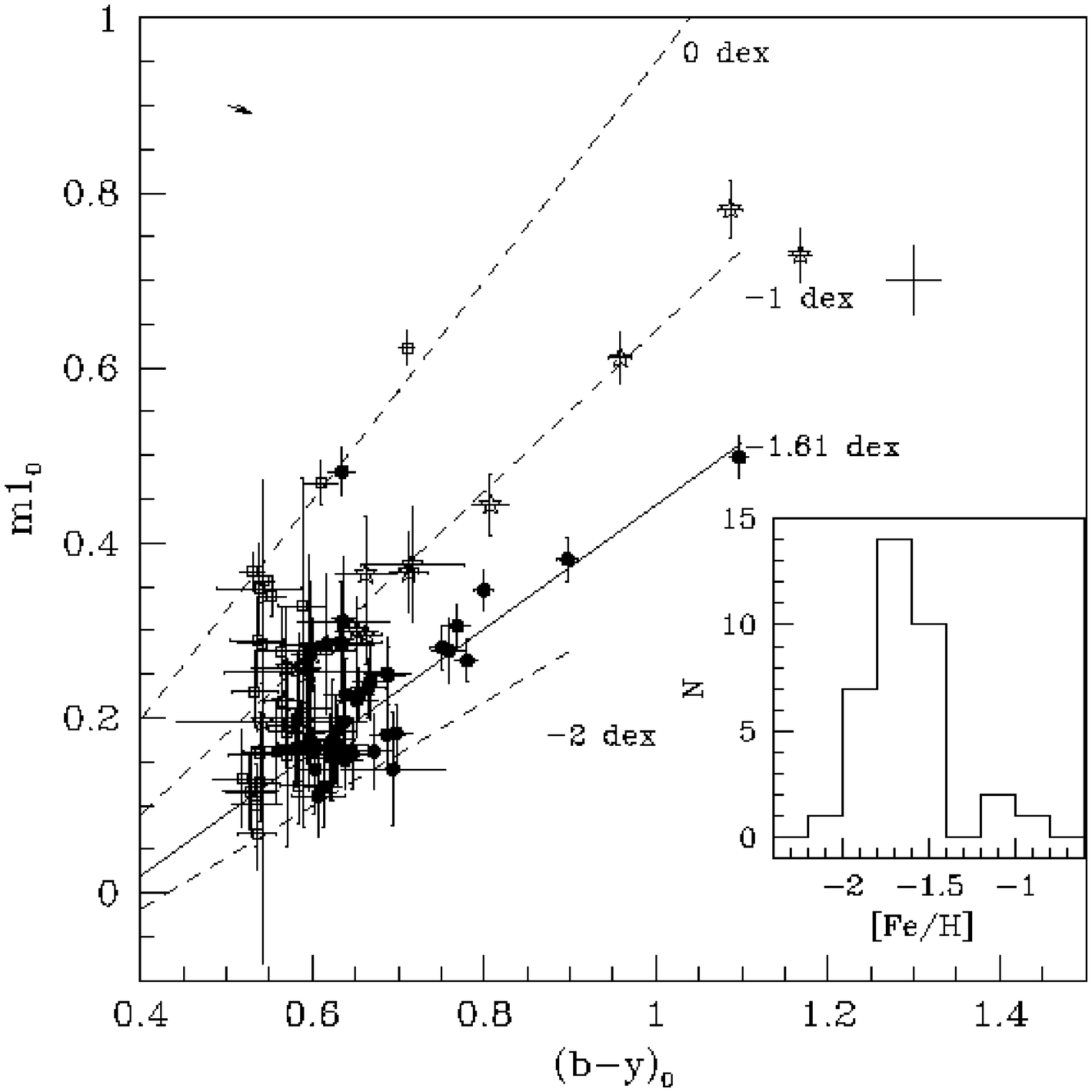}}
\caption{
  Two-colour diagram of the stars in the field of NGC~2257. Since no
  radial selection criterion has been applied, stars have only been
  selected in the CMD, according to colour and luminosity. Stars being
  most probable cluster members are shown as filled circles, stars
  fainter the mean RGB are shown as open star symbols and brighter
  stars as squares.  Stars bluer $(b-y)=0.6$ for which the calibration
  is valid, but have been discarded are shown as open squares as well.
  For the selection criterion also refer to Fig.~\ref{fig:2257cage}.}
\label{fig:2257cme}
\end{minipage}
}
\put(9,3){
\begin{minipage}[t]{8.5cm}
\resizebox{\hsize}{!}{\includegraphics{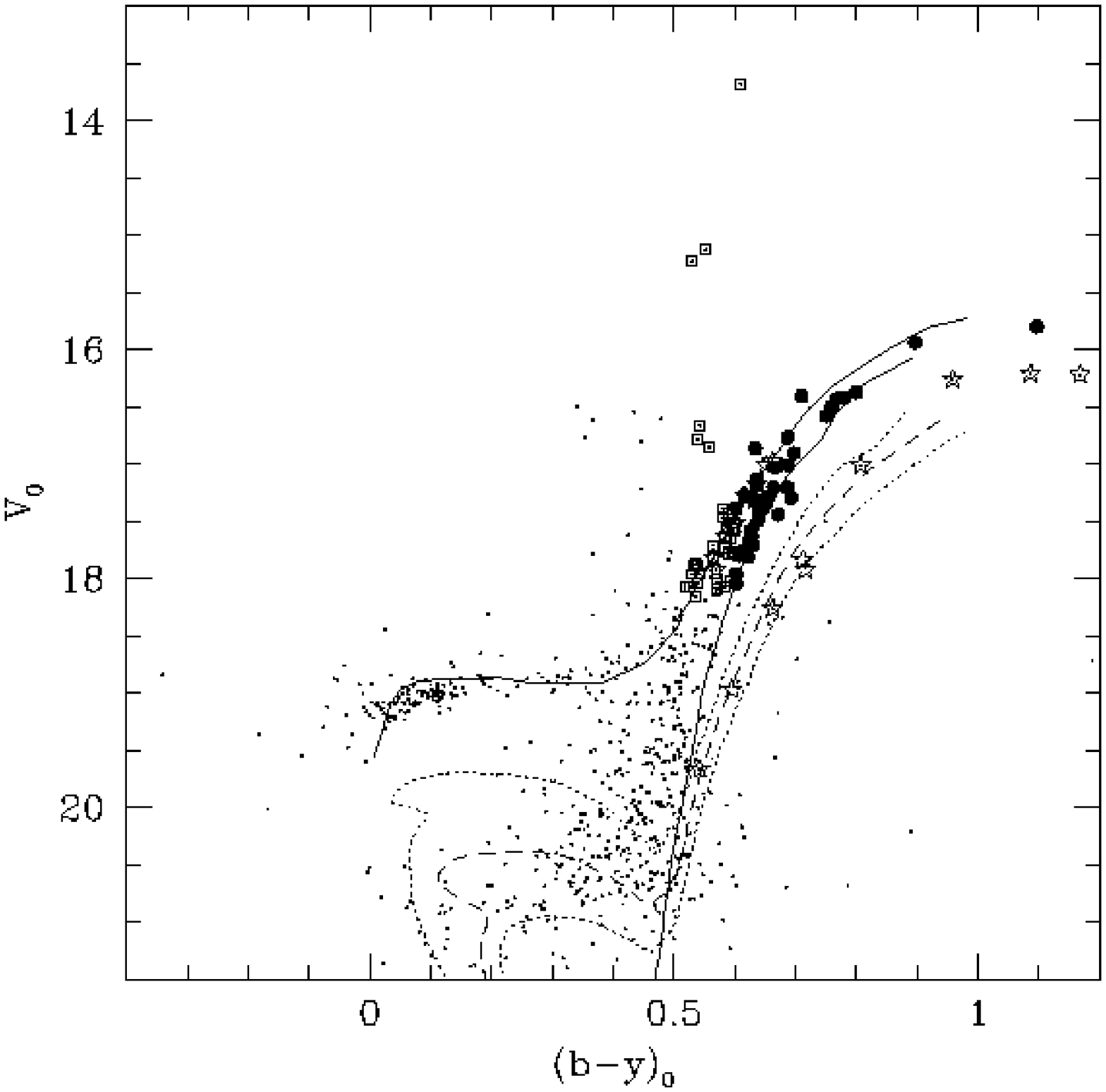}}
\caption{
  CMD of the stars around NGC~2257. The solid Padua isochrone has a
  metallicity of $-1.7$ dex and an age of $10^{10.24}$ yr. The dashed
  Geneva isochrone has a metallicity of $-1$ dex and an age of
  $10^{9.4}$ yr and the dotted Padua isochrones have $-1.3$ dex and
  $10^{9.2}$~yr, $10^{9.6}$~yr. Symbols as in Fig.~\ref{fig:2257cme}.}
\label{fig:2257cage}
\end{minipage}
}
\end{picture}
\end{figure*}
The age has been determined like in the previous cases, however since
no Geneva isochrone with such an age and metallicity had been
available we used Padua isochrones. The best matching (-1.7 dex)
isochrone is overlaid on the CMD of this field in
Fig.~\ref{fig:2257cage}. The RGB of an old population evolves slowly
and thus its location does not differ much between different ages.
Therefore, we can only define an age range from $10^{10}$ yr to
$10^{10.3}$ yr for this cluster, when limiting the age determination
to the RGB. If the isochrones represent well the colour of the HB,
then the age uncertainty will drop severely, since the HB of younger
clusters is much redder (around $(b-y)\simeq0.3$) than that of older
ones. In this case the allowed age range is merely $10^{10.24} -
10^{10.30}$ yr. The linear age uncertainty is much larger than for the
younger clusters, discussed above. The reason for this is not an
increased photometric error, but rather an intrinsic effect of the
age-luminosity evolution: the brightness (in magnitudes) of the RGB
decreases with age approximately logarithmically, therefore the error
in the age of young and old stars should be in first order the same in
the {\it exponent}. For the younger cluster this error was around 0.1,
comparable to the error in case of NGC~2257.
 
Testa et al. (\cite{testa95}) obtained deep B and V HST observations
and concluded from the Turn-Off location and an assumed metallicity of
-1.7 dex that the age is $\simeq 10^{10.1}$ yr. Geisler et al.
(\cite{geisler97}) found an age of $10^{10.07}$ yr. Recently,
  Johnson et al. (\cite{johnson99}) used deep HST observations of
  three true LMC globular clusters including NGC~2257 to derive their
  relative ages compared to Galactic globulars.  They found that these
  old clusters have an age that is not distinguishable from that of
  M~3 and M~92.

\subsection{The surrounding field population} 

In spite of the low stellar density, a few field stars could have been
identified due to their deviating age and metallicity. In the
two-colour diagram (Fig.~\ref{fig:2257cme}) stars having a metallicity
around $-1$ dex and being redder than $b-y=0.6$ have been marked with
open star symbols.  These stars form in the CMD
(Fig.~\ref{fig:2257cage}) a RGB lying below the cluster RGB. With a
metallicity of $[Fe/H] = -1$ dex the age of the population is
$10^{9.4}$ yr.  No other field component (except probably some
Galactic foreground stars around $(b-y)=0.5$) have been found.

The age and metallicity obtained for the field stars agrees well with
the values found for the field population around the investigated
clusters that are closer to the LMC centre. This indicates that, if a
radial metallicity gradient exists, it cannot be very pronounced. This
is consistent with the result of Olszewski et al. \cite{olszewski91}
who did not find a radial metallicity gradient in the cluster
population of the same age. However, there is a metallicity gradient
in the sense that younger clusters and field stars tend to be more
concentrated (Santos Jr. et al. \cite{santos99}) and thus the mean
metallicity of all stars should show a metallicity gradient. In
contrast Kontizias et al. \cite{kontizias93} found a metallicity
gradient in the outer cluster system while none was observable for the
inner clusters.

The field population in the vicinity of NGC~2257 has been studied by
Stryker (\cite{stryker84}). Her analysis revealed that the metallicity
of the field population is larger than that of NGC~2257 and that {\it
  ``star formation occurred in the field long after the formation of
  the cluster''.} She estimated the age of the field component to be
$6-7$ Gyr old. We recalculated the age of this population based on her
photographic CMD: we estimated the luminosity difference between the
turn-off of this population and its red HB to be $\Delta V=2 \pm 0.3$.
Walker et al. (\cite{walker99}) also found a younger field population
around NGC~2257 with HST observations. The field population around
this cluster is best fitted by a Padua isochrone 
with an age of $3.5$~gyr and a metallicity of $-0.6$~dex.
Using $[Fe/H]=-1$ and applying the calibration given in Binney \&
Merrifield (\cite{binney98}) we end up with an age of $4\pm1 $ Gyr for
this population which is comparable to our derived age.

\section{The method to derive an AMR and SFH of the field population}

In order to extract detailed information concerning an AMR and/or the
chemical enrichment history of the field, it is necessary to use a
more sophisticated analysis than just deriving a mean metallicity and
its standard deviation.  Thus we estimated an age for each star with a
measured metallicity. For these measurements a set of isochrones with
different ages for each metallicity has to be available. Therefore we
interpolated Geneva isochrones linearly to generate isochrones that
are continuously distributed in metallicity.  For each star the
isochrone of the appropriate abundance with the minimal luminosity
difference to the star has been identified and the corresponding age
has been assigned to the star. This method obviously resulted in
discrete age binning for the field stars (the size of the age bins is
$log(t)=0.1$). We did not interpolate in age, since the age
uncertainty due to the discrete age sampling is smaller than the error
due to the photometric error, the calibration error and the errors
connected to reddening and blending.

Our method is not free from ambiguity, since for most stars one can
derive two solutions, one for the RGB and one for the AGB. For stars
older than $10^{9.3}$ yr one can neglect this effect because {\bf a)}
the luminosity difference between the AGB and RGB and thus the
inferred age difference is small compared to the other uncertainties
in determining an age for these evolved stars and {\bf b)} the
fraction of AGB to RGB stars in our colour range is small due to the
lifetime difference. For younger ages the AGB stars play a
considerable role: they dominate in the used colour range for
populations with an age between $10^{8.5}$ yr and $10^{9.1}$ yr (see
e.g. the overlayed isochrone in Fig.~\ref{fig:1806cage}).  To account
for this problem we used the following approach: we determined an AGB
age and a RGB age for each star.  If the AGB {\it and} the RGB age
were older than $10^{9.3}$ yr we assigned only a RGB age to the star.
In case that we found a RGB age between $10^{8.4}$ yr and $10^{8.7}$
yr we used the AGB age to account for the dominance of the AGB stars
in our used colour range. For the other ages we used either a mean AGB
\& RGB age, if none of the isochrones had a luminosity difference of
less than $0.1$ mag or we used the age that correspond to the
isochrone with a luminosity difference of less than $0.1$ mag to the
star.

The uncertainties in the extension of the isochrone towards the red is
not critical for this investigation as long as one is not concerned
with the number of stars with a certain age/metallicity.  More severe
are possible problems in the shape of the theoretical models, which
might result in a systematic shift or distortion in the age scale,
thus all our results on the field population is valid for the only
currently available Str\"omgren isochrone set. However, since these
isochrones fit reasonably well to the studied clusters, we are
confident that the AMR is quite robust concerning the applied
isochrones.  We cannot circumvent this problem and it is necessary to
get Str\"omgren isochrones for more recent stellar models with which
the results can be compared.

Since the age resolution on the RGB is not very good the derived age
for an individual star is not more precise than a factor of $2$ for
older stars, therefore all results can only be interpreted
statistically.

The applied procedure resulted in an AMR and an age number
distribution (AND). The latter can be used to derive the Star
Formation History (SFH) of the combined population. However, one has
to bear in mind that the AND is not just the star formation rate (SFR)
counted in logarithmic bins. First and most obvious is the fact, that
stars of different ages have different masses on the RGB. Therefore
one has to account for the IMF to get the SFH. Secondly, systematic
shifts, for example due to differential reddening or binaries, have to
be considered. Finally one has to be aware of the fact that the
conclusions depend sensitively on the used set of isochrones. Taking
all these effects into account is a highly complex problem and there
is little hope to disentangle them analytically. To get nevertheless a
handle on these effects and an idea about the accuracy of our method,
we created synthetic CMDs of field populations with different ages and
metallicities using a Monte Carlo algorithm and Geneva isochrones. The
program allows one to include the (measured) errors, differential
reddening, depth structure, a binary fraction and arbitrary SFHs.
Binaries have been chosen randomly (according to a white random
distribution) and are not important for the further investigation that
regards only red giants. Even apart from the problems in interpreting
the AND, the age distribution and thus the SFH depends rather strongly
on the assumed reddening and possible differential reddening. Thus the
results should be regarded more in comparison with the SFH derived
with deep photometry (e.g. Gallagher et al. \cite{gallagher96}, than
as an independent measurement of the starformation history.  Holtzman
et al. \cite{holtzman99}, Elson et al. \cite{elson97}, Romaniello et
al. \cite{romaniello99}) and serve as a consistency check.

\subsection{Tests with simulated data}

To test our applied method we generated several artificial data sets.
In Fig.~\ref{fig:AMRsim}{\bf a,b} we show the resulting AMR and AND of
two of these simulations. The left one ({\bf a}) consists of three
populations with $10^{8.4}$ yr, $-0.2$ dex and $10^{9.1}$ yr, $-0.4$
dex and $10^{9.8}$ yr, $-0.7$ dex, respectively. The number ratio
old-to-young-stars is $1:1:10$. No error was applied. In the right two
panels (Fig.~\ref{fig:AMRsim}{\bf b}) we have included the photometric
errors, a binary fraction of 70\% and differential reddening of
$\Delta E_{B-V}=0.04$. The IMF used in both simulations (a Salpeter
IMF down to $0.8 M_{\odot}$) is not important for the resulting AMR
because of the small mass interval on the RGB. The open circles show
the input age and metallicity, while filled circles are used for the
extracted mean values for the metallicity.  However, the IMF is an
essential input parameter when deriving a SFH. Fortunately, it is not
very critical for relative number ratios as long as the IMF did not
change with time. This is an ad hoc assumption in our simulation since
we cannot constrain any mass function with our method.  The AMR of the
input population in the shown simulation follow the AMR proposed by
Pagel \& Tautvai\u{s}vien\.e (\cite{pagel98}). Thus this simulation
demonstrates our ability to be able to recover the shape of the AMR
proposed by these authors (on the basis of the given isochrones).  It
is impossible to see potential bursts in the SFH and it is clear that
it is very difficult to derive the SFH from this procedure, only rough
estimations of the SFH can be made.

\begin{figure}[t]
\resizebox{\hsize}{!}{\includegraphics{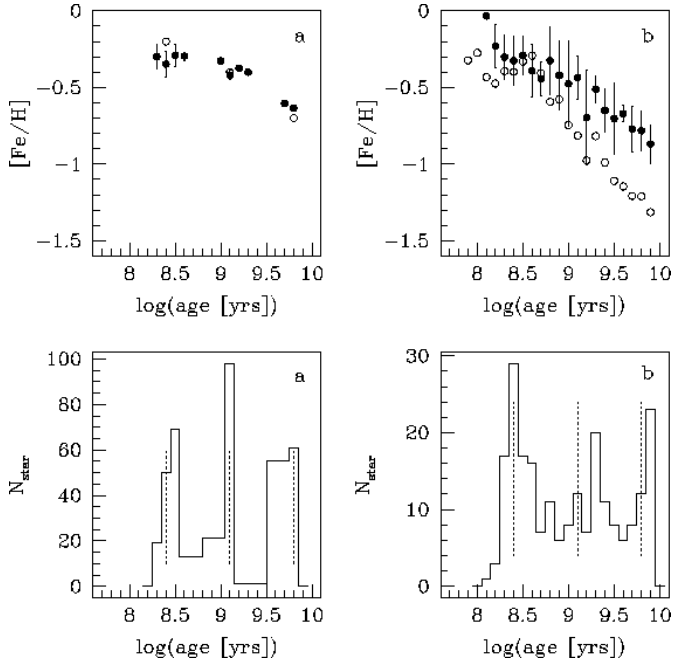}}
\caption{
  The age-metallicity relation (AMR) and the age-number diagram (AND)
  of two simulations are shown to demonstrate the effect of selection
  effects, photometric errors and systematic uncertainties. The input
  of the simulation consists of three discrete populations with
  $10^{8.4}$ yr, $-0.2$ dex, $10^{9.1}$ yr, $-0.4$ dex and $10^{9.8}$
  yr, $-0.7$ dex as age and metallicity, respectively. The number
  ratio is $1:1:10$.  In the left panel {\bf a} no error was applied,
  in the right panel {\bf b} we assumed the same photometric errors
  as for NGC~1711, a binary fraction of 70\% and a differential
  reddening of $\Delta E_{B-V} = 0.04$ (peak to peak). The open circle
  indicate in {\bf a} the input population's age and metallicity and
  in {\bf b} our measured mean AMR. Filled circles are used for the
  mean metallicity we derived from the simulated measurements.  The
  dotted lines in the lower graphs illustrate the input population's
  age.}
\label{fig:AMRsim}
\end{figure}

\subsection{Tests with the observed fields}

We tested our method also with the aid of the observed cluster stars.
We used radially selected samples around the cluster centre and
determined for them automatically the mean age and metallicity.  The
results are shown in Table~\ref{table:result_auto} where also the
error of the mean metallicity and the standard deviation (
$\sigma$([Fe/H]) and $\sigma$(log(Age [y]))) is given for each
cluster.  The results for all clusters (except for NGC~2257, see
below) are in good agreement with the ages that had been found with
isochrone "fitting". For NGC~2257 we did not expect to find the
cluster's age and metallicity since or method applies only to stars
having an age of less than $\approx 10$ Gyr and a metallicity of more
than $-1.5$ dex. The standard deviations around the mean values are
considerable, however, systematically $\approx 25\%$ smaller than for
the field populations.

\begin{table*}
\begin{center}
\caption{
Automatically determined age and metallicity of the investigated clusters}
\label{table:result_auto}
\begin{tabular}{ccccc}
\hline
Cluster &  [Fe/H] & $\sigma$([Fe/H]) & log(Age [y]) & $\sigma$(log(Age [y]))  \\[0.5ex] \hline
NGC~1651$^1$ & $-0.63\pm0.04$ & $0.26$ & $9.24\pm0.06$ & $0.35$ \\ 
NGC~1806 & $-0.56\pm0.04$ & $0.32$ & $8.66\pm0.05$ & $0.38$ \\
NGC~2031 & $-0.45\pm0.05$ & $0.19$ & $8.16\pm0.07$ & $0.26$ \\ 
NGC~2136/37 & $-0.56\pm0.03$ & $0.10$ & $8.16\pm0.05$ & $0.16$ \\ 
NGC~2257$^2$ & $-0.85\pm0.10$ & $0.29$ & $9.27\pm0.14$ & $0.40$ \\ \hline
\end{tabular}
\begin{list}{}{}
\item[]{$^1$ A reddening of $E_{b-y}=0.03$ has been used.
$^2$ The whole field of NGC~2257 is used.}
\end{list}
\end{center}
\end{table*}

\begin{figure}
\resizebox{\hsize}{!}{\includegraphics{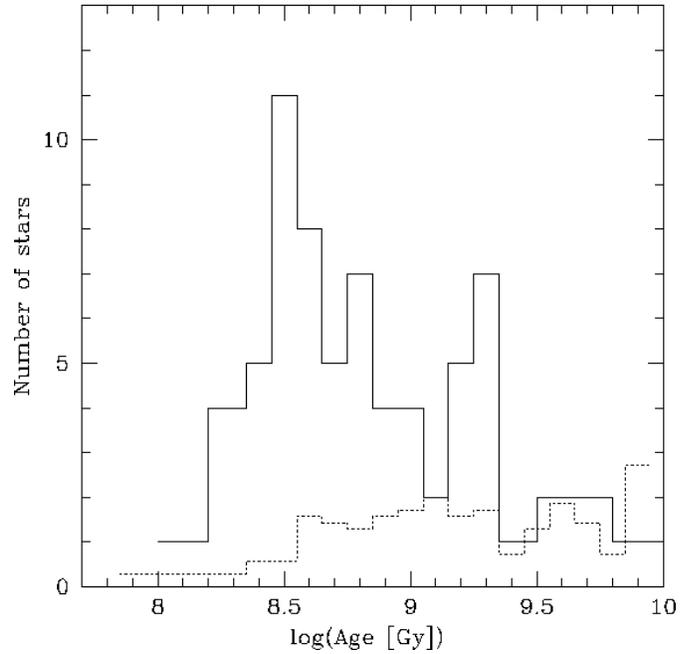}}
\caption{
  The age distribution of stars around the combined fields of NGC~1806
  and NGC~1651. The solid lines shows the number of stars recovered
  when stars are used that are less than $75 \arcsec$ away from the
  centre of NGC~1651 and $60"$ from the centre of NGC~1806. We used a
  smaller selection radius for NGC~1806 since this cluster appears
  considerably larger than NGC~1651 and we wanted to get comparable
  results. Stars nearer than $20 \arcsec$ to the cluster centre have
  been excluded.  The dotted line are the number of field stars around
  these clusters scaled to the same area.}
\label{fig:agexem}
\end{figure}

In Fig.~\ref{fig:agexem} we show for two clusters one with a
pronounced RGB and the other with a AGB the result of this method. We
use the stars up to a distance of $60 \arcsec$ and $75 \arcsec$ around
NGC~1806 and NGC~1651, respectively, as a combined input. The clusters
can be seen as peaks at the corresponding age ($10^{9.2}$ yr,
$10^{8.7}$ yr). The selection radius for NGC~1651 was two times larger
than the one we used to derive the cluster's age and metallicity via
isochrone fitting. In Fig.~\ref{fig:agexem} the age distribution
of the field stars scaled to the same area as the cluster stars is
plotted with the dotted line.

\section{The AMR of the combined field population}

A major problem in deriving a reliable AMR and SFH is the small number
statistic of field stars available for such an investigation after
applying the selection criteria. To overcome the statistical problems,
we have summed up all the CCD fields to work on a ``global'' LMC field
population. With this approach, we are able to present an overall
picture of the field AMR of the observed regions.  Our sample of field
stars (comprising 693 RGB stars, for which an age and a metallicity
has been measured) enabled us to derive a "global" AMR and AND, where
"global" rather means an average over our pointings.  The distances of
our inner clusters from the LMC centre are between $1.6^0$ and $4^0$.
This corresponds to a projected distance difference of approximately
$2$ kpc between the investigated fields. However, we did not weight
these different fields and thus our results are more influenced by the
stellar population around NGC~2031 and NGC~1806 than by for example
the field stars around NGC~1651.  Since after $1$ Gyr the stars should
be well mixed within the LMC (Gallagher et al. \cite{gallagher96};
they assumed $1$ km/sec as velocity dispersion of a typical LMC star).
Thus this sample has a global meaning at least for the stars being
older than $1$ Gyr.

The CMD and two-colour diagram of this combined population is shown in
Fig.~\ref{fig:allfme} and Fig.~\ref{fig:allfage}, respectively.

\begin{figure}[t]
\resizebox{\hsize}{!}{\includegraphics{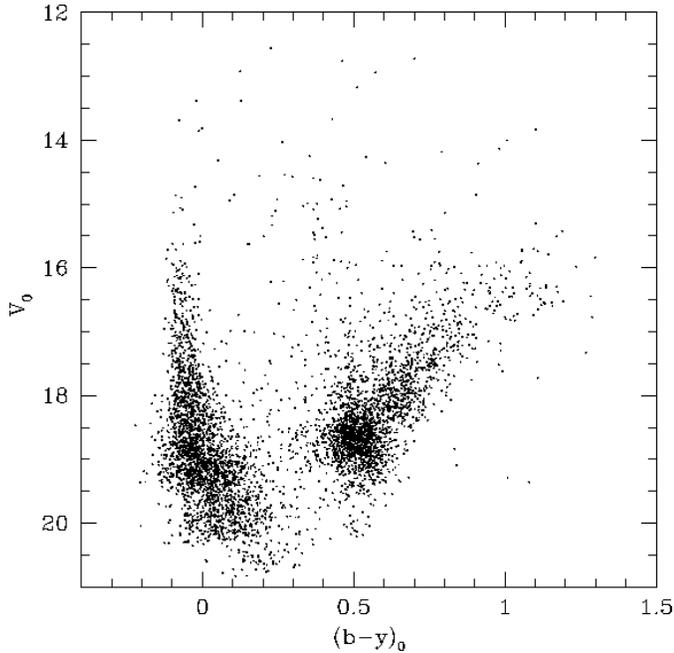}}
\caption{
CMD of all field stars found around the investigated clusters. 
Only stars are plotted that fulfil the error selection 
criteria (see Sect.4.2).}
\label{fig:allfage}
\end{figure}

\begin{figure}[t]
\resizebox{\hsize}{!}{\includegraphics{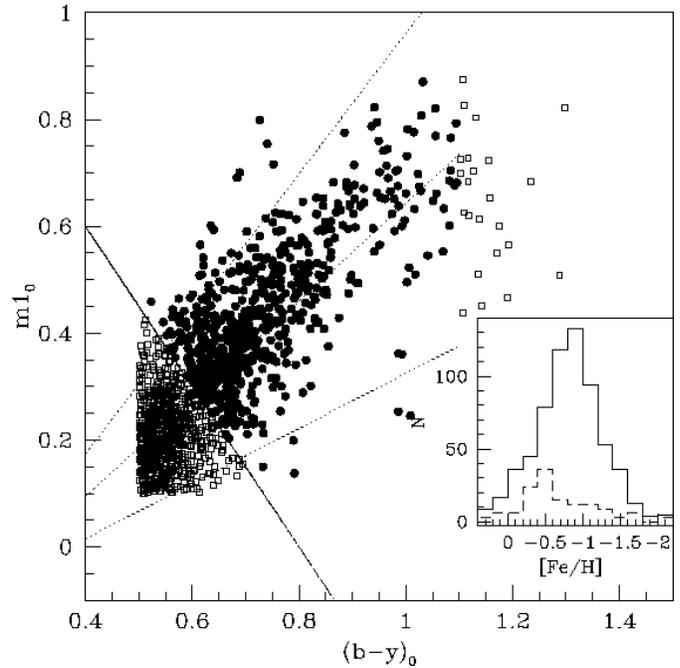}}
\caption{
  Two-colour diagram of all field stars found around the investigated
  clusters. Solid circles are used to mark stars with a reliable
  metallicity measurement.  Stars being redder than $(b-y)=1.1$ have
  been excluded, as usual.  The solid histogram in the insert shows
  the metallicity distribution of all stars, the dashed line the
  distribution of stars being younger than $10^{8.4}$ yr. }
\label{fig:allfme}
\end{figure}

The derived AMR is shown in Fig.~\ref{fig:AMRallf} and tabulated in
Table\ref{tab:AMRallf}. The plotted ``error'' bars give the standard
deviation in metallicity of stars with the same age and is not an
error of the mean metallicity, since we do not believe that in each
age bin all stars have the same metallicity even if we could measure
the age with much higher accuracy. The age resolution on the giant
branch drops considerably for stars that are older than $10^{9.2}$ yr
because the spacing in luminosity between isochrones of different ages
shrinks. This could cause the flat appearance of the AMR for these old
ages, which is compatible with the shown "error" bars.

The upper age limit of our investigation is $10^{9.9}$ yr because no
older Geneva isochrones were available, the lowest metallicity of the
Geneva isochrone set is $-1.3$ dex. For stars that are more metal poor
we used the $-1.3$ dex isochrones to derive the age (if the luminosity
difference between the star and the isochrone was $< 0.1$ mag), which
means an underestimation. With these stars included, the oldest bin of
the AMR contains stars, which in reality are older and more metal poor
than the above stated limits. Therefore the number of stars in the age
range $10^{9.5}$ yr - $10^{9.9}$ yr is slightly overestimated.

\begin{figure}[t]
\resizebox{\hsize}{!}{\includegraphics{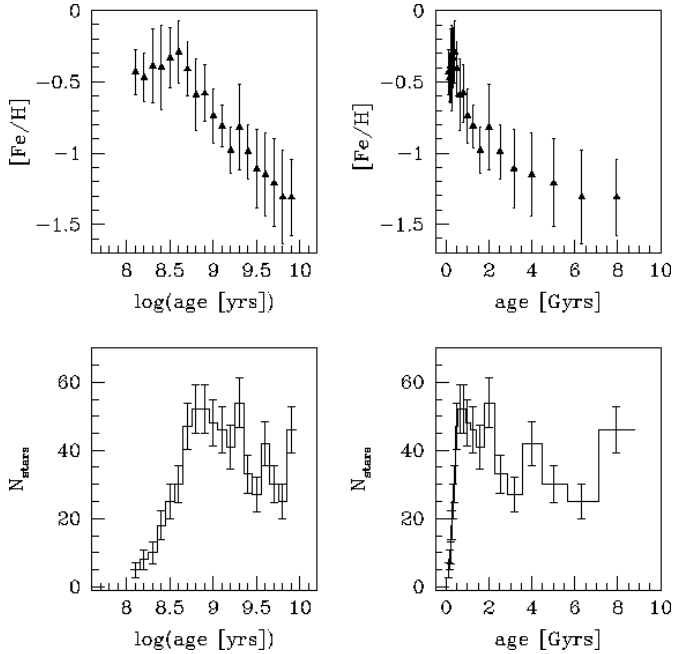}}
\caption{
  The AMR and AND in logarithmic and linear age representation of all
  field stars around the investigated cluster.}
\label{fig:AMRallf}
\end{figure}

The colour selection can introduce a bias, since the RGB of a metal
rich population lies completely in the employed colour range, while
for example only half of a $-1.3$~dex RGB extends so far red which can
be seen from the employed isochrones. The fraction of the RGB within
the selected colour range is in good approximation independent of the
age as long as stars older than $1$ Gyr are considered (concluded from
visual inspection of the employed Geneva isochrones).  For younger
ages the problem becomes less severe, since the red supergiants extend
further red.  Since all our conclusions on the metallicity are based
on simple means this introduces a bias towards higher metallicities if
equal aged stars with different metallicity exist. To estimate the
amount of the most extreme shift, we assume two populations with the
same age but one having solar abundance and the other having a
metallicity of $-1.3$~dex.  The difference in the mean metallicity, if
only half of the metal poor giants are observed compared to the mean
metallicity of the whole sample is $\Delta [Fe/H]_{mean} = -1.3 (1/2 -
1/3) = 0.2$~dex. Thus we conclude that deviations due to this problem
are well less than $0.2$~dex.

The CN anomaly leads to an overestimation of the metallicity and hence
to an underestimation of the age. The deviation of the age according
to the deviation of an overestimation of the photometric metallicity
is nearly parallel to the observed AMR for ages larger than
$10^{8.5}$~yr.  Even with CN anomalous stars we should be able to
distinguish between the proposed AMR and for example the AMR propposed
by Pagel \& Tautvai\u{s}vienn\.e (\cite{pagel98}): the influence of CN
anomalous stars on the second AMR would result in an even more
pronounced difference as it is already seen, since metal poorer stars
would be shifted to even more metal rich and younger locations.

\section{Discussion}

\subsection{The Age-Metallicity Relation}

\begin{table*}[t]
\begin{center}
\caption{
Results for the 5 clusters investigated in this work. The error
includes the calibration uncertainty.}
\label{table:result}
\begin{tabular}{ccccc}
\hline
Cluster & $E_{B-V}$ & Metallicity [dex] & log(Age [y]) & Remarks \\[0.5ex] \hline
NGC~1651 & $0.01$ to $0.05$ & $-0.65$ to $-0.45 $  & $9.4$ to $9.1$ & reddening problematic\\
NGC~1711 & $0.09 \pm 0.03$ & $-0.57\pm0.17$ & $ 7.7\pm0.05 $ & reddening of the field is larger\\ 
NGC~1806 & $0.16 \pm 0.06$ & $-0.71\pm0.24$ & $8.7\pm0.1$ & \\
NGC~2031 & $0.09 \pm 0.05$ & $-0.52\pm0.21$ & $8.2\pm0.1$ & \\
NGC~2136/37 & $ 0.09 \pm 0.05$ & $-0.55\pm0.23$ & $8.0\pm0.1$ & no differences between the two clusters\\
NGC~2257 & $0.04 \pm 0.04$ & $-1.63\pm0.21$ & $10.2\pm0.1$& \\[0.5ex]\hline
\end{tabular}
\end{center}
\end{table*}

In Table\ref{table:result} we summarise the resulting ages and
metallicities of the investigated clusters.

In order to compare our results with the literature, we compiled a
list of clusters with ages and metallicities according to various
sources (all published after 1989).  The data is tabulated in
Table\ref{tab:LitData}. In addition, we used the compilation of Sagar
\& Pandey (\cite{sagar89}), from which only clusters have been
selected with a limiting magnitude below $V\!=\!21$. This limit shall
serve as a rough quality criterion that is comparable to the more
recent data and explains why most (photographic) papers cited by Sagar
\& Pandey are excluded.  These clusters are plotted together with the
newly investigated clusters and our field AMR in
Fig.~\ref{fig:LitData}. The solid line is the field AMR accompanied by
two dotted lines which mark 1$\sigma$ borders. If the older clusters
are excluded, a weak correlation appears for the clusters: clusters
younger $1$ Gyr have a mean metallicity of $[Fe/H]=-0.34$ with a
standard deviation of 0.14 and in the age range $1-2.5$ Gyr the mean
metallicity is $[Fe/H]=-0.71$ with a standard deviation of $0.17$ (11
cluster).

\begin{figure*}[t]
\resizebox{12cm}{!}{\includegraphics{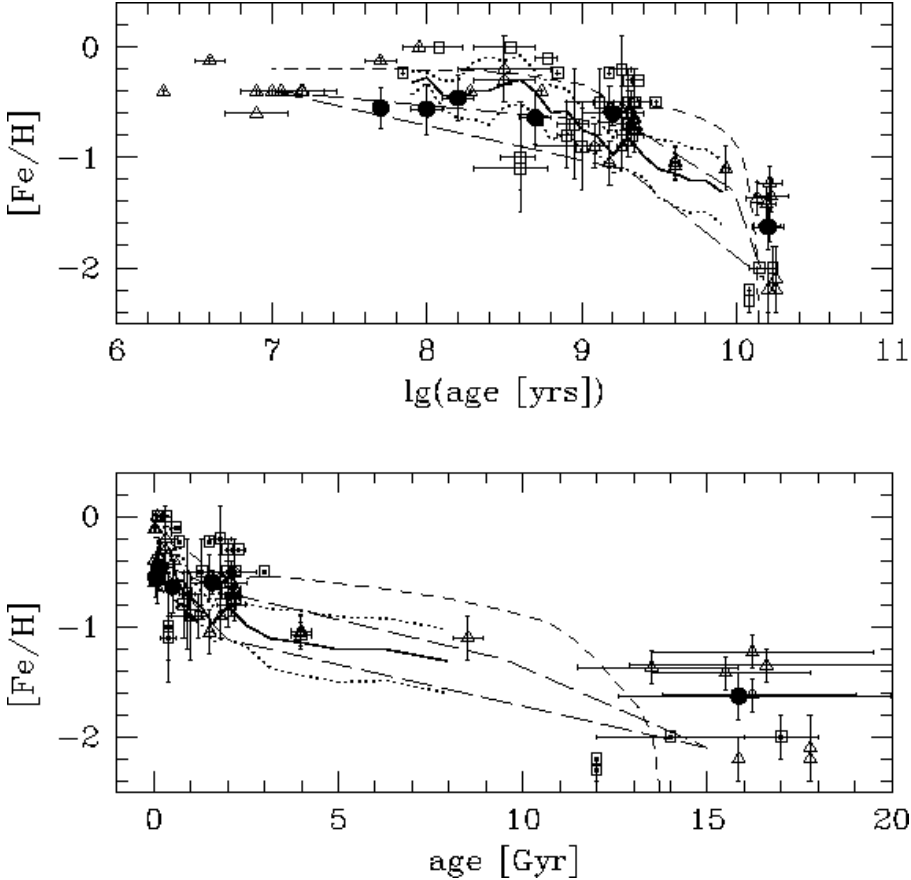}}
\hfill
\parbox[b]{55mm}{
\caption{Published ages and metallicities for LMC clusters, given in Table\ref{tab:LitData} 
  (open triangles) and in the compilation by Sagar \& Pandey
  (\cite{sagar89}) with a limiting magnitude of fainter than $V=20$
  (open squares). The cluster data from this investigation are marked
  with solid circles. The solid line connects the points in our
  derived AMR for the field population. The dotted line surrounding
  the solid line marks the standard deviation of the metallicity
  around a given age. In addition three models for the AMR are
  plotted: Pagel \& Tautvai\u{s}vien\.e (\cite{pagel98}) as short
  dashed line and two models calculated by Geha et al. (\cite{geha98})
  as long dashed line.}
\label{fig:LitData}
}
\end{figure*}

The mean metallicity of our young clusters ($<10^{9.0}$ yr) is $-0.57
\pm 0.04$ dex and thus lower than what we found using the newer
cluster sample from the literature. The field of the same age has a
mean metallicity of $-0.4\pm0.2$ dex and is in good agreement with
spectroscopic measurements ($-0.38\pm0.11$ dex) of young field stars.
The latter comparison is reasonable since {\bf a)} most of the stars
for which high resolution spectroscopy has been obtained are located
at a similar radial distance and {\bf b)} no radial gradient can be
seen in the spectroscopic sample. We compiled a list of high
resolution spectroscopic measurements of LMC stars in
Table~\ref{tab:fieldme}.

Bica et al. (\cite{bica98}) obtained ages and metallicities of 13
outer clusters in the LMC using Washington photometry.  The mean
metallicity of all the surrounding field stars is
$\langle[Fe/H]\rangle\simeq-0.6\pm0.1$. The mean metallicities of our
field populations seem to be systematically more metal poor than this
value thus indicating a possible zero point difference of the order of
0.2 dex and comparable to the probable shift between our cluster
metallicities and the ones taken from the literature.  However, such a
difference between cluster and field stars has not been seen in a
study by Korn et al. (\cite{korn00}) who employed high resolution
spectroscopy of supergiants.

The AMR for stars older than $3$ Gyr is consistent with little or even
no enrichment until 8 Gyr ago.  The AMR in this age range agrees well
with the $4$ Gyr old clusters studied by Sarajedini
(\cite{sarajedini98}) and also ESO 121 SC03 is in agreement with the
derived field star AMR, especially when taking the systematic
underestimation of the metallicity for older stars on the order of
$0.05-0.1$ dex into account (see Sect.~3). Therefore, we do not see
the necessity that ESO 121 SC03 belongs to a dwarf galaxy that is in
the process of merging with the LMC as proposed by Bica et al.
{\cite{bica98}).
  
  The field population of NGC~1651 and NGC~2257 is considerably
  different from that around the other clusters: around NGC~1651 we
  find two distinct field populations, around NGC~2257 only one, thus
  these fields cannot be compared to the other fields, where a mixture
  of populations have been detected. These fields contain a
  significantly larger fraction of old stars than the other fields,
  what is expected from their location in the LMC (e.g. Santos Jr. et
  al. \cite{santos99}).
  
  If cluster and field are compared it becomes apparent that our AMR
  does not argue for an extremely decoupled enrichment history between
  cluster and field stars, only hints can be seen that the younger
  clusters are slightly more metal poor than the surrounding field
  population of the same age. Bica et al. (\cite{bica98}) found the
  same behaviour for several of their (young) clusters and the
  surrounding field population.  One has to consider the possibility,
  that these low mean cluster abundances are a result of the
  statistically larger effect of blending towards the cluster. This
  has been proposed by Bessell (\cite{bessell93}) to explain the low
  Str\"omgren metallicity of NGC~330 measured by Grebel \& Richtler
  (\cite{grebel92}). In this work of Grebel \& Richtler
  (\cite{grebel92}) the mean metallicity found for the surrounding
  field population ($-0.74$ dex) agreed well with later on performed
  spectroscopic measurements ($-0.69$ dex, Hill \cite{hillV99}) (
    for a more comprehensive discussion on NGC~330 the reader is
    referred to the work by Gonzalez \& Wallerstein
    \cite{gonzalez99}).  Having this agreement in mind, one can
  estimate from the difference of mean field and cluster metallicity
  that the contamination has a minor effect on our derived
  metallicities, accounting possibly for a systematic deviation of
  less than $<0.15$ dex.
  
  Our AMR is inconsistent with a recent calculation presented by Pagel
  \& Tautvai\u{s}vien\.e (\cite{pagel98}) based on LMC clusters
  and on planetary nebulae observed by Dopita et al.
  (\cite{dopita97}), that predicts a steeper increase of the
  metallicity in earlier time, thus older stars should have a higher
  metallicity than what we observe (see Fig.~\ref{fig:LitData}). The
  AMR is more consistent with closed box model calculations performed
  by Geha et al. (\cite{geha98}). They present theoretical enrichment
  models for the two SFHs put forward by Holtzman et al. (1997) and by
  Vallenari et al. (1996a,b). These SFHs agree in the sense, that a
  long period of low star formation activity was followed by a sudden
  increase about 2 Gyr ago.  With the Vallenari et al.- SFH, the
  metallicity increased by a factor of five during the last 2 Gyr,
  while a modest increase of a factor of three resulted from the
  Holtzman et al.-SFH.
  
  Dopita et al. (\cite{dopita97}) published an AMR for the LMC based
  on planetary nebulae and found that the metallicity only doubled in
  the last 2-3 Gyr which is seen in our AMR as well. Another common
  feature is that a distinct enrichment (if any) between 4 and 9 Gyr
  cannot be seen. A comparison of the Dopita et al.-values with ours
  is made difficult by the fact that they measured $\alpha$-element
  abundances instead of $[Fe/H]$, but they stated that {\it "there is
    no evidence in this sample of any "halo" abundance object"}. If we
  would apply a constant shift of $-0.35$ on the [O/H]-abundance, to
  correct approximately the [O/Fe] overabundance in the LMC in
  comparison to the Milky Way the metallicity of the PNs with an age
  of $10^{8.8}-10^{9.8}$ yr would nicely be in coincidence with our
  measurements.  However, the $[O/Fe]$ variation in dependence of
  $[Fe/O]$ is still under discussion (see e.g. Russell \& Dopita
  \cite{russell92} or Pagel \& Tautvai\u{s}vien\.e \cite{pagel98}).
  
  Judging from the field around NGC~2257 we find that no radial
  metallicity gradient can be seen, since the field stars are
  consistent with the AMR derived from the inner fields. Taking also
  NGC~1651 into account we find that in fields where no recent star
  formation happened the stellar population is dominated by a
  population with an age between $2$ and $4$ Gyr. Thus deriving a
  global SFH on a limited sample is quite uncertain.

\subsubsection{The Star Formation History}

The manner in which we derived the field star SFH contains several
points that may induce biases. One reason is that the isochrones have
only a crude spacing in the parameters age and metallicity. Therefore,
simulations are helpful for a discussion of the SFH of the field
population as described above (Sect.~12).  To derive a SFH from our
data is more difficult than to derive an AMR, since one has to know
not only the age of a star with a given metallicity, but the amount of
stars with a given age has to be quite precise. As a result the AMR is
quite robust against for example reddening variations compared to the
AND.

Two SFHs are shown that illustrate how to interpret the AND
(Fig.~\ref{fig:bestSFH}). One SFH has a constant SFR during the whole
LMC evolution and thus serves to give an impression how the selection
effects behave (left two panels in Fig.~\ref{fig:bestSFH}).The SFR of
the second SFH was constant until $10^{9.7}$ yr ago, then it increased
by a factor of $5$ until $10^{8.4}$ yr ago, before the SFR dropped to
its old low level. The AND \& AMR resulting from this SFH is plotted
in the right panel of Fig.~\ref{fig:bestSFH}.

The constant SFR is marginally inconsistent with our data, which holds
for a different reddening correction of $E_{b-y} = \pm 0.02$. This is
not true for exact behaviour of the SFH: for example a decrease in the
reddening of $E_{B-V}=0.02$ results in a SFH in which a much larger
increase (around a factor of $10$) is necessary to describe the
observations.  However, the general trend, namely, the increase of the
SFR around $10^{9.5\pm0.2}$ yr ($2-5$~Gyr) ago and the necessary
declining SFR some $10^{8.5}$~yr ago in these fields is more robust.
Since stars in the LMC should be mixed (at least azimuthally) after
$\approx 1$ Gyr (Gallager et al. \cite{gallagher96}) the SFH of the
older stars should be a measure for the average SFH of the LMC in the
radial distance of the investigated clusters.  As a rule of thumb an
increase in the applied reddening correction of $E_{b-y}=0.1$ results
for a single age population in a decrease in age by a factor of $0.7$.

\begin{figure}[t]
\resizebox{\hsize}{!}{\includegraphics{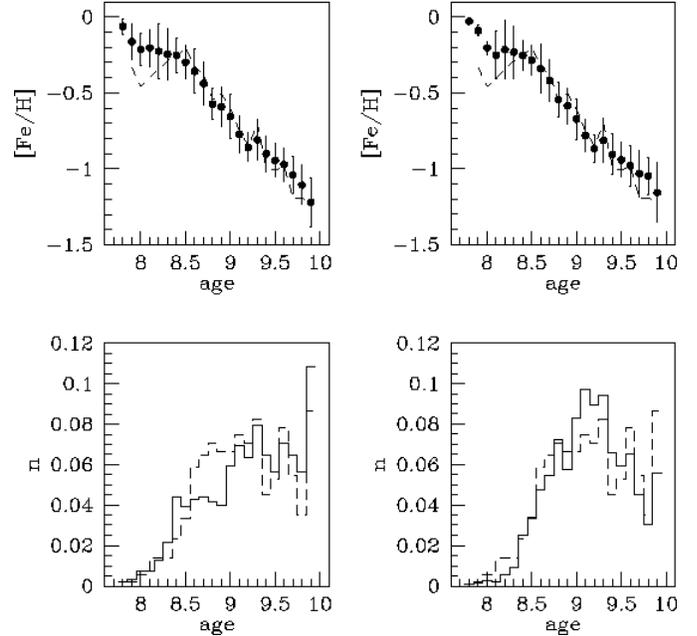}}
\caption{
  Simulation with a quasi continuous SFR. The left panel shows the AMR
  and AND for a constant SFR over the last $10$ Gyr.  In the right
  graph the symbols and solid line correspond to a SFH which was
  constant until $10^{9.7}$ yr, at $10^{9.6}$ yr it increased by a
  factor of $4$ until $10^{8.6}$ yr ago after that the SFR dropped to
  the same level as in the beginning.  The dashed line shows the
  result of our composite field.}
\label{fig:bestSFH}
\end{figure}

Vallenari et al. (1996a,b) proposed, on the basis of ground based
  observations, a SFH in which the SFR increased about a factor of
ten 2 Gyr ago, thus only around 5 \% of the stars should be older than
4 Gyr. This has recently also been found by Elson et al.
(\cite{elson97}) with HST observations. A different SFH was
advanced by Holtzman et al. (\cite{holtzman97}), Geha et al.
(\cite{geha98}) and Holtzman et al.  (\cite{holtzman99}) also
  based on HST observations. In their model, approximately half of
the stars are older than 4 Gyr.  In our data the fraction of stars
older than 4 Gyr is $40 \pm 20$\%, but we note that already a small
additional reddening of of $E_{b-y}=0.015$ leaves only $\approx 15$\%
of the stars older than 4 Gyr. Olsen (\cite{olsen99}) used
  Washington photometry of the LMC field population and derived a SFH
  which is compatible with the one proposed by Holtzman et al.
  (\cite{holtzman99}).

Summarizing, despite the uncertainty in the amount of the increase,
the SFH is consistent with an increased SFR that started roughly $3\pm
1$ Gyr ago.  Interestingly the sparsely populated outer fields are
tentatively populated by mainly a population with ages between $2$ and
$4$ Gyr. If this result will hold for a larger sample of outlying
fields this could mean that the stellar body of the inner part of the
LMC contains more younger {\it and} older stars compared to these
intermediate age stars than the more remote parts of this galaxy.
However, spectroscopic studies of several of these candidate stars are
needed, especially because the tilted red clump of NGC~1651 could be
due to a He overabundance and thus isochrones might be misleading.

The observed SFH is inconsistent with a starformation history in which
no star has been born between $4$ and $8$~Gyr. The simulations showed
that virtually no star should have been recovered with an age of more
than $10^{9.3}$~yr, even if the differential reddening is as large as
$E_{B-V}=0.07$ and the binary fraction is $70\%$. However, a large
amount of old ($>10$~Gyr) CN anomal stars could mimic starformation
between $4$ and $8$ Gyr.

A last remark on the cluster formation rate: it has been noted several
times that there apparently was a long period in the LMC where no
clusters (or a few) have been formed.  Recently, Larsen \& Richtler
(1999) performed a search for bright star clusters in 21 face-on
galaxies. They found a correlation of the specific cluster frequency
with parameters indicating the SFR. The age gap of the LMC cluster
thus could reflect the low SFR during this period, where the condition
for cluster formation where not present.

\section{Summary}

We tried to determine the Age Metallicity Relation (AMR) and the Star
Formation History (SFH) in the LMC on the basis of metallicities and
ages of red giants, measured by Str\"omgren photometry. Our stars are
located both in star clusters and in the respective surrounding
fields. While statements regarding the AMR are relatively robust, the
SFH is much more difficult to evaluate because of the incompleteness
effects, for which we can only approximately correct.

Between $8$ to $3$ Gyr ago the metallicity of the LMC was constant or
varied only very slow, after this period the speed of the rate of the
enrichment grew: starting $3$ Gyr ago the metallicity of the stars
increased by a factor of six. The cluster and field AMR during this
time was coupled, however a possibility remains that the cluster have
on the average a slightly smaller metallicity than the field stars of
the same age.  Our field star AMR is also consistent with the $4$ Gyr
old clusters recently studied by Sarajedini (\cite{sarajedini98}) and
with ESO 121 SC03. Thus for the latter there is no need to explain
this cluster as a recent merger remnant. Good agreement can be found
with the photometrically determined metallicity of the young stars and
the abundances measured with high resolution spectroscopy.  The star
formation rate increased around $3$ Gyr ago, however it is not
possible to constrain the SFH further due to uncertainties in
reddening, the CN anomaly and the difficult completeness
considerations.

\begin{acknowledgements}
  The authors gratefully acknowledge observing time at La Silla and
  the aid by the staff of the European Southern Observatory.  We are
  grateful to E. Grebel and J. Roberts for the opportunity to use
  their isochrones.  We thank the referee for his/her comments and
  suggestions which greatly helped to improve the paper. We also thank
  Doug Geisler and Antonella Vallenari for their valuable comments.
  BD was supported through the DFG Graduiertenkolleg "The Magellanic
  Clouds and other dwarf galaxies" (GRK 118).  MH thanks Fondecyt
  Chile for support through `Proyecto FONDECYT 3980032'.  WPG
  gratefully acknowledges support received by Fondecyt grant No.
  1971076.  TR thanks the Uttar Pradesh State Observatory, Nainital,
  for warm hospitality and financial support.  This research has made
  use of the Simbad database, operated at CDS, Strasbourg, France.
\end{acknowledgements}
\appendix
\section{Appendix}
\begin{figure*}[t]
\resizebox{15cm}{!}{\includegraphics{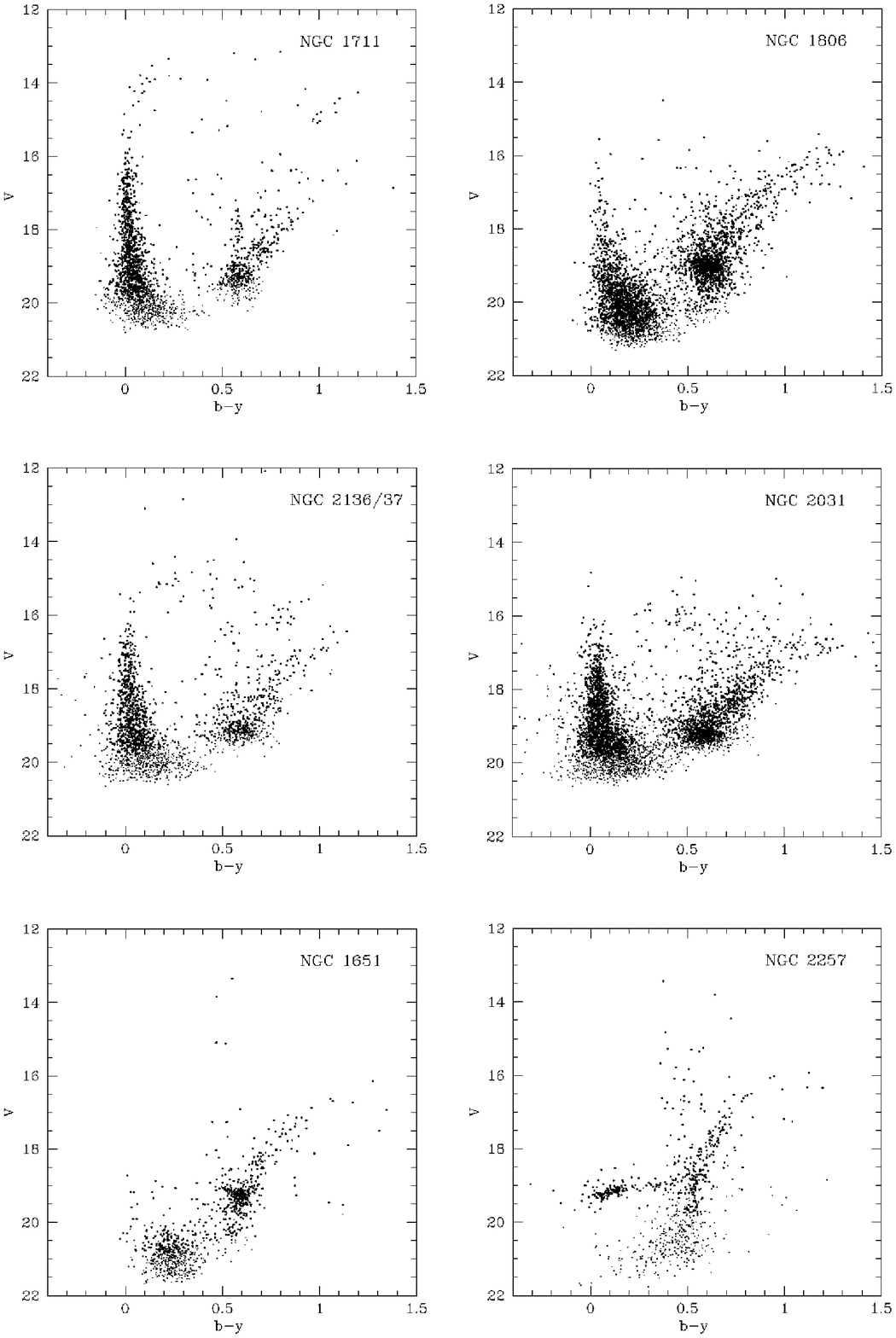}}
\caption{The CMDs of the whole fields containing the observed clusters. No reddening correction 
has been applied. The points in bold face are stars with an error $\Delta(b-y)<0.1$ and $\Delta m1<0.1$; 
the other dots are used for stars found with larger errors.}
\label{fig:cmd}
\end{figure*}
\begin{table}[t]
\caption{Observing log}
\label{tab:log}
\begin{tabular}{ccccc}
\hline
Object & Filter & Night & Exposure time & Seeing \\\hline
NGC~1711 & y & 8.1.94 & 300 sec & 1.4" \\[-0.5ex] 
         & y & 8.1.94 & 300 sec & 1.3" \\[-0.5ex]
         & b & 8.1.94 & 600 sec & 1.2" \\[-0.5ex]
         & b & 8.1.94 & 600 sec & 1.3" \\[-0.5ex]
         & v & 8.1.94 & 900 sec & 1.5" \\[-0.5ex]
         & v & 8.1.94 & 900 sec & 1.2" \\ \hline 
NGC~1806 & y & 6.1.94 & 300 sec & 1.1" \\[-0.5ex]
         & y & 6.1.94 & 300 sec & 1.1" \\[-0.5ex]
         & y & 6.1.94 & 900 sec & 1.5" \\[-0.5ex]
         & b & 6.1.94 & 900 sec & 1.1" \\[-0.5ex]
         & y & 7.1.94 & 600 sec & 1.0" \\[-0.5ex]
         & v & 7.1.94 & 1200 sec & 1.1" \\[-0.5ex]
         & b & 7.1.94 & 900 sec & 1.0" \\[-0.5ex]
         & y & 8.1.94 & 600 sec & 1.3" \\[-0.5ex]
         & b & 8.1.94 & 900 sec & 1.2" \\ [-0.5ex]
         & v & 8.1.94 & 900 sec & 1.3" \\[-0.5ex]
         & v & 8.1.94 & 900 sec & 1.3" \\ \hline
NGC~1651 & y & 6.1.94 & 300 sec & 1.4" \\[-0.5ex]
         & y & 6.1.94 & 600 sec & 1.1" \\ [-0.5ex]
         & b & 6.1.94 & 1200 sec & 1.1" \\[-0.5ex]
         & v & 6.1.94 & 1200 sec & 1.2"\\[-0.5ex]
         & v & 6.1.94 & 1200 sec & 1.1"\\[-0.5ex]
         & y & 7.1.94 & 600 sec & 0.9" \\[-0.5ex]
         & b & 7.1.94 & 1200 sec & 1.0" \\[-0.5ex]
         & v & 7.1.94 & 1200 sec & 1.0" \\[-0.5ex]
         & v & 7.1.94 & 1200 sec & 1.0" \\ \hline
NGC~2031 & y & 12.11.92 & 300 sec & 1.1" \\[-0.5ex]
         & b & 12.11.92 & 600 sec & 1.3" \\[-0.5ex]
         & v & 12.11.92 & 240 sec & 1.3" \\[-0.5ex]
         & v & 12.11.92 & 1200 sec & 1.2" \\[-0.5ex]
         & y & 14.11.92 & 600 sec & 1.2"\\[-0.5ex]
         & y & 14.11.92 & 120 sec & 1.4" \\[-0.5ex]
         & b & 14.11.92 & 120 sec & 1.3" \\[-0.5ex]
         & b & 14.11.92 & 400 sec & 1.1" \\[-0.5ex]
         & v & 14.11.92 & 240 sec & 1.4" \\[-0.5ex]
         & v & 14.11.92 & 1140 sec & 1.3" \\ \hline
NGC~2136/37 & y & 12.11.92 & 120 sec & 1.1" \\[-0.5ex]
         & b & 12.11.92 & 120 sec & 1.5" \\[-0.5ex]
         & v & 12.11.92 & 120 sec & 1.1" \\[-0.5ex]
         & y & 13.11.92 & 600 sec & 1.2" \\[-0.5ex]
         & b & 13.11.92 & 900 sec & 1.2" \\[-0.5ex]
         & v & 13.11.92 & 1200 sec & 1.2" \\\hline
NGC~2257 & v & 6.1.94 & 1200 sec & 1.2" \\  [-0.5ex]
     & v & 6.1.94 & 600 sec & 1.2" \\[-0.5ex]
     & b & 6.1.94 & 600 sec & 1.2" \\[-0.5ex]
     & y & 6.1.94 & 300 sec & 1.2" \\[-0.5ex]
     & y & 7.1.94 & 600 sec & 1.1" \\[-0.5ex]
     & b & 7.1.94 & 900 sec & 1.1" \\ [-0.5ex]
     & b & 8.1.94 & 300 sec & 1.0" \\[-0.5ex]
     & b & 8.1.94 & 300 sec & 1.0" \\[-0.5ex]
     & b & 8.1.94 & 300 sec & 1.0" \\[-0.5ex]
     & y & 8.1.94 & 300 sec & 1.4" \\[-0.5ex]
     & y & 8.1.94 & 600 sec & 1.1" \\[-0.5ex]
     & v & 8.1.94 & 900 sec & 1.2" \\[-0.5ex]
     & v & 8.1.94 & 900 sec & 1.2" \\ \hline
\end{tabular}
\end{table}
%
%
\begin{table}[t]
\caption{The age of NGC~1711 given in the literature.}
\label{tab:1711lit}
\begin{tabular}{ccc}
\hline
Author &  Age & Method \\[0.5ex]\hline
Cassatella et al. \cite{cassatella96}  & $10^{7.41}$ yr & UV spectra \\
Girardi et al. \cite{girardi95} &  $10^{7.58}$ yr & $M_V$ of MSTO \\
                                &  $10^{7.99}$ yr & $M_V$ of CHeB stars \\
Barbaro \& Olivi \cite{barbaro91} & $10^{7.57}$ 
&UV colours \\ 
Sagar \& Richtler \cite{sagar91} & $10^{7.18}$ yr &Isochrones \\ 
Mateo \cite{mateo88} &  $10^{7.7}$ yr  &Isochrones \\[0.5ex]
\hline
\end{tabular}
\end{table}
%
%
\begin{table}[t]
\caption{The derived AMR for all field stars around the investigated clusters.}
\label{tab:AMRallf}
\begin{tabular}{c|ccc}
\hline
log(age [y])& Metallicity & standard deviation & number of \\ 
                        &                               & of metallicities      & stars         \\[+0.5ex] \hline
 7.9 & -0.26 & 0.11 & 2 \\[-0.5ex] 
 8.0 & -0.27 & 0.04 & 3 \\[-0.5ex]
 8.1 & -0.43 & 0.16 & 5 \\[-0.5ex]
 8.2 & -0.47 & 0.17 & 8\\[-0.5ex]
 8.3 & -0.39 & 0.26 & 10 \\[-0.5ex]
 8.4 & -0.40 & 0.30 & 18 \\[-0.5ex]
 8.5 & -0.33 & 0.21 & 25 \\[-0.5ex]
 8.6 & -0.29 & 0.22 & 30 \\[-0.5ex]
 8.7 & -0.41 & 0.19 & 47 \\[-0.5ex]
 8.8 & -0.59 & 0.25 & 52 \\[-0.5ex]
 8.9 & -0.58 & 0.20 & 52 \\[-0.5ex]
 9.0 & -0.74 & 0.19 & 48 \\[-0.5ex]
 9.1 & -0.81 & 0.15 & 46 \\[-0.5ex]
 9.2 & -0.98 & 0.16 & 41 \\[-0.5ex]
 9.3 & -0.82 & 0.30 & 54 \\[-0.5ex]
 9.4 & -0.99 & 0.19 & 33 \\[-0.5ex]
 9.5 & -1.11 & 0.28 & 27 \\[-0.5ex]
 9.6 & -1.15 & 0.29 & 42 \\[-0.5ex]
 9.7 & -1.21 & 0.31 & 30 \\[-0.5ex]
 9.8 & -1.31 & 0.33 & 25 \\[-0.5ex]
 9.9 & -1.31 & 0.27 & 46 \\[+0.5ex]
\hline
\end{tabular}
\end{table}
%
%
\begin{table}[t]
\caption{Metallcity measurements of LMC field stars with high resolution spectroscopy.}
\label{tab:fieldme}
\begin{tabular}{cccc}
\hline
Sample and size & [Fe/H]$^2$ & Reference \\[0.5ex]
\hline
SG/Cepheids     - $7$ & $-0.37$ &  Luck \& Lambert \cite{luck92}\\
SG              - $8$ & $-0.47 / -0.33$ &  Russell \& Bessell \cite{russell89}\\
SG              - $6$ & $-0.51 / -0.29$ &  Mc~William \& Williams \cite{mcwilliam90}\\
SG              - $9$ & $-0.27$ &  Hill et al. \cite{hillV95}\\
SG              - $2$ & $-0.42^1$ & J\"uttner et al. \cite{juettner93}\\
SG              - $1$ & $-0.3^1$ &  Fry \& Aller \cite{fry75}\\
SG              - $2$ & $-0.5^1$ & Korn et al. \cite{korn00}\\[0.5ex]\hline
\end{tabular}
\begin{list}{}{}
\item[] {SG stands for supergiants.}
\item[]{$^1$ transformed to a solar iron abundance of $7.50$.
$^2$ If two values are given, the first was derived with the Fe I line the second with the Fe II line.
The stars measured by different authors are not always different stars: for example for 4 stars
in the list of Mc~William \& Williams (\cite{mcwilliam90}) also Hill et al. \cite{hillV95} obtained
a metallicity. We nevertheless use a simple mean for the mean field star abundance in our paper.}
\end{list}
\end{table}
%
%
%
\begin{table*}[t]
\caption{Literature values for age, metallicity and reddening derived or assumed in photometric observations
resulting in an CMD.}
\label{tab:LitData}
\begin{tabular}{ccccc}
\hline
Cluster & $E_{B-V}$ & $[Fe/H]$ & $\log$(Age [y]) & Authors \\[0.5ex] \hline
ESO~121-SC03 & $0.03$           & $-1.1\pm0.2$          & $9.93\pm0.01$         & Bica et al. \cite{bica98}\\
R~136    & $0.38$               & $-0.4$                & $7.0$                 & Hunter et al. \cite{hunter95} \\
OHSC~33  & $0.09$               & $-1.05\pm0.2$         & $9.18\pm0.03$         & Bica et al. \cite{bica98}\\
OHSC~37  & $0.15$               & $-0.7\pm0.2$          & $9.32\pm0.03$         & Bica et al. \cite{bica98}\\
LH~47/48 & $0.11$               & $-0.4$                & $6.3$                 & Oey \& Massey \cite{oey95} \\
LH~52/53 &                      & $-0.4$                & $7.0$                 & Hill et al. \cite{hill95} \\
LH~72    & $0 - 0.17$           & $-0.6$                & $6.7 - 7.18$ (age spread) & Olsen et al. \cite{olsen97}\\
LH~77    & $-0.06\pm0.01$       & $-0.4$                & $7.20 \pm 0.14$       & Dolphin \& Hunter \cite{dolphin98}\\
SL~8     & $0.04$               & $-0.55\pm0.2$         & $9.26\pm0.03$         & Bica et al. \cite{bica98}\\
SL~126   & $0.01$               & $-0.5\pm0.2$          & $9.34\pm0.03$         & Bica et al. \cite{bica98}\\
SL~262   & $0.00$               & $-0.6\pm0.2$          & $9.32\pm0.03$         & Bica et al. \cite{bica98}\\
SL~388   & $0.03$               & $-0.7\pm0.2$          & $9.34\pm0.03$         & Bica et al. \cite{bica98}\\
SL~451   & $0.1$                & $-0.75\pm0.2$         & $9.34\pm0.03$         & Bica et al. \cite{bica98}\\
SL~503   & $0.04\pm0.01$        & $-0.4$                & $7.20 \pm 0.22$       & Dolphin \& Hunter \cite{dolphin98}\\
SL~509   & $0.03$               & $-0.9$                & $9.08$                & Bica et al. \cite{bica98}\\
SL~663   &                      & $-1.05\pm0.16$        & $9.60 \pm 0.03$       & Sarajedini \cite{sarajedini98}\\
SL~817   & $0.07$               & $-0.55\pm0.2$         & $9.18\pm0.03$         & Bica et al. \cite{bica98}\\
SL~842   & $0.03$               & $-0.65\pm0.2$         & $9.34\pm0.03$         & Bica et al. \cite{bica98}\\
SL~862   & $0.09$               & $-0.9\pm0.2$          & $9.26\pm0.03$         & Bica et al. \cite{bica98}\\
NGC~1711 & $0.09$               & $0$                   & $7.8$                 & Richtler \& Sagar \cite{richtler91}\\
NGC~1754 & $0.09\pm0.02$        & $-1.42\pm0.15 (-1.54 ^a)$ & $10.19 \pm 0.06$  & Olsen et al. \cite{olsen98}\\
NGC~1786 & $0.09\pm0.05$        & $-2.1\pm0.3$          & as old as M68         & Brocato et al. \cite{brocato96} \\
NGC~1835 & $0.08\pm0.02$        & $-1.62\pm0.15 (-1.72 ^a)$ & $10.21 \pm 0.07$  & Olsen et al. \cite{olsen98}\\
NGC~1841 & $0.20\pm0.03$        & $-2.2\pm0.2$          & as old as M68         & Brocato et al. \cite{brocato96} \\
         & $0.18\pm0.02$        & $-2.3 \pm 0.4$        & as old as M92         & Walker \cite{walker90} \\
NGC~1848 & $0.2$ $^3$           & $-0.4$                & $6.7 - 7.0$           & Will et al. \cite{will96} \\
NGC~1850A & $0.18\pm0.02$       & $-0.12\pm0.03 ^1$     & $7.7\pm0.1$           & Gilmozzi et al. \cite{gilmozzi94} \\
        & $0.18$                & $-0.4$                & $7.7\pm0.1$           & Vallenari et al. \cite{vallenari94b} \\
NGC~1850B & $0.18\pm0.02$       & $-0.12\pm0.03 ^1$     & $6.6\pm0.1$           & Gilmozzi et al. \cite{gilmozzi94} \\
        & $0.18$                & $-0.4$                & $7.0$                 & Vallenari et al. \cite{vallenari94b} \\
NGC~1858 & $0.15$               & $-0.4$                & $6.9$                 & Vallenari et al. \cite{vallenari94b}\\
NGC~1866 & $0.03$               & $-0.43\pm0.18$        & $8$                   & Hilker et al. \cite{hilker95b}\\
NGC~1898 & $0.07\pm0.02$        & $-1.37\pm0.15 (-1.37^a)$ & $10.13\pm0.07$     & Olsen et al. \cite{olsen98}\\ 
NGC~1955 & $0.09\pm0.01$        & $-0.4$                & $7.19\pm0.15$         & Dolphin \& Hunter \cite{dolphin98}\\
NGC~1978 & $0.08$               & $-0.4$                & $9.34$                & Bomans et al. \cite{bomans95} \\
NGC~2004 & $0.07\pm0.01$        & $-0.4$                & $7.19\pm0.15$         & Dolphin \& Hunter \cite{dolphin98}\\
         & $0.06$               & $0$                   & $6.9$                 & Benicivenni et al. \cite{beniciveni91}\\ 
NGC~2005 & $0.1\pm0.02$         & $-1.35\pm0.15 (-1.92 ^a)$ & $10.22\pm0.11$    & Olsen et al. \cite{olsen98}\\
NGC~2019 & $0.06\pm0.02$        & $-1.23\pm0.15 (-1.81 ^a)$ & $10.21\pm0.08$    & Olsen et al. \cite{olsen98}\\  
NGC~2027 & $0.05\pm0.01$        & $-0.4$                & $7.06\pm0.14$         & Dolphin \& Hunter \cite{dolphin98}\\
NGC~2121 &                      & $-1.04\pm0.13$        & $9.60\pm0.03$         & Sarajedini \cite{sarajedini98}\\
NGC~2134 & $0.22$               & $-0.4$                & $8.28$                & Vallenari et al. \cite{vallenari94a} \\
NGC~2155 &                      & $-1.08\pm0.12$        & $9.60\pm0.03$         & Sarajedini \cite{sarajedini98}\\
NGC~2164 & $0.10$               & $0$-$-0.4$            & $8$                   & Richtler \& Sagar \cite{richtler91}\\ 
NGC~2210 & $0.09\pm0.03$        & $-2.2 \pm 0.2$        & as old as M68         & Brocato et al. \cite{brocato96} \\
NGC~2214 & $0.07$               & $0$                   & $7.95$                & Bhatia \& Piotto \cite{bhatia94}\\
         &                      & $0$                   & $7.78\pm0.1$          & Lee \cite{lee92}\\
         & $0.07$               & $0$-$-0.4$            & $8$                   & Richtler \& Sagar \cite{richtler91}\\
NGC~2249 & $0.25$               & $-0.4$                & $8.74$                & Vallenari et al. \cite{vallenari94a} \\
\hline
\end{tabular}
\begin{list}{}{}
\item[]{We collected literature which has been published after the work by Sagar \& Pandey
(\cite{sagar89}). 
The metallicities are in several cases from other sources:
$^1$ Jasniewicz \& Thevenin (\cite{jasniewicz94}), $^2$ Olszewski et al. (\cite{olszewski91}), 
$^3$ Schwering \& Israel (\cite{schwering91}). $^a$ The authors quote two values for the cluster, one 
derived with the method by Sarajedini (\cite{sarajedini94}) and one obtained earlier by 
Olszewski et al. (\cite{olszewski91}); the authors prefer the photometrically determined metallicities and
thus only ages according to these metallicities are stated.}
\end{list}
\end{table*}

\begin{thebibliography}{}
\bibitem[1995]{andersen95} Andersen M.J., Freyhammer L., Storm J., 1995,
        Technical Reports on the 1.54m Telescope,\\ 
        www.ls.eso.org/lasilla/Telescopes/2p2T/D1p5M/RepsFinal/
\bibitem[1997]{ardeberg97} Ardeberg A., Gustafsson B., Linde P., et al. 1997,
        A\&A 322, L13
\bibitem[1991]{barbaro91} Barbaro, G., Olivi F.M., 1991,
        AJ 101, 922
\bibitem[1998]{beaulieu98} Beaulieu J.-P., Sackett P.D., 1998
        AJ 116, 209
\bibitem[1978]{bell78} Bell R.A., Gustafsson B., 1978,
        A\&AS 34, 229
\bibitem[1991]{beniciveni91} Bencivenni D., Brocato E., Buonanno R., et al., 1991,
        AJ 102, 137
\bibitem[1994]{padua94} Bertelli G., Bressan A., Chiosi C., et al., 1994,
        A\&AS 106, 275
\bibitem[1993]{bessell93} Bessell M.S., 1993, In
        New Aspects of Magellanic Cloud research, Springer Berlin Heidelberg, 
	ed Baschek B., Klare G., Lequeux J., p. 321
\bibitem[1994] {bhatia94} Bhatia R., Piotto G., 1994,
        A\&A 283, 424
\bibitem[1996]{bica96} Bica E., Clari\'a J. J., Dottori, H., et al., 1996,
        ApJS 102, 57
\bibitem[1998]{bica98} Bica E., Geisler D., Dottori H., et al., 1998,
        AJ 116, 723
\bibitem[1998]{binney98} Binney J., Merrifield M., 1998, 
        Galactic Astronomy, Princeton University Press, Princeton, New Jersey, p.~347 
\bibitem[1995] {bomans95} Bomans D.J., Vallenari A., de Boer K.S., 1995,
        A\&A 298, 427
\bibitem[1996] {brocato96} Brocato E., Castellani V., Ferraro F.R., et al.,  1996,
        MNRAS 282, 614
\bibitem[1982]{burstein82} Burstein A., Heiles C., 1982,
        AJ 87, 1165
\bibitem[1987]{cassatella87} Cassatella A., Barbero J., Geyer E.H., 1987,
        ApJS 64, 83
\bibitem[1996]{cassatella96} Cassatella A., Barbero J., Brocato E., et al.,  1996,
        A\&A 306, 125
\bibitem[1997]{strobel97} Cayrel de Strobel G., Soubiran C., Friel E.D., et al.,  1997,
        A\&AS 124, 299
\bibitem[1970]{crawford70} Crawford D.L., Barnes J.V., 1970,
        AJ 75, 978
\bibitem[1998]{dieball98} Dieball A., Grebel E.K., 1998,
        A\&A 339, 773
\bibitem[1998]{dolphin98} Dolphin A.E., Hunter D.A., 1998,
        AJ 116, 1275
\bibitem[1997]{dopita97} Dopita M.A., Vassiliadis E., Wood P. R., et al., 1997,
        ApJ 474, 188
\bibitem[1987]{elson87} Elson R.A.W., Fall S.M., Freeman K.C., 1987,
        ApJ 323, 54
\bibitem[1997]{elson97} Elson R.A.W., Gilmore G.F., Santiago B.X., 1997,
        MNRAS 289, 157
\bibitem[1990]{frogel90} Frogel J.A., Blanco V.M., 1990,
        ApJ 365, 168
\bibitem[1975]{fry75} Fry M.A., Aller L.H., 1975,
        ApJS 29, 55
\bibitem[1996]{gallagher96} Gallagher J.S., Mould J.R., de Feijter E., 1996,
        ApJ 466, 732
\bibitem[1998]{geha98} Geha M.C., Holtzman J.A., Mould J.R., et al., 1998,
        AJ 115, 1045
\bibitem[1997]{geisler97} Geisler D., Bica E., Dottori H., et al.,  1997,
        AJ 114, 1920
\bibitem[1994]{gieren94} Gieren W.P., Richtler T., Hilker M., 1994,
        ApJ 433, L73
\bibitem[1998]{gieren98} Gieren W.P., Fouqu\'e P., G\'omez M., 1998,
        ApJ 496, 17
\bibitem[1994]{gilmozzi94} Gilmozzi R., Kinney E.K., Ewald S.P., et al., 1994,
        ApJ 435, L43
\bibitem[1999]{girardi99} Girardi L., 1999, astro-ph/9907086
\bibitem[1995]{girardi95} Girardi L., Chiosi C., Bertelli G., et al., 1995,
        A\&A 298, 87
\bibitem[1999]{gonzalez99} Gonzalez G., Wallerstein G., 1999,
        AJ 117, 2286
\bibitem[1992]{grebel92} Grebel E.K., Richtler T., 1992,
        A\&A 253, 359 
\bibitem[1995a]{grebel95a} Grebel E.K., Roberts W.J., 1995a,
        AAS 186, 0308
\bibitem[1995b]{grebel95b} Grebel E.K., Roberts W.J., 1995b,
        A\&AS 109, 293
\bibitem[1979]{gustafsson79} Gustafsson B., Bell R.A., 1979,
        A\&A 74, 313
\bibitem[1999]{hilker99} Hilker M., 1999,
        astro-ph/9911387 
\bibitem[1995a]{hilker95a} Hilker M., Richtler T., Stein D., 1995a,
        A\&A 299, L37
\bibitem[1995b]{hilker95b} Hilker M., Richtler T., Gieren W.P., 1995b,
        A\&A 294, 37 
\bibitem[1999]{hillV99} Hill V., 1999, A\&A 345, 430
\bibitem[1995]{hillV95} Hill V., Andrievsky S., Spite M., 1995,
        A\&A 293, 347
\bibitem[1995]{hill95} Hill R.S., Cheng K.P., Bohlin R.C., et al., 1995,
        ApJ 446, 622 
\bibitem[1997]{holtzman97} Holtzman J.A., Mould J.R., Gallagher J.S. III, et al., 1997,
        AJ 113, 656
\bibitem[1999]{holtzman99} Holtzman J.A., Gallagher III J.S., Cole A.A., et al., 1999,
        astro-ph/9907259
\bibitem[1995] {hunter95} Hunter D.A., Shaya E.J., Holtzman J.A., et al., 1995,
        ApJ 448, 179
\bibitem[1999] {ivans99} Ivans I.I., Sneden C., Kraft R.P., et al.,  1999,
        astro-ph9905370
\bibitem[1994]{jasniewicz94} Jasniewicz G., Th\'evenin F., 1994,
        A\&A 282, 717
\bibitem[1999]{johnson99} Johnson J.A., Bolte M., Stetson P.B., et al., 1999,
        ApJ 527, 199
\bibitem[1993]{jonchsorensen93} J{\o}nch-S{\o}rensen H., 1993,
        A\&A 102, 637
\bibitem[1993]{juettner93} J\"uttner A., Stahl O., Wolf B. et al.,  1993 In
        New Aspects of Magellanic Cloud research, Springer, Berlin Heidelberg, ed. Baschek B., Klare G., Lequeux J., p.337
\bibitem[1993]{kontizias93} Kontizias M., Kontizias E., Michalitsiano A.G., 1993, 
        A\&A 269, 107
\bibitem[2000]{korn00} Korn A.J., Becker S.R., Gummersbach C.A., et al., 2000,
        A\&A 353, 655
\bibitem[1999]{larsen99} Larsen S.S., Richtler T., 1999,
        A\&A 345, 59
\bibitem[1992]{lee92} Lee M.G., 1992,
        ApJ 399, L133
\bibitem[1992]{luck92} Luck R.E., Lambert D.L., 1992,
        ApJS 79, 303
\bibitem[1998]{madore98} Madore B.F., Freedman W.L., 1998,
        ApJ 492, 110
\bibitem[1988]{mateo88} Mateo M., 1988,
        ApJ 331, 261
\bibitem[1986]{mateo86} Mateo M., Hodge P., Schommer R.A., 1986,
        ApJ 311, 113
\bibitem[1984]{mcgregor84} Mc~Gregor P.J., Hyland A.R., 1984
        ApJ 277, 149
\bibitem[1990] {mcwilliam90} Mc~William A., Williams R.E., 1990,
        The Magellanic Clouds, IAUS 148, eds. Haynes R., Milne D., p.~391
\bibitem[1990]{meurer90} Meurer G.R., Freeman K.C., Cacciari C., 1990,
        AJ 99, 1124
\bibitem[1993]{mould93} Mould J.R., Xystus D.A., Da Costa G.S., 1993,
        ApJ 408, 108
\bibitem[1997]{mould97} Mould J.R., Han M., Stetson P.B., Gibson B., et al., 1997,
        ApJ 483, L41 
\bibitem[1995] {oey95} Oey M.S., Massey P., 1995,
        ApJ 452, 210
\bibitem[1999]{olsen99} Olsen K.A.G., 1999,
        AJ 117, 2244
\bibitem[1997]{olsen97} Olsen K.A.G., Hodge P.W., Wilcots E.M., et al., 1997,
        ApJ 475, 545
\bibitem[1998]{olsen98} Olsen K.A.G., Hodge P.W., Mateo M., et al., 1998,
        MNRAS 300, 665
\bibitem[1991]{olszewski91} Olszewski E.W., Schommer R.A., Suntzeff N.B., et al., 1991,
        AJ 101, 515
\bibitem[1998]{pagel98} Pagel B.E.J., Tautvai\u{s}vien\.e G., 1998,
        MNRAS 299, 535
\bibitem[1991]{panagia91} Panagia N., Gilmozzi R., Macchetto F., et al., 1991,
        ApJ 380, L23    
\bibitem[1999]{piatti99} Piatti A.E., Geisler D., Bica E., Clari\'a J.J., et al., 1999,
        astro-ph/9909475
\bibitem[1996] {pilachowski96} Pilachowski C.A., Sneden C., Kraft R.P., et al., 1996,
        AJ 112, 545
\bibitem[1985]{ratnatunga85} Ratnatunga K.U., Bahcall J.N., 1985,
        ApJS 59, 63
\bibitem[1990]{reitermann90} Reitermann A., Stahl O., Wolf B., Baschek B., 1990,
        A\&A 234, 109
\bibitem[1999]{richter99} Richter P., Hilker M., Richtler T., 1999,
        astro-ph/9907200
\bibitem[1990]{richtler90} Richtler T., 1990,
        A\&AS 86, 103
\bibitem[1989]{richtler89} Richtler T., Spite M., Spite F., 1989,
        A\&A 225, 351
\bibitem[1991]{richtler91} Richtler T., Sagar R., 1991,
        A\&A 250, 324
\bibitem[1999]{romaniello99} Romaniello M., Panagia N., Scuderi S., 1999,
        astro-ph/9908144
\bibitem[1989]{russell89} Russell S.C., Bessell M.S., 1989,
        ApJS 70, 865
\bibitem[1992]{russell92} Russell S.C., Dopita M.A., 1992
        ApJ 384 , 508
\bibitem[1989]{sagar89} Sagar R., Pandey A.K., 1989,
        A\&AS 79, 407
\bibitem[1991]{sagar91} Sagar R., Richtler T., 1991,
        A\&A 250, 324
\bibitem[1999]{santos99} Santos Jr., J.,F.C., Piatti A.E., Clari\'a J.J., et al., 1999,
        AJ 117, 2841
\bibitem[1998] {santiago98} Santiago B.X., Elson R.A.W., Sigurdsson S., et al. , 1998,
        MNRAS 295, 860
\bibitem[1994] {sarajedini94} Sarajedini A., 1994,
        AJ 107, 618
\bibitem[1998] {sarajedini98} Sarajedini A., 1998,
        AJ 116, 738
\bibitem[1993]{geneva93} Schaerer D., Charbonnel C., Meynet G., et al., 1993,
        A\&AS 102, 339
\bibitem[1998]{schlegel98} Schlegel D., Finkbeiner D., Davis M., 1998,
        ApJ 500, 252
\bibitem[1991]{schwering91} Schwering P.B.W., Israel F.P., 1991,
        A\&A 246, 231
\bibitem[1984]{stryker84} Stryker L.L., 1984,
        ApJS 55, 127
\bibitem[1993]{subramaniam93} Subramaniam A., Sagar R., 1993
        A\&A 273, 100
\bibitem[1995]{testa95} Testa V., Ferraro F.R., Brocato E., et al., 1995,
        MNRAS 275, 454
\bibitem[1992]{thevenin92} Th\'evenin F., Jasniewicz G., 1992,
        A\&A 266, 85
\bibitem[1994a] {vallenari94a} Vallenari A., Aparicio A., Fagotto F., et al., 1994a,
        A\&A 284, 424
\bibitem[1994b] {vallenari94b} Vallenari A., Aparicio A., Fagotto F., et al., 1994b,
        A\&A 284, 447
\bibitem[1996a]{vallenari96a} Vallenari A., Chiosi C., Bertelli G., et al., 1996a,
        A\&A 309, 367
\bibitem[1996b]{vallenari96b} Vallenari A., Chiosi C., Bertelli G., et al., 1996b,
        A\&A 309, 358
\bibitem[1990]{walker90} Walker A.R., 1990,
        AJ 100, 1532
\bibitem[1993]{walker93} Walker A.R., 1993,
        AJ 106, 999
\bibitem[1993]{walker99} Walker A.R., Schommer R.A., Suntzeff N.B., et al., 1993, In
        New Views of the Magellanic Cloud, IAUS 190, eds. Y.-H. Chu, N.B. Suntzeff, J.E. Hesser, et al., p.341
\bibitem[1997] {westerlund97} Westerlund B.E., 1997, The Magellanic Clouds,
        Cambridge University Press, Cambridge 
\bibitem[1996] {will96} Will J.-M., Bomans D.J., Vallenari A., et al., 1996,
        A\&A 315,125
\bibitem[1997]{zaritsky97} Zaritsky D., Lin D.N.C., 1997,
        AJ 114, 2545
\end{thebibliography}
\end{document}